\documentclass[11pt]{article}
\usepackage[utf8]{inputenc}
\usepackage{amsmath, booktabs, tabularx, mathtools}
\usepackage[left=1in,top=1in,right=1in,bottom=1in]{geometry}
\usepackage{titlesec}
\usepackage{placeins}
\usepackage{enumitem}
\usepackage{verbatim}
\usepackage{amsfonts}
\usepackage{textcomp}

\usepackage{hhline}
\usepackage{multirow}

\usepackage[left=1in,top=1in,right=1in,bottom=1in]{geometry}
\usepackage{titlesec}
\usepackage{enumitem}

\titleformat{\section}
  {\normalfont\fontsize{11}{15}\bfseries}{\thesection}{1em}{}
\titleformat{\subsection}
  {\normalfont\fontsize{11}{15}\bfseries}{\thesubsection}{1em}{}
\titleformat{\subsubsection}
  {\normalfont\fontsize{11}{15}\bfseries}{\thesubsubsection}{1em}{}
\titlespacing*\section{0pt}{0pt}{0pt}
\titlespacing*\subsection{0pt}{0pt}{0pt}
\titlespacing*\subsubsection{0pt}{0pt}{0pt}

\usepackage{accents}

\DeclareMathOperator*{\argmin}{arg\,min}
\newcommand{\E}{\mathbb{E}}

\renewcommand{\thesubsection}{\arabic{section}.\arabic{subsection}}
\renewcommand{\thesection}{\arabic{section}}
\renewcommand{\thesubsubsection}{\arabic{section}.\arabic{subsection}.\alph{subsubsection}}

\usepackage{authblk}
\title{A Machine Learning Approach to Measuring Climate Adaptation}
\author{Max Vilgalys\footnote{ {\tt max.vilgalys@erg.com} Eastern Research Group, Inc. }}

\date{}   
\usepackage[bibencoding=auto,backend=biber,natbib, style=authoryear, maxbibnames=9, maxcitenames=3, isbn=false, doi=false, url=false ]{biblatex}
\usepackage[hidelinks]{hyperref}
\usepackage{cleveref}
\usepackage{caption,subcaption}
\usepackage{setspace}
\begin{document}

\setlength{\parskip}{0pt}
\setlength{\abovedisplayskip}{0pt}
\setlength{\belowdisplayskip}{0pt}
\setlength{\abovedisplayshortskip}{0pt}
\setlength{\belowdisplayshortskip}{0pt}

\maketitle
\begin{abstract}
% TODO: make every sentence better.
% Is it possible to phrase this so that it doesn't just sound like you're conducting robustness checks on the literature? Only one way to find out... 
I measure adaptation to climate change by comparing elasticities from short-run and long-run changes in damaging weather. I propose a debiased machine learning approach to flexibly measure these elasticities in panel settings. In a simulation exercise, I show that debiased machine learning has considerable benefits relative to standard machine learning or ordinary least squares, particularly in high-dimensional settings. I then measure adaptation to damaging heat exposure in United States corn and soy production. Using rich sets of temperature and precipitation variation, I find evidence that short-run impacts from damaging heat are significantly offset in the long run. 
I show that this is because the impacts of long-run changes in heat exposure do not follow the same functional form as short-run shocks to heat exposure.

\paragraph{Keywords:} Climate change, directional derivative,  machine learning, panel data 
\paragraph{JEL Classification:} C14, C33, C55, Q51, Q54
\end{abstract}
% Without a structural model of adpatation, it is not clear which of these estimates is preferred.

\providecommand{\myoutputs}{}
\renewcommand{\myoutputs}{"./outputs/"}

\section{Introduction}
Measuring adaptation to recent climate change is important for informing climate policy and projecting damages of future climate change. 
The extent of recent adaptation can signal when policy interventions are needed and give a sense of how much adaptation will be possible to projected changes. 
Researchers can study this by modeling how weather shocks impact economic outcomes, and examining how impacts change with more exposure to the weather shock (e.g. \citealt{Burke2016}) or over time (e.g. \citealt{Barreca2016}). 
These studies depend on accurate models of the relationship between weather and economic outcomes. 

Machine learning (ML) can help model these relationships when researchers have rich data, but lack expert guidance to suggest a functional form. 
When domain experts provide a model of how weather shocks impact economic outcomes, researchers can use this to accurately measure and compare impacts. 
Otherwise the researcher must learn this relationship from data. 
Economists typically use classic statistical tools to model weather-economic relationships without imposing strong functional form assumptions \citep{Hsiang2016}. 
Such tools work well with low-dimensional weather variation, but can lead to high variance or inconsistent estimates as the dimensionality of covariates increases.
Even a model with only temperature and precipitation can become high-dimensional if a researcher flexibly models each variable and the interactions between them. 
% Even considering only temperature and precipitation, classical tools can require massive datasets to flexibly model each variable and the interactions between them.
% As an example, these tools can model the nonlinear impacts of temperature \citep{Schlenker2009,Deschenes2011,Barreca2016} but 
% One common approach is to flexibly measure the impacts of temperature by separate bins
ML is well suited for flexible modeling in such settings \citep{Mullainathan}, and can help researchers take full advantage of high-dimensional weather variation in modern panel datasets. 

I introduce an ML approach to study adaptation to damaging heat exposure in United States (U.S.) corn and soy production. 
High temperatures are generally damaging for crop growth, although adaptation may offset some of these damages in the long run. 
\citet{Schlenker2006} introduce a parsimonious model of how an annual shock of heat exposure impacts crop yield. 
They model crop yields as a piecewise-linear function of heat exposure below and above a crop-specific temperature threshold, where heat exposure below the threshold is beneficial and heat exposure above the threshold is damaging. 
By estimating the parameter on damaging heat using different sources of variation, \citet{Schlenker2009}, \citet{Burke2016}, and \citet{Lemoine2018} argue that the observed degree of adaptation to damaging heat exposure will not be sufficient to offset projected climate damages.

I measure the degree of adaptation after learning the crop yield-weather relationship from data. 
I compare impacts from short-run and long-run changes in damaging heat exposure. 
With a low-dimensional linear model, this is equivalent to the method from \citet{Burke2016}. 
The approach allows me to incorporate more flexible models and high-dimensional weather variation, making the method suitable for applications with rich weather variation and little expert guidance on how that variation impacts economic outcomes.
I use ML to model the crop yield-weather relationship, specifically Least Absolute Shrinkage and Selection Operator (Lasso) and a neural network (NNet). 
My ML models account for additive fixed effects while flexibly modeling temperature, precipitation, and interactions between them.

To estimate the degree of adaptation, I estimate the elasticity of crop yields with respect to damaging heat exposure. 
This elasticity summarizes how damaging heat exposure impacts crop yields, and can be computed for each model. 
There is no single parameter I can compare across models, as in \citet{Burke2016}.
Instead, I find the elasticity by fitting a regression function of log crop yields on temperature and precipitation and computing the average directional derivative in the direction of a marginal increase in damaging heat exposure.
I estimate these regression functions via ordinary least squares (OLS) and ML. 
% This allows me to measure the impact of damaging heat exposure for a general function and a rich set of weather variation. 

I implement a debiasing procedure to address bias from standard ML models. 
ML approaches can induce bias from overfitting or regularization, but there are approaches to reduce this bias when estimating causal parameters or statistics based on regression functions \citep{Chernozhukov2018,chernozhukov2022automatic}.
I adapt the estimator from \citet{chernozhukov2022automatic}. 
This approach uses double machine learning (DML), where standard ML estimates are corrected using a second ML algorithm. 
I adjust the second ML algorithm to suit the panel setting. 
I then apply this DML procedure to debias the elasticity estimates. 

I compare elasticities to estimate the degree that short-run impacts from damaging heat exposure are offset in the longer term.
I construct panel datasets of long-run and short-run variation in crop yield and weather variables from 1990-2019. 
I then estimate the elasticity of crop yield with respect to damaging heat exposure in each dataset. 
I compute this elasticity using OLS, DML, and ML without bias correction. 
Comparing these estimates, I find the extent that short-run impacts from damaging heat exposure are offset in the longer term.

Before taking this approach to the data, I conduct a simulation exercise to evaluate DML estimates of an elasticity relative to OLS and standard ML approaches. 
Each simulation trial uses the empirical distribution of temperature and precipitation from U.S. counties, but simulates outcome variables. 
I examine the performance of my DML procedure relative to OLS and ML without bias correction, comparing the bias and variance of recovering the true elasticity. 
I use three potential sets of weather variables: a simple, commonly used set of annual temperature and precipitation variables, a set that includes richer variation in annual temperature, and a set that includes monthly observations of precipitation and the rich variation in temperature. 

This simulation exercise clearly highlights the benefits of using debiased machine learning for high-dimensional settings.
Both NNet and Lasso have advantages over OLS in high-dimensional cases, although NNet has lower bias. 
With the rich set of annual temperature variation, DML estimates have significantly lower variance than OLS and lower bias than standard ML estimates. 
With the set of monthly temperature and precipitation, DML estimates have significantly less bias and variance than OLS, and lower bias than standard ML.

I then apply the approach to a dataset of U.S. crop yields, and find evidence that adaptation is offsetting impacts from damaging heat exposure. 
I implement Lasso, NNet, and OLS estimators on a county-level dataset of corn and soy yields from 1990 to 2019. 
I compare estimates of the elasticity using short-run and long-run variation.
I take 500 bootstrap trials of the estimated elasticity, using the three sets of weather variation as in the simulation exercise. 
Using the simple annual set of weather variables as in \citet{Schlenker2009} and \citet{Burke2016}, I find that there has been little to no significant adaptation to climate change in corn or soy production.
This confirms the results from \citet{Burke2016}. 

Using more flexible sets of weather variation, I find that a large share of short-run impacts from damaging heat exposure are offset in the long run.
With short-run variation, I find statistically and economically significant declines in yield from a marginal increase in damaging heat exposure. 
However, I do not find evidence of such declines when using long-run variation.
These results hold for both corn and soy. 
The primary difference comes from using a richer set of weather variables. 
I make this same conclusion using OLS with the more flexible set of weather variation, although the DML approach results in smaller confidence intervals. 
This shows that a substantial degree of the short-run impacts from damaging heat exposure are offset in the long run, suggesting substantial adaptation to this heat exposure.

% Flexible modeling shows that there is more adaptation to damaging heat exposure than was previously believed. 
This result differs dramatically from the conclusions by \citet{Burke2016} and other analyses \citep{Schlenker2009,Lemoine2018}. 
This is likely explained by model misspecification for the impact of a long-run shift in heat exposure on crop yields. 
% Rather than only say that you can replicate their results with their model and get different results with your model, can you isolate the portion of your DML flexible model that is linear, and tell us, WITHIN YOUR MODEL, how much of the results is from the linear portion that is the same as Burke & Emerick, and how much of the results come from additional flexible terms. After that, perhaps you can explain what those additional flexible weather variables are and whether it is sensible for them to produce these different results?
I show that in the panel with long-run variation, the simple model from \citet{Schlenker2006} does not adequately summarize the flexible role of temperature variables.
While damaging heat exposure is correlated with declines in crop yield, other temperature variation is better able to explain these declines.
This suggests that there is limited adaptation to some damaging feature of climate change, but not to the specific feature of marginal increase in damaging heat exposure.

This paper is related to several literatures. First is a literature on estimating the degree of adaptation to climate change. 
Measuring adaptation to climate change requires understanding how weather influences economic outcomes. 
For a review of economics literature on measuring the economic impacts of the weather, see \citet{Dell2014}.
\citet{Hsiang2016} provides an overview of econometric approaches to measuring these impacts. 
Much of this literature focuses on agriculture, as this sector is directly exposed to weather and hence is particularly vulnerable to climate change \citep{shukla2019ipcc}.
The first approach to studying impacts of climate change used the Ricardian approach, where researchers compare the value of agricultural land in cross sections.
\citet{Mendelsohn1994} forecast the impacts of climate change by regressing average temperature and agricultural property value in a cross section of U.S. counties.
This approach is susceptible to omitted variable bias, and subsequent work has focused on addressing specific omitted variables such as endogenous changes in farmer technology \citep{kurukulasuriya2011adaptation} or nonfarm income \citep{ortiz2020role}. 
Other approaches to estimate the potential for adaptation in agriculture involve economy-wide simulations \citep{Costinot2016}, production changes in historical migrations \citep{Sutch2011,Olmstead2011}, natural experiments \citep{Hornbeck2012,hagerty2021adaptation}, or panel approaches.

Panel approaches address omitted variable bias by identifying adaptation from annual or long-term variation within a panel dataset. 
\citet{Schlenker2009} uses panel data to estimate the elasticity of crop yields with respect to extreme heat exposure, and conclude that there is limited potential to adapt to climate change because these damages are similar in the southern and northern U.S. despite climatic differences.
\citet{Barreca2016} use a flexible model of temperature exposure to document how the mortality consequences of extreme heat declined over the 20$^\text{th}$ century. 
\citet{Burke2016} show that panel variation can be used to estimate adaptation to recent climate change by using separate sources of variation to identify the impacts of weather shocks and shifts in average temperature. 
\citet{Lemoine2018} provides an alternate approach that partially identifies the degree of possible adaptation by considering the role of ex-ante and ex-post adaptation to heat exposure shocks.

My paper is most closely related to \citet{Burke2016}.
Like their paper, I estimate the degree that damages to corn and soy yields from short-run changes in weather are offset over longer exposures. 
I also use crop and weather data from U.S. agriculture. 
My approach differs because I consider richer sets of weather variables, and use DML to model learn the relationship between these data and crop yields.
I conclude that there has been a higher degree of adaptation to damaging heat exposure. 

Second is a growing literature on applying ML methods in economics. 
\citet{Kleinberg2015} discuss applications of predictive machine learning in economics, and \citet{varian2014big} and \citet{Mullainathan} provide a practical guide to algorithms.
Several recent papers have used ML to measure important outcomes in environmental economics. 
\citet{Crane-Droesch2018} proposes a semi-parametric NNet and uses it to study the impact of climate change on corn yields.
\citet{Deryugina2019} uses a ML approach to measure the costs of air pollution.
\citet{burlig2020machine} use ML to refine estimates of energy efficiency improvements. 
\citet{stetter2022using} use a DML approach to measure effectiveness of an agricultural intervention, and \citet{klosin2022} introduce a DML approach to measure elasticities in a panel setting. 
There are also numerous applications within agriculture; for a review, \citet{Liakos2018}.

My paper is most related to \citet{Crane-Droesch2018} and \citet{klosin2022}.
\citet{Crane-Droesch2018} estimates a NNet that accounts for unobservable county-level fixed effects, and uses the model to predict yield under counterfactual climate change scenarios.
% \citeauthor{Crane-Droesch2018} compares the performance of this estimator to standard ML approaches and to OLS, and concludes that there are significant benefits for predicting crop yields. 
I use the same technique to address county-level fixed effects, although I modify the algorithm in order to recover derivatives from the network and to ensure that the network is differentiable.
% , and to weight the penalty term per observation. 
\citeauthor{Crane-Droesch2018} studies the impacts of future climate change, while I use this tool to understand adaptation to recent climate change. 
Like \citet{klosin2022}, I estimate the elasticity of crop yield with respect to an increase in damaging heat exposure. 
I consider more flexible representations of temperature variation, and I apply the estimator to measure adaptation to recent climate change. 

Within the literature on machine learning, this paper applies results from the emerging field of DML.
\citet{Chernozhukov2018} prove that sample splitting and constructing Neyman-orthogonal moment conditions can yield approximately debiased machine learning estimates in certain settings. 
\citet{Semenova2021} extend the Neyman-orthogonal moment condition approach to several other statistical targets, including structural derivatives.
In an alternate approach, \citet{chernozhukov2022automatic,chernozhukov2022debiased} give an approximately debiased estimator for a more general class of linear functionals based on the Riesz representation theorem.
\citet{klosin2022} adopt this approach in panel settings and prove asymptotic normality of the estimator for the average derivative. 
Like \citet{klosin2022}, my approach applies the result from \citet{chernozhukov2022automatic} in panel settings; I use a different approach to address fixed effects and consider a more flexible representation of temperature variation. 

The rest of this paper proceeds as follows. 
In \Cref{sec:Data}, I describe the data used for this project. I illustrate the degree of climate change and describe the transformations required to generate the growing degree days. 
In \Cref{sec:Methods}, I describe the methods used; this includes details on how to compute average derivatives and an explanation of the debiased machine learning estimation approach. 
In \Cref{sec:Simulation}, I give details and results of the simulation exercise. 
In \Cref{sec:Results}, I present and discuss results from using the estimation procedure to measure the degree of adaptation to climate change. 
\Cref{sec:Discussion} concludes. 

\FloatBarrier
\section{Data}
\label{sec:Data}
For the empirical application, I use weather and crop data from U.S. corn and soy production from 1990-2019. 
I consider counties east of the 100\textdegree  West meridian, which defines an agricultural region of significant corn and soy cultivation.
From 1990-2019, this region produced over 93\% of the nation's corn and over 99\% of the nation's soy. 
Crop data are annual yield (bushels per acre) of corn and soy, from the U.S. Department of Agriculture's Survey of Agriculture.
These data also include the area planted (acres) in each county. 

\begin{figure}[t]
     \centering
     \begin{subfigure}[b]{0.25\textwidth}
         \centering
         \includegraphics[width=\textwidth]{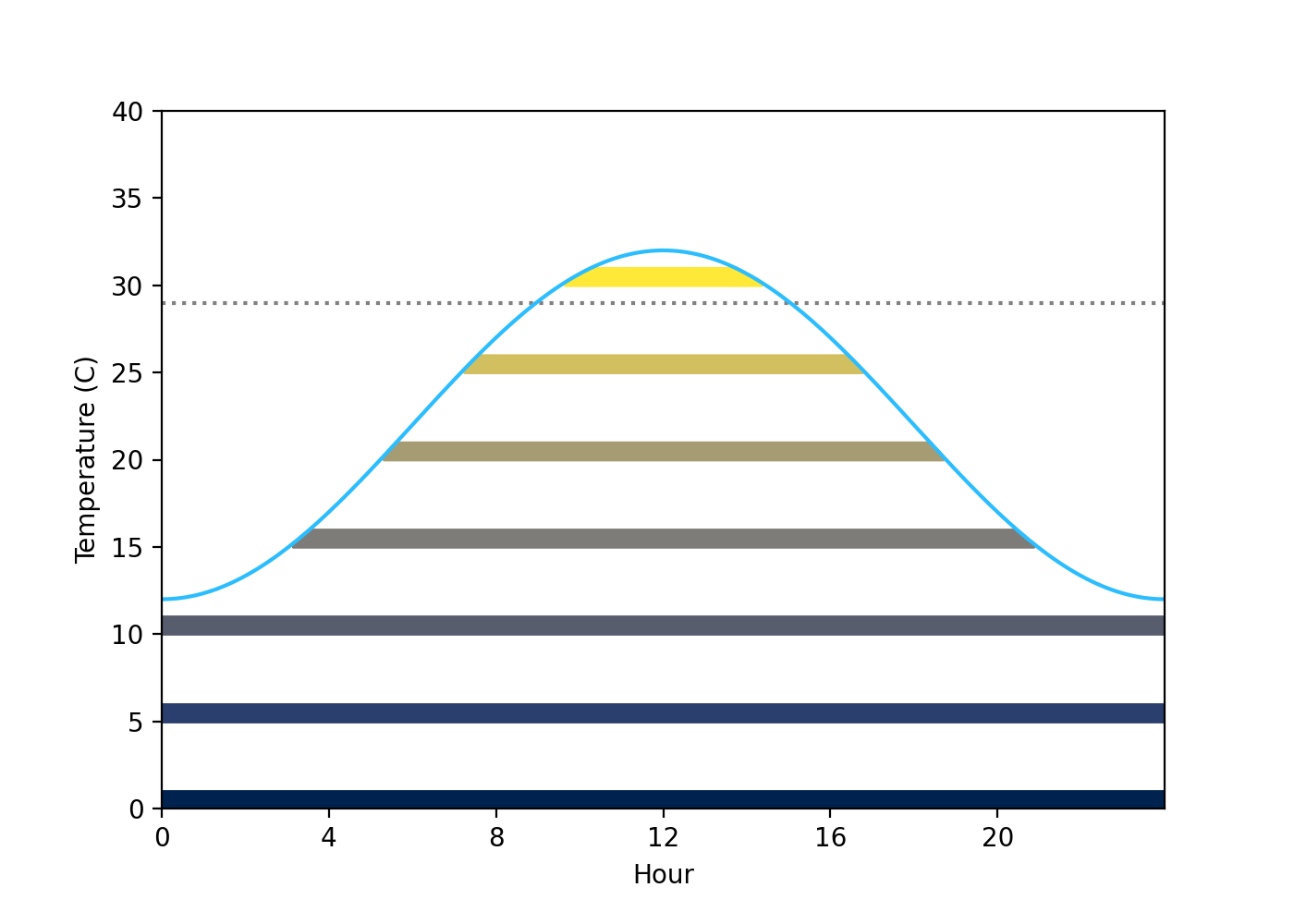}
         \caption{}
         \label{fig:temp 1 day}
     \end{subfigure}
     \hfill
     \begin{subfigure}[b]{0.25\textwidth}
         \centering
         \includegraphics[width=\textwidth]{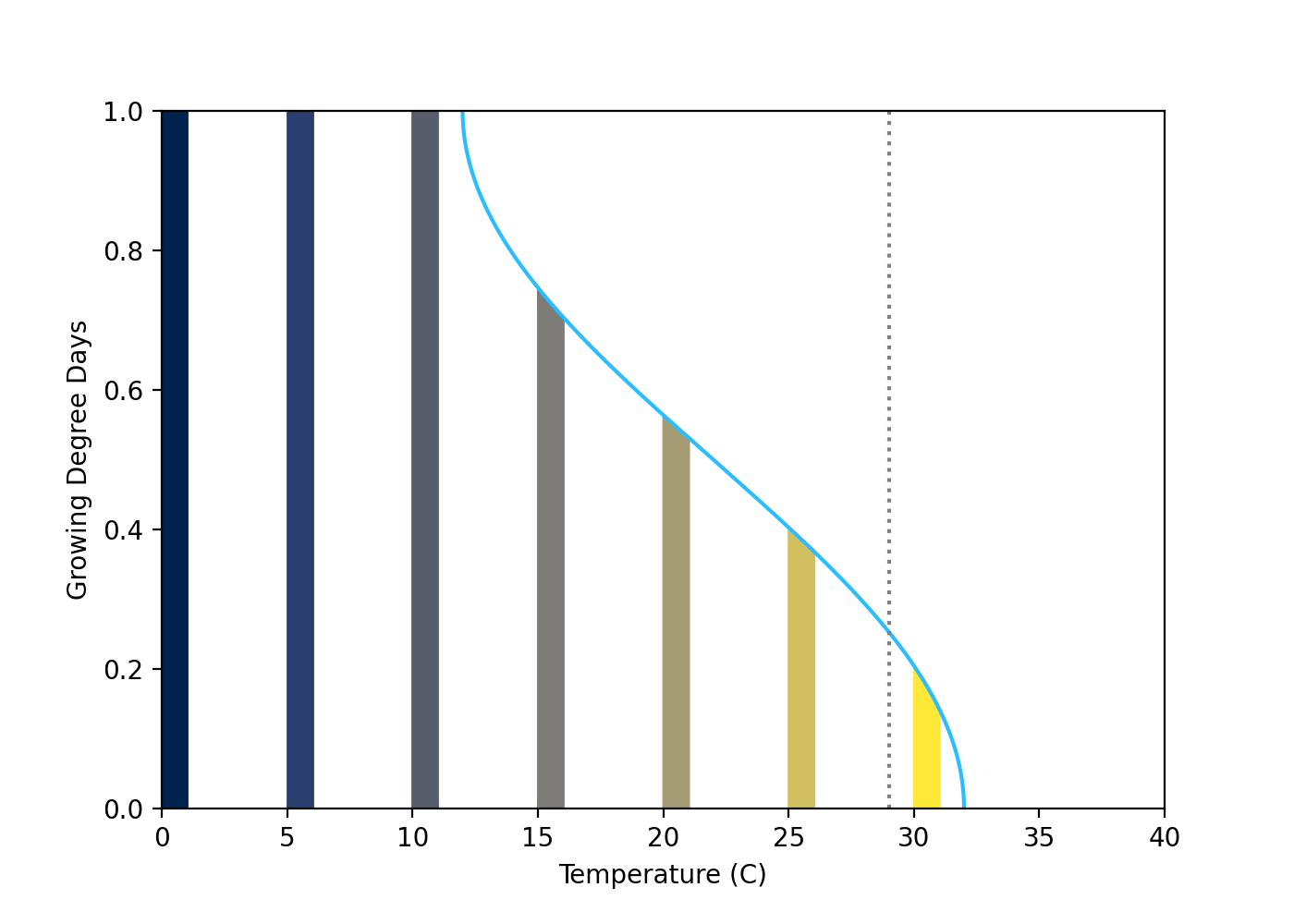}
         \caption{}
         \label{fig:temp 1 day stacked}
     \end{subfigure}
     \hfill 
     \begin{subfigure}[b]{0.45\textwidth}
         \centering
         \includegraphics[width=\textwidth]{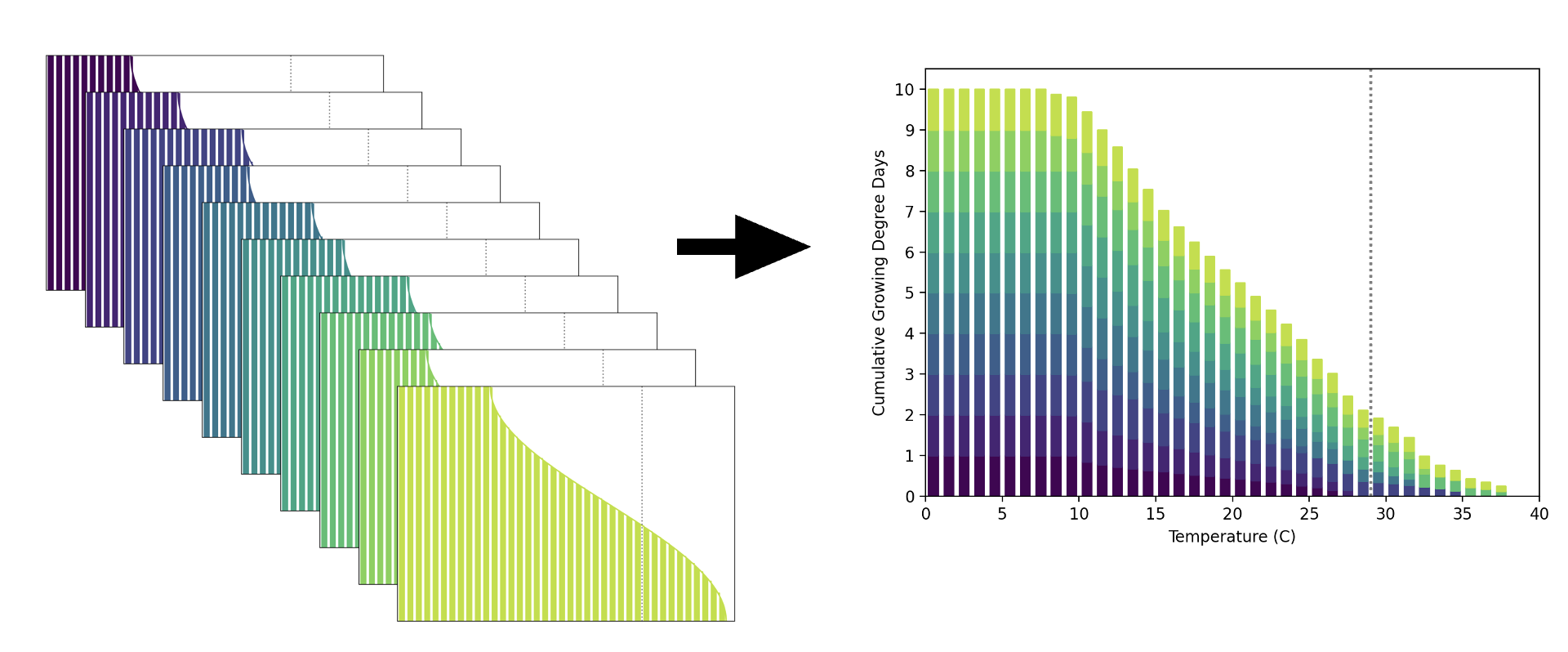}
         \caption{}
     \end{subfigure}
     \newline
      \begin{subfigure}[b]{0.305\textwidth}
         \centering
         \includegraphics[width=\textwidth]{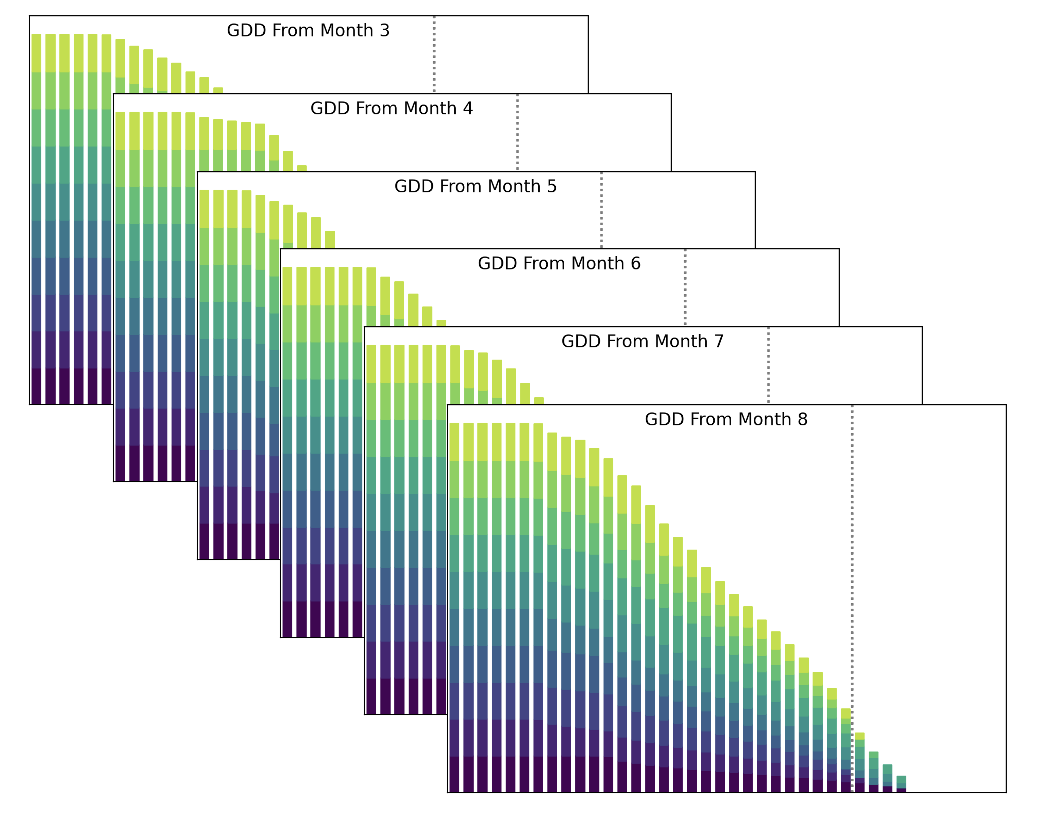}
         \caption{Monthly Flexible}
     \end{subfigure}
           \begin{subfigure}[b]{0.305\textwidth}
         \centering
         \includegraphics[width=\textwidth]{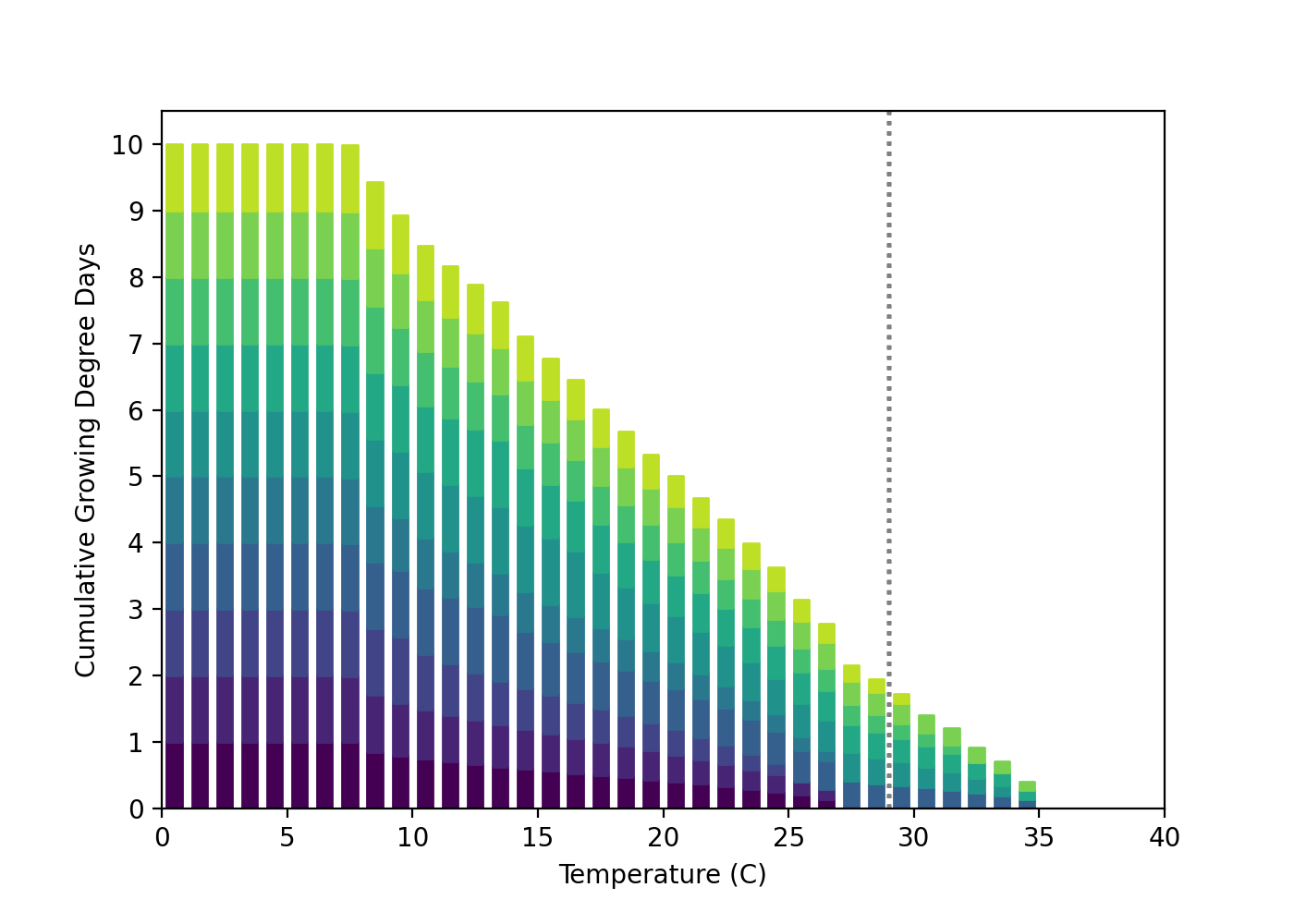}
         \caption{Yearly Flexible}
     \end{subfigure}
           \begin{subfigure}[b]{0.305\textwidth}
         \centering
         \includegraphics[width=\textwidth]{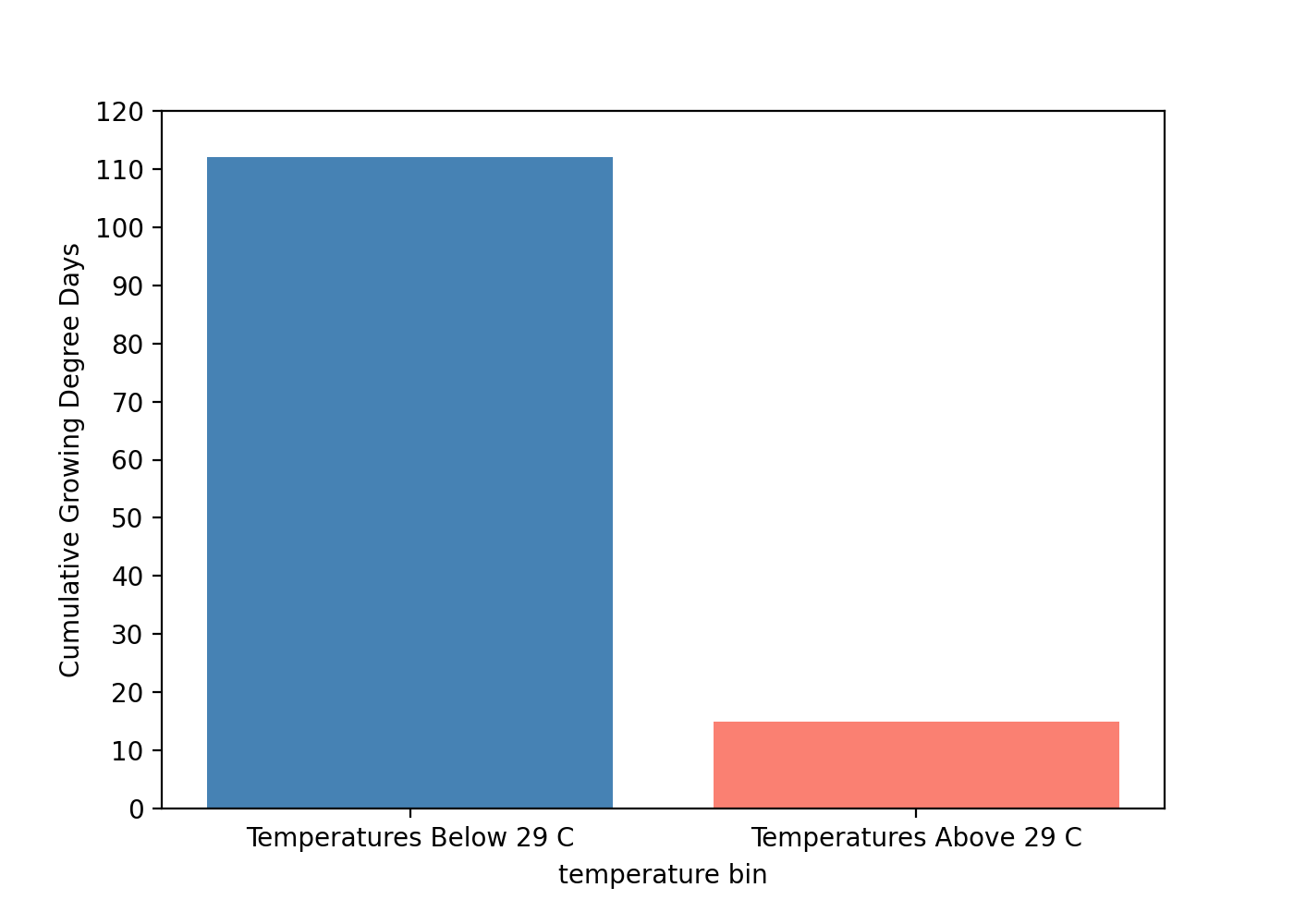}
         \caption{Yearly Linear}
         \label{fig:Yearly Linear}
     \end{subfigure}
        \caption{Transformations from daily minimum and maximum temperature records into the weather variables used in the analysis. (a) shows daily temperature within one day. In (b), this temperature is translated into growing degree days (GDD). (c) illustrates how daily GDD observations are aggregated into cumulative GDD exposure. (d) is the Monthly Flexible variable, where daily temperature is aggregated into GDD in each temperature bin for each month. In (e), this is aggregated into total growing season heat exposure in each temperature bin. In (f), this is further aggregated into total growing season heat exposure above and below the dotted line, 29\textdegree C. }
\label{fig:GDD_explainer}
\end{figure}

Weather data are generously shared by \citeauthor{Schlenker2009}, who provide a gridded dataset of daily temperature and precipitation from a network of consistently reporting weather stations. 
As in \citet{Schlenker2009} and \citet{Ortiz-Bobea2013}, I consider weather during the March-August growing season.
I aggregate the gridded dataset to a county-level dataset of daily maximum temperature, minimum temperature, and precipitation.
I then transform daily temperature exposure into growing degree days (GDD) at a monthly level and for the growing season.
\Cref{fig:GDD_explainer} illustrates the transformation from daily temperature observations to GDD.

\providecommand{\myimages}{}
\renewcommand{\myimages}{"./images/"}

Throughout this paper, I consider three sets of weather variables per county: total growing season heat exposure above and below 29\textdegree C plus total growing season precipitation (Yearly Linear), total growing season heat exposure in each 1\textdegree C temperature bin plus total growing season precipitation (Yearly Flexible), and monthly heat exposure in each 1\textdegree C bin and monthly precipitation for each month of the growing season (Monthly Flexible). 
\Cref{fig:GDD_explainer} illustrates the transformations of temperature variables.
The Yearly Linear transformation, shown in \Cref{fig:Yearly Linear}, is widely used in economic analysis. 
\citet{Schlenker2009} demonstrate that, in a panel setting, regression based on heat exposure above and below a crop-specific damaging threshold explains as much variation in yields as more flexible models. 
\citet{Burke2016} identify 29\textdegree C as the threshold for damaging heat exposure in the long run, for both corn and soy. 
I use the 29\textdegree C threshold throughout the analysis, and refer to heat exposure above 29\textdegree C as damaging heat exposure. 

\begin{figure}[t]
\centering
     \begin{subfigure}[b]{0.31\textwidth}
         \centering
         \includegraphics[width=\textwidth]{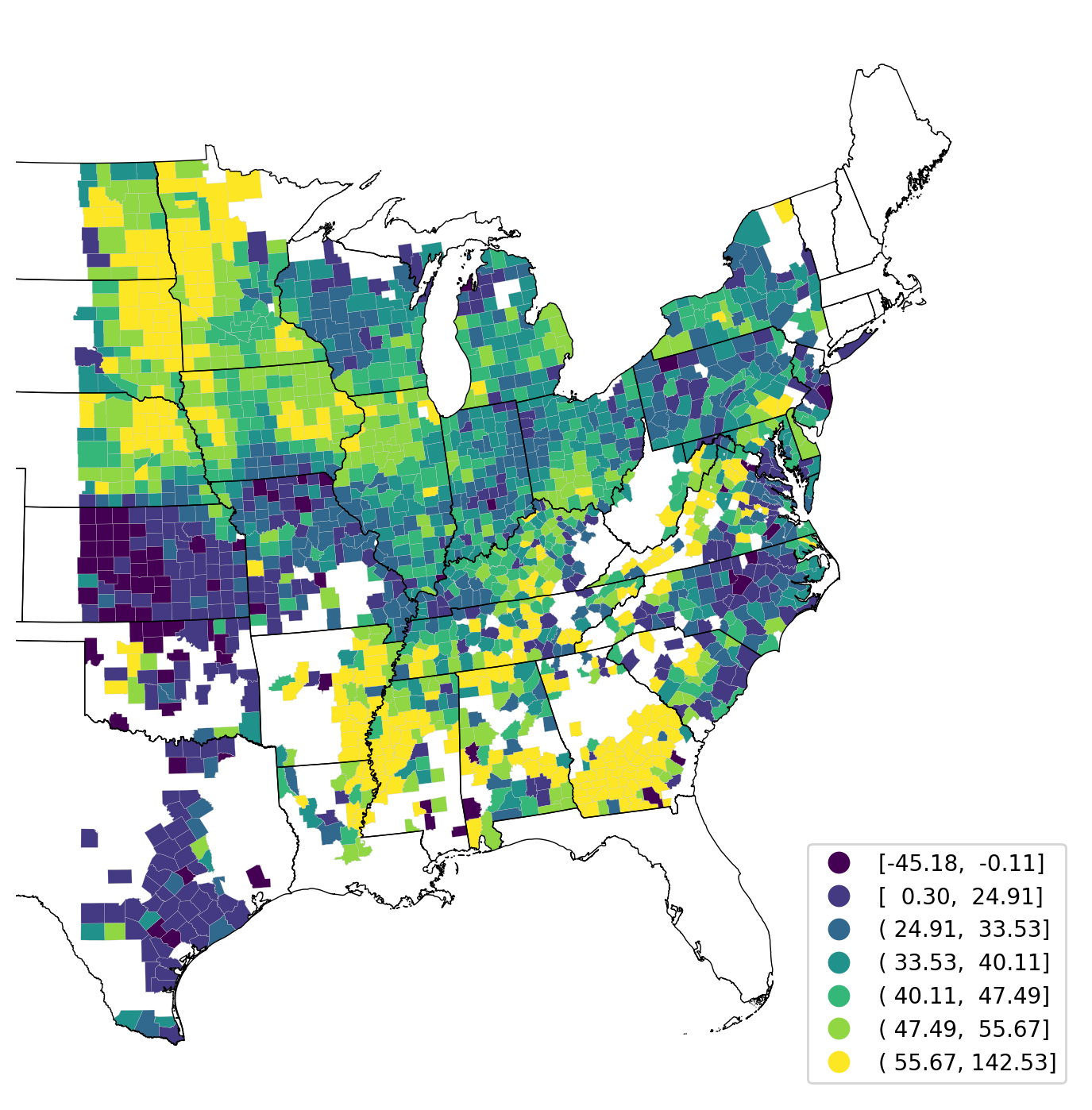}
         \caption{Corn yields}
         \label{fig:corn diff map}
     \end{subfigure}
     \begin{subfigure}[b]{0.31\textwidth}
         \centering
         \includegraphics[width=\textwidth]{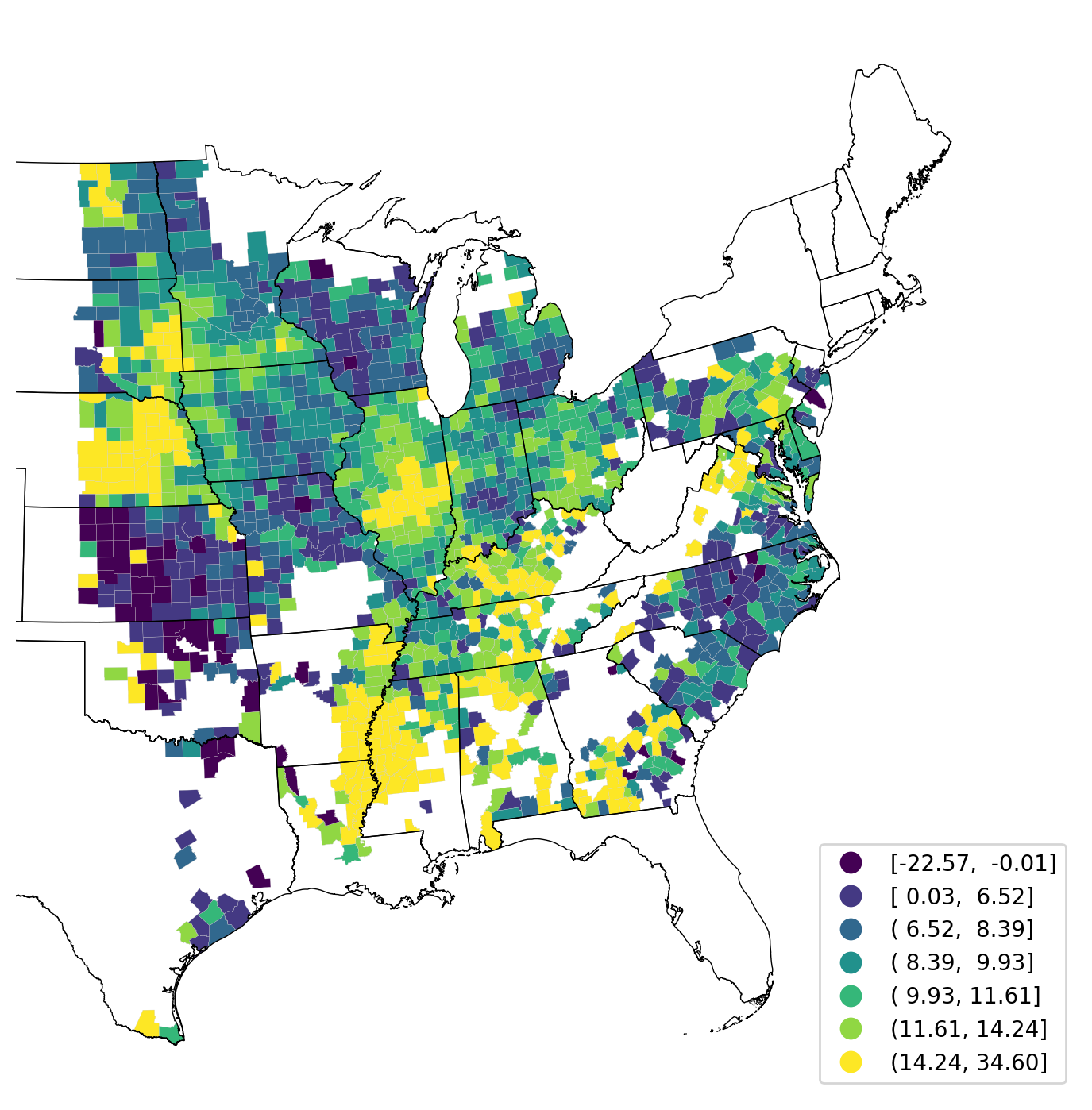}
         \caption{Soy yields}
         \label{fig:soy diff map}
     \end{subfigure}
          \begin{subfigure}[b]{0.31\textwidth}
         \centering
         \includegraphics[width=\textwidth]{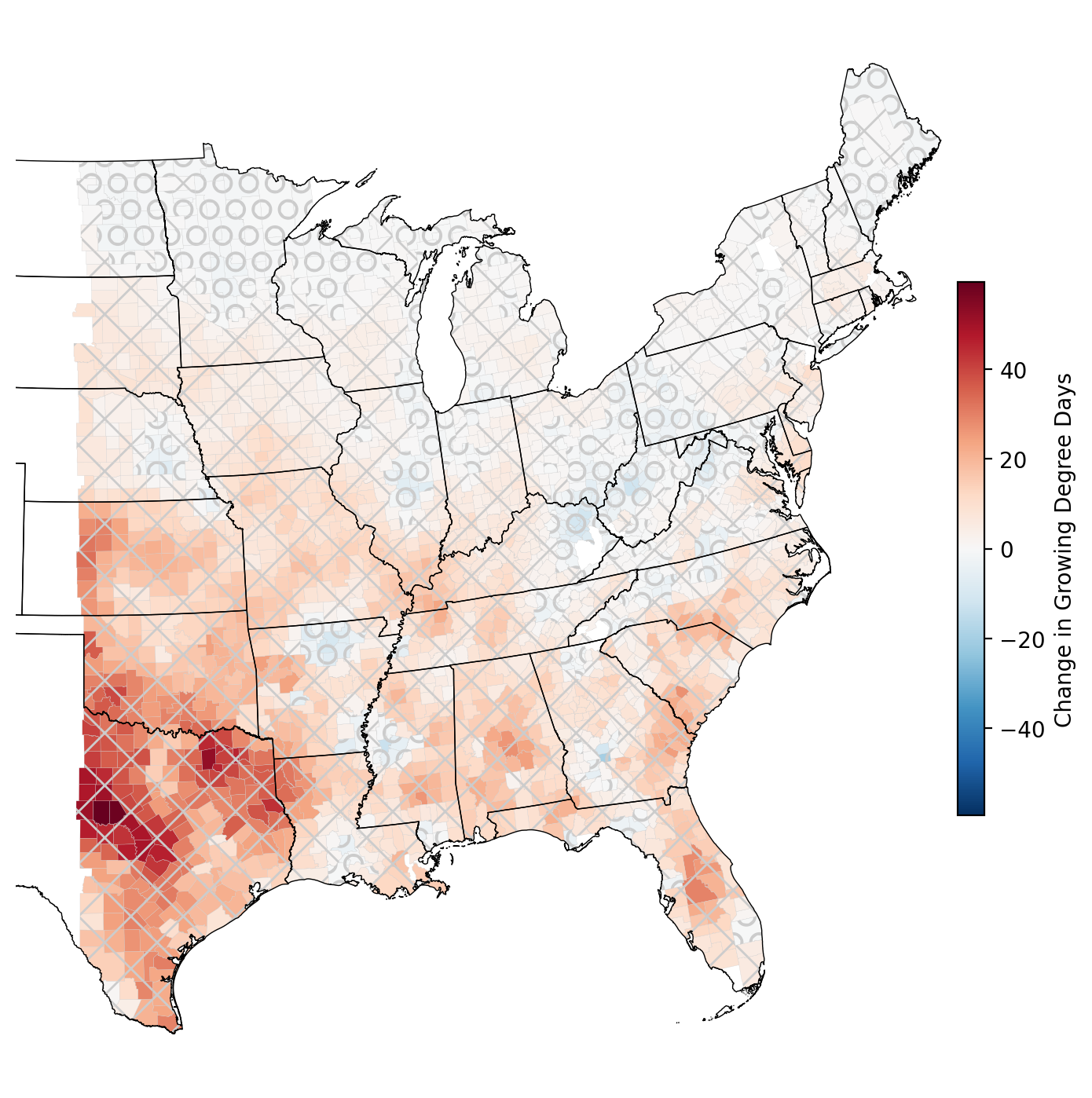}
         \caption{GDD above 29\textdegree C}
         \label{fig:higher diff map}
     \end{subfigure}
        \caption[Maps of differences]{Differences in county-level values of corn yields, soy yields, and growing-season GDD above 29\textdegree C,  between 1990-1999 averages and 2010-2019 averages. 
        Maps include counties east of the 100 \textdegree W meridian. 
        In (a) and (b), the first color bin includes all counties with yield declines (roughly 4\% of counties) and the remaining distribution is divided evenly among the remaining bins.  In (c), `X' (`O') hatching indicates counties with an increase (decrease) in heat exposure above 29\textdegree C. }
\label{fig:Difference maps}
\end{figure}

Variation in the degree of climate change between counties is used to estimate the long-run elasticities in our sample. 
\Cref{fig:higher diff map} shows illustrates the difference in GDD exposure above 29\textdegree C per US county, from the period 1990-1999 to the period from 2010-2019. 
Heat exposure increases in 80\% of counties in the sample, but there is substantial variation in the degree of warming even within states. 
There are many pairs of neighboring counties where one experienced an increase in damaging heat exposure and the other experienced a decrease.
This variation is plausibly uncorrelated with unobservable factors such as soil quality or per-state fixed effects. 
See \citet{Burke2016} for a detailed argument that this type of variation can identify the effects of climate change. 

While many counties experienced damaging heat exposure increases and decreases, only a handful had average yields decline between these time periods.
\Cref{fig:corn diff map} and \Cref{fig:soy diff map} show changes in average yields of corn and soy over this time period. 
Only 4.03\% of counties saw corn yields decline and only 3.96\% saw soy yields decline, among counties that grew corn or soy in both 1990-1999 and 2010-2019. 
This suggests that improved farming technology increased yields throughout the sample, and that these improvements exceeded damages from increased damaging heat exposure. 

\FloatBarrier
\section{Methods}
\label{sec:Methods}
In this section, I discuss the methods used to estimate the degree of adaptation. 
I first define adaptation in terms of average directional derivatives, and describe how to measure these average directional derivatives. 
I then introduce the ordinary least squares (OLS), Lasso and neural network (NNet) procedures, including details on how to account for fixed effect terms in each model and the cross-folds training procedures. 
I also describe the double machine learning (DML) approach used to adjust for bias in the standard machine learning (ML) estimates. 

\subsection{Adaptation}
\label{sec:adaptation}
I define adaptation as the amount of short-term impact to yield from damaging heat exposure that is offset in the longer term.
This definition encompasses all adaptation behaviors an agent makes to their production technology, management practices, or variety choice within each crop as they are exposed to climate change. 
It does not capture some other important margins of adaptation such as crop switching or exit from agriculture. \citet{Burke2016} provide evidence that these margins of adaptation are limited. 
This definition also does not capture changes that alter the impact of damaging heat independently from an individual's exposure to climate change, such as the decline in heat-related mortality as studied by \citet{Barreca2016}.
% The definition does not capture other margins of adaptation such as crop switching or exit from agriculture.
% It also does not capture technological change that shifts all players, such as introduction of air conditioners to reduce heat-related death \citep{Barreca2016}.
% By this definition, adaptation to extreme heat can be measured by comparing the elasticity from two linear regressions: a panel regression (capturing short-run weather variation) and a long-difference regression (capturing variation in climate) of log yields versus heat exposure and precipitation.
First I define the estimation target, as introduced by \citet{Burke2016}.
I generalize this definition in terms of elasticities, and discuss how to estimate those elasticities using flexible functional forms and/or richer sets of weather variation.

Following \citet{Burke2016}, I measure adaptation by comparing the elasticity of crop yields with respect to extreme heat using short-run and long-run variation.
Short-run variation comes from year-to-year changes in an annual panel of weather observations and log crop yields. 
To capture long-run variation, I first average weather and crop yield data over a long period (I use 10-year periods) and then construct a two-period panel.
The elasticity computed using short run variation captures the extent that a weather shock of extreme heat in a single year impacts yields, while the elasticity computed using long run variation captures the extent that exposure to a long period of increased heat exposure will impact average crop yields. 
Estimates using long-run variation arguably capture the extent of damages from climate change, because there is a change in average heat exposure over a relatively long period where farmers have time to adjust to those changes. 
\Cref{fig:Difference maps} illustrates long-run variation in the sample, comparing the differences in long-run average crop yields and damaging temperature exposure over the decades I study.
Comparing impacts using these sources of variation, I can conclude whether the short-term damages are offset in the long term. 

It is straightforward to recover these elasticities in a model where log yield is linear in total growing season heat exposure.
\citet{Burke2016} use such a model, which I adapt below: 
\begin{equation}
\label{eqn:linear model}
y_{it} = a_i + \beta_1 lower_{it} + \beta_2higher_{it} + g(prec_{it}) + \varepsilon_{it} 
\end{equation}
where $higher_{it}$ ($lower_{it}$) is total growing season heat exposure above (below) the damaging temperature threshold, $g$ is some function of precipitation, $a_i$ is an additive per-county fixed effect term, and $\varepsilon_{it}$ is an additive error term. 
\citet{Burke2016} estimate this equation with OLS, after using the within transformation to remove the fixed effect term\footnote{\citet{Burke2016} advocate using first differences to remove the fixed effect in the long-run variation panel; I use a within transformation approach as this is numerically equivalent to taking first differences in a two-period panel. }.
The key parameter here is $\beta_2$, which captures the extent that log crop yield changes with marginal increase in damaging heat exposure.
This is equivalent to the elasticity of crop yield with respect to damaging heat exposure. 
Note that $\beta_2$ is expected to be negative in the short run, as temperature shocks in this range are damaging to crop growth \citep{Schlenker2006,Schlenker2009}. 
The estimate of the share of short-run damages that are offset in the longer term is therefore $(\hat{\beta}_2^{SR} - \hat{\beta}_2^{LR})/\hat{\beta}_2^{SR} = 1 - \hat{\beta}_2^{LR}/\hat{\beta}_2^{SR}$, where $\hat{\beta}_2^{SR}$ ($\hat{\beta}_2^{LR}$) is the estimate using the short-run (long-run) variation.
\citet{Burke2016} take bootstrap samples of this ratio and fail to reject the null hypothesis that the ratio is different from 0.

To recreate this ratio with a more flexible functional form, I replace the parameter estimates with average directional derivatives. 
Consider a more general functional form: 
\begin{equation}
y_{it} = a_i + \gamma(X_{it}) + \varepsilon_{it}
\end{equation}
Here, $X_{it}$ is a collection of weather variables, $\gamma$ is a general function, $a_i$ is an additive fixed effect term, and $\varepsilon_{it}$ is an additive error term.
I use three different collections of weather variables for $X_{it}$, as described in \Cref{fig:GDD_explainer}.
As $y_{it}$ is the log of crop yields, the elasticity of crop yield with respect to some variable is equivalent to the average directional derivative of $\gamma$ with respect to that variable.

The analogue to $\beta_2$ from \Cref{eqn:linear model} is therefore the average directional derivative of $\gamma$ with respect to $higher_{it}$. 
Let $\theta^{SR}$ ($\theta^{LR}$) be this average directional derivative using short-run (long-run) variation. 
I then take bootstrap samples of the ratio $1 - \hat{\theta}_2^{LR}/\hat{\theta}_2^{SR}$ to test the hypothesis that damages from short-run heat exposure are offset in the longer run. 

This average directional derivative is equivalent to an average partial derivative when using the Yearly Linear set of weather variables. 
Specifically, take $\hat{\theta} = \E[\partial \hat{\gamma}(X_{it}) / \partial higher_{it}]$. 
When using a linear specification, this average derivative is equivalent to the parameter estimate from OLS.
When using NNet, this can be recovered after training the network (see \Cref{sec:Neural Network}).
When the specification involves basis functions (such as OLS with polynomial basis functions or Lasso), I recover this derivative by taking the average of the dot product of the gradient of the basis functions and the estimated coefficients. I describe this procedure in \Cref{sec:OLS}.

When I use another set of weather variation, it is also necessary to account for the extent that each weather variable contributes to the total growing season heat exposure.
I find this using the chain rule:
\begin{equation}
\hat{\theta} = \sum_{X_{it} \in X_{it}^{higher}} \frac{\partial \hat{\gamma}(X_{it})}{\partial X_{it}} \frac{\partial X_{it}}{\partial higher_{it}}
\end{equation}
Where $X_{it}^{higher}$ is the set of weather variables that are summed to reach total growing season heat exposure above the temperature threshold.
I set $\partial X_{it}/\partial higher_{it} = X_{it} / higher_{it}$; this captures the assumption that additional marginal heat will be distributed proportionally to the empirical heat exposure distribution.

This allows me to measure adaptation to climate change from recent panel variation while using a flexible model of high-dimensional weather variation.
In the following sections, I describe how I estimate the regression function and average directional derivative using various estimation techniques. 

\subsection{Ordinary Least Squares methods}
\label{sec:OLS}
Ordinary Least Squares (OLS) is a classical statistics approach to estimating this elasticity. 
I use methods with a linear functional form (OLS Linear), and after applying a basis function transformation of polynomial functions and interactions (OLS Poly). 
In both cases, I use the within transformation to remove a county-level fixed effect term, and then include yearly fixed effect terms via dummy variables. 

I use basis functions that include interactions and flexible functional forms of the original data. 
% Basis functions allow me to specify a flexible set of transformations for the original data.
I specify polynomial basis functions of all terms, as well as interactions between these polynomial expansions.
To produce a tractable model, I limit the space of potential interactions to interactions between heat exposure and precipitation variables within the same time period. 
For example, in the Monthly Flexible specification, I consider cumulative GDD in July between 28 and 29 C interacted with precipitation in July, as well as squared values of each term and the interactions between those polynomial expansions, but do not consider interactions of that variable with cumulative GDD in any other temperature bin, or precipitation in any other month. 
I use 3rd order polynomials for the Yearly Linear and Yearly Flexible variable sets, and 2nd order polynomials for the Monthly Flexible variable set. 
I then scale each flexible basis function so that it has mean zero and variance 1.

For OLS Linear, I use the identity set of basis functions; that is, $b(X_{it}) = X_{it}$. 
For Yearly Linear, $X_{it}$ has 3 covariates; for Yearly Flexible $X_{it}$ has 41; and for Monthly Flexible $X_{it}$ has 246.  
For OLS Poly, I use the set of basis functions described above. 
For Yearly Linear, $b(X_{it})$ has 30 covariates; for Yearly Flexible $b(X_{it})$ has 486; and for Monthly Flexible $b(X_{it})$ has 1464.

OLS assumes that the following is a true model of the relationship: 
\begin{equation}
\label{eqn:true OLS}
y_{it} = a_i + b(X_{it})\beta_0 + \varepsilon_{it}
\end{equation}
As is common in economics, I assume that we have relatively short panels where it is not possible to consistently estimate $a_i$ by including dummy variables. 
I therefore use the within transformation to remove county-level fixed effect terms:
\begin{equation}
\label{eqn:ddot ols}
\ddot{y}_{it} = \ddot{b}(X_{it})\beta_0 + \varepsilon_{it}
\end{equation}
where the double dot denotes the within transformation, i.e. $\ddot{y}_{it} := y_{it} - \text{mean}(\{y_{it} \forall t\})$ and $\ddot{b}(X_{it}) := \text{mean}(\{b(X_{it} \forall t\})$. 
Note that to construct $\ddot{b}(X_{it})$, the mean of all observations in that panel unit is subtracted after applying the basis function transformation. 
This ensures that $\beta_0$ is the same parameter vector between models. 
I use OLS to estimate $\hat{\beta}$.
 % when observations are weighted equally, and weighted least squares (WLS) otherwise. I use the terms OLS Linear and OLS Poly to refer to both methods.

I then compute the average directional derivative by projecting my estimate of $\beta_0$ on partial derivatives of the basis functions.
Define the gradient of the basis function as $b_{higher}$; this is a 1 by $p$ dictionary of the derivative of each basis function with respect to $higher_{it}$.
The true average directional derivative is then $\theta_0 = \E[b_{higher}(X_{it})\beta_0]$ and its estimate is $\hat{\theta} = \E[b_{higher}(X_{it})\hat{\beta}]$.

In specifications in my main analysis, I include per-year fixed effects as dummy variables. I assume that there are enough observations per time period to consistently estimate these variables separately. The derivative of each per-year fixed effect is zero, so including these terms does not change how I estimate the average directional derivative. 

\subsection{Machine Learning Methods}
\label{sec:ML methods}
% In this section, I specify the training and estimation procedure for the machine learning (ML) approach. 
Machine Learning (ML) methods include a range of estimation techniques that allow consistent function approximation in high-dimensional settings, when the number of covariates is large relative to the number of observations. 
Such settings poses challenges for classical statistical methods such as OLS, binning, or kernel regression.
I use Lasso and Neural Networks (NNets), although the same procedure could be used for another algorithm such as random forests, support vector machines, or other methods. 
I focus on these two machine learning algorithms because each enables the researcher to incorporate linear fixed effects and to evaluate derivatives without numerical differentiation. 
Standard ML methods can induce bias in regression analysis; I overcome this bias by using a procedure from \citet{chernozhukov2022automatic}.

The average derivative is computed from a ML regression of the output variable on weather inputs. 
Write $\gamma(\cdot; \lambda)$ to denote the flexible machine learning function, emphasizing the dependence on the hyperparameter $\lambda$.
The hyperparameter is a researcher-specified value that influences the behavior of the model, such as the regularization penalty in Lasso or the network width in NNet.
The machine learner is estimated as a regression function: $\E[\ddot{y}_{it} | \ddot{X}_{it}] =\hat{\ddot{\gamma}}(X_{it}; \lambda)$. 
\Cref{sec:Lasso} and \Cref{sec:Neural Network} give details on each estimation procedure. 
Let $m(\gamma, X_{it}; \lambda)$ denote the directional derivative of $\gamma(\cdot; \lambda)$ evaluated on observation $X_{it}$. 
% The average directional derivative is $\E[m(\gamma, X_{it}; \lambda)]$.

I use the double machine learning (DML) procedure from \citet{chernozhukov2022automatic} to find an approximately debiased estimate of the average directional derivative.
I discuss this procedure in more detail in \Cref{sec:Double Machine Learning}.
Briefly, I estimate a second machine learner and use this to construct an estimate of the average directional derivative that is robust to errors in estimating either the first or second machine learner. 
Let $\alpha(X; \kappa)$ denote this second machine learner, emphasizing the dependence on hyperparameter $\kappa$.
\citet{chernozhukov2022automatic} show that the expression $\E[m(\hat{\gamma}, X_{it}; \lambda) + \hat{\alpha}(X_{it}; \kappa)\hat{\varepsilon}_{it}]$ is an approximately unbiased estimate of the true average directional derivative, where $\hat{\varepsilon}_{it}$ is the residual from estimating $y_{it}$.
I modify the form of the doubly robust estimator in \citet{chernozhukov2022automatic} to account for the panel structure of the data; details are in \Cref{sec:Double Machine Learning}.

I use a data-driven process to determine the value of the hyperparameters.
First split each panel unit (a U.S. county) into one of the $k$ folds for cross validation.
Grouping observations from each panel unit into the same fold reduces correlation between training and test data, as counties share unobservable characteristics that likely influence the distribution of weather and crop yields. 
Let $\mathcal{I}_\ell$ denote the set of indices in fold $\ell$, for $\ell \in \{1, 2, \dots, k\}$.
For each of the $k$ folds, I train the ML and DML algorithm on data not in fold $\ell$, and evaluate the algorithm only on indices in fold $\ell$.
Let $\hat{\gamma}_{\ell}$ and $\hat{\alpha}_{\ell}$ denote the ML and DML estimators trained on the set of indices not in fold $\ell$.
Then select a hyperparameter value by searching over a grid of potential values, and selecting the value that minimizes a loss function. 
Let $\mathcal{L}_\gamma(\gamma_\ell, \mathcal{I}_{\ell}; \lambda)$ be the mean squared error of the function $\gamma_\ell$ with the hyperparameter $\lambda$ on the data in $\mathcal{I}_{\ell}$.
Let $\mathcal{L}_\alpha(\alpha_\ell, \mathcal{I}_{\ell}; \kappa)$ be the loss function of the function $\alpha_\ell$ with the hyperparameter $\kappa$ on the data in $\mathcal{I}_{\ell}$; I describe this loss function in \ref{sec:Appendix Auto DML}.

Once the hyperparameter is selected, I evaluate the (debiased) score on the test sets using the same folds defined above.
The full estimation procedure for the debiased score is below. 
To use this procedure without the debiasing correction,  omit the bias correction term $\hat{\alpha}_\ell(X_{it}; \hat{\kappa})(\ddot{y}_{it} - \hat{\ddot{\gamma}}_\ell(X_{it}; \hat{\lambda}))$ from each step. 
% To use this procedure with per-observation weights, multiply the loss function of each observation by the respective weight and take the weighted average when computing the score and asymptotic variance.  
\begin{enumerate}
\item Select hyperparameter $\hat{\lambda}$ that minimize test-set mean squared error of the regression: 
\[
\hat{\lambda} = \argmin_{\lambda} \sum_{\ell = 1}^k \mathcal{L}_\gamma(\hat{\gamma}_\ell, \mathcal{I}_{\ell}; \lambda)
\]
\item Select hyperparameter $\hat{\kappa}$ that minimize test-set loss of the double machine learner: 
\[
\hat{\kappa} = \argmin_{\kappa}  \sum_{\ell = 1}^k \mathcal{L}_\alpha(\hat{\alpha}_{\ell}, \mathcal{I}_{\ell}; \kappa)
\]
\item Evaluate the debiased score using these hyperparameters: 
\[
\hat{\theta} = \frac{1}{N} \sum_{\ell = 1}^k \sum_{it \in \mathcal{I}_\ell} m(\hat{\gamma}_\ell, X_{it}; \hat{\lambda}) + \hat{\alpha}_\ell(X_{it}; \hat{\kappa})(\ddot{y}_{it} - \hat{\ddot{\gamma}}_\ell(X_{it}; \hat{\lambda}))
\]
\item Find the asymptotic variance of the estimator, after adjusting for within-panel-unit correlations.
% A derivation is provided in \Cref{sec:clustering}.
Let $\hat{\theta}_{\ell; it} := m(\hat{\gamma}_\ell, X_{it}; \hat{\lambda}) + \hat{\alpha}_\ell(X_{it}; \hat{\kappa})(\ddot{y}_{it} - \hat{\ddot{\gamma}}_\ell(X_{it}; \hat{\lambda}))$ and $ \hat{\theta}_{\ell; i} := 1 / T \sum_{t = 1}^T \hat{\theta}_{\ell; it}$. 
Then the asymptotic variance is: 
\[
    \hat{V} = \frac{1}{N}   \sum_{\ell=1}^{k} \sum_{i \in \mathcal{I}_\ell} \left\{ \sum_{t = 1}^{T} (\hat{\theta}_{\ell; it} - \hat{\theta})^2 +   2 \sum_{t = 1}^{T - 1}  \sum_{t' = t + 1}^{T} (\hat{\theta}_{\ell; it} - \hat{\theta}_{\ell; i})(\hat{\theta}_{\ell; it'} - \hat{\theta}_{\ell; i}) \right\}
\]

\end{enumerate}

In the following subsections, I describe how to train and evaluate the Lasso and NNet estimators, and introduce the DML procedure.

\subsubsection{Lasso}
\label{sec:Lasso}
Least absolute shrinkage and selection operator (Lasso) is a regression procedure that selects a sparse linear model from a researcher-specified set of basis functions. 
The procedure finds this sparse linear combination by minimizing squared error of estimation, while penalizing more complex models via regularization. 
The procedure is similar to OLS Poly, but estimates can differ greatly because of this penalization. 
The hyperparameter $\lambda$ is the magnitude of the regularization term, which I determine via the cross fitting procedure outlined above. 

As with OLS, I assume that there is a true linear model of the form \Cref{eqn:true OLS} and take within transformations to remove county-level fixed effects to result in \Cref{eqn:ddot ols}. 
I use the same flexible set of basis functions used in OLS Poly, as defined in \Cref{sec:OLS}.
Unlike in OLS Poly, I assume that the true parameter vector is sparse and find an estimate by solving a regularized optimization problem. 

For each cross-validation fold, I find the estimate $\hat{\beta}_\ell$ via the following minimization problem: 
\begin{equation}
\hat{\beta}_\ell = \argmin_{\beta} \sum_{i \in \mathcal{I}_\ell} \sum_{t = 1}^{T} (\ddot{y}_{it} - \ddot{b}(X_{it})\beta)^2 + \lambda |\beta|_1
\end{equation}
where $|\beta|_1$ is $\ell1$ penalty or the sum of the absolute value of each component of $\beta$. 
To incorporate yearly fixed effects, I include dummy variables and do not apply the regularization penalty to the coefficients on those dummy variables. This reflects the assumption that while coefficients in $\gamma$ are sparse, coefficients in the fixed effects terms are not \citep{belloni2016inference}.
Additional details on the Lasso procedure are included in \ref{sec:Appendix Lasso}.

Once I have estimated $\hat{\beta}_\ell$, I compute the directional derivative by projecting my estimate of $\beta_0$ on partial derivatives of the basis functions. 
This is the same procedure to compute the derivative using OLS, as described in \Cref{sec:OLS}.

\subsubsection{Neural Network}
\label{sec:Neural Network}
% In this subsection, I describe the implementation of the neural network (NNet) and how to recover partial derivatives from the network. 
A neural network (NNet) learns a relationship between inputs and outputs through an iterative process, where the best fitting model is selected from a large space of flexible transformations of all potential interactions of input features. 
NNets have several advantages for my application. 
First, it is straightforward to compute a gradient of the output of the entire network with respect to each input feature. 
Second, NNets can incorporate fixed effects in panel models, as demonstrated by \citet{Crane-Droesch2018}. 
Third, estimating NNets does not require the researcher to specify basis functions.

I compute derivatives of the NNet by extracting gradients computed during training the network. 
A NNet is a weighted composition of activation functions (user-specified transformations) applied to an input vector. 
After a random initialization of internal parameters, the model goes through many iterations of predicting the output variable, calculating a loss from training data, and updating the parameter values. I use the mean squared error for this loss function. 
While training or evaluating the network, the algorithm computes the derivative of the output with respect to each input. 
This is commonly used to adjust parameter values during the training procedure, in a process known as back propagation. 
I also use these automatic derivatives to compute the partial derivative of the prediction with respect to each weather input. 

I follow \citet{Crane-Droesch2018} to estimate the model with nonparametric treatment of weather inputs and additive linear fixed effects.
Training this network involves an iterative procedure that can be interpreted as selecting basis functions from a large space of candidate functions.
After this training procedure is completed, I use standard econometric techniques to account for linear terms, treating the nonlinear transformation of inputs as a feature in a linear model. 
Specifically, I use the within transformation to remove individual fixed effects and use OLS to find other linear terms such as yearly fixed effects.
The iterative procedure is performed on the set of training data, and the OLS step is performed on the test set. 
This requires an architecture with a top layer that is linear in all fixed effect components and the output of the nonlinear network transformation. 
See \citet{Crane-Droesch2018} for more details on such networks and their performance relative to linear models or fully nonparametric models in estimating agricultural yields.
Additional details of the NNet architecture and training procedure are included in \ref{sec:Appendix Neural Network}.

\subsubsection{Double Machine Learning}
\label{sec:Double Machine Learning}
Double machine learning (DML) is an approach to remove bias from standard machine learning algorithms. 
I use an approach introduced by \citet{chernozhukov2022automatic}, relying on the statistical theory of the Riesz representer. 

There are two related, yet orthogonal, problems in estimating the average derivative. 
First is the regression function, and second is the derivative operator on a regression function.
\citet{chernozhukov2022automatic} show how to conduct this second step without estimating a regression function, and that these two methods can be combined to construct an approximately debiased estimate of the original target derivative. 

This second function can be estimated from data using the Riesz representation theorem.
I denote this second function $\alpha_0$, and its estimate $\hat{\alpha}$.
The Riesz representation implies that $\E[\alpha_0(X_{i,t}) \ddot{\gamma}_0(X_{i,t})] = m(\ddot{\gamma}_0, X_{i,t})$.  \citet{chernozhukov2022automatic} show how to use this fact to estimate $\alpha_0$ from data.
In a panel setting, I use the within-transformed function $\ddot{\gamma}$; I use $m(\ddot{\gamma}_0, X_{i,t}) = m(\gamma_0, X_{i,t})$ because the functional $m$ is a directional derivative and the derivative of the mean value is 0. 
I assume that $\alpha_0$ is linear in the set of basis functions $b$. 
The true Riesz representer is then $\alpha_0(X_{it}) := \ddot{b}(X_{it}) \rho_0$ and its estimate is $\hat{\alpha}(X_{it}) := \ddot{b}(X_{it}) \hat{\rho}$.
I estimate $\hat{\rho}$ using an optimization package, based on the moment conditions identified by \citet{chernozhukov2022automatic}. \cite{klosin2022} demonstrate that this procedure is effective in panel settings. 
\ref{sec:Appendix Auto DML} includes more details on the motivation of this Riesz representer and the estimation procedure. 

This estimate is then used to construct the following doubly robust score:
\[
\hat{\theta} = \frac{1}{N} \sum_{\ell = 1}^k \sum_{it \in \mathcal{I}_\ell} m(\hat{\gamma}_\ell, X_{it}; \hat{\lambda}) + (\ddot{y}_{it} - \hat{\ddot{\gamma}}_{\ell}(X_{it})) \hat{\alpha}_\ell(X_{it}; \hat{\kappa})
\]

\FloatBarrier
\section{Simulation Exercise}
\label{sec:Simulation}
I conduct a simulation exercise to compare the performance of the estimation procedure to OLS while varying the set of weather variables used. 
To match the setting as closely as possible, I use the empirical distribution of weather covariates in all trials. 
I then apply a simulated production function of the the piecewise-linear functional form from \citet{Schlenker2009}. 

The simulation exercise focuses on comparing ML, DML, and OLS in a regression setting with high-dimensional variation. 
OLS is correctly specified in all trials. 
In cases where OLS is not correctly specified, ML and DML would likely perform better because they are able to represent a richer set of flexible functional forms. 
Simulation exercises by \citet{Chernozhukov2018} demonstrate this property. 

\begin{figure}[p]
     \centering
     % \hfill 
     \begin{subfigure}[b]{0.8\textwidth}
         \centering
         \includegraphics[width=\textwidth]{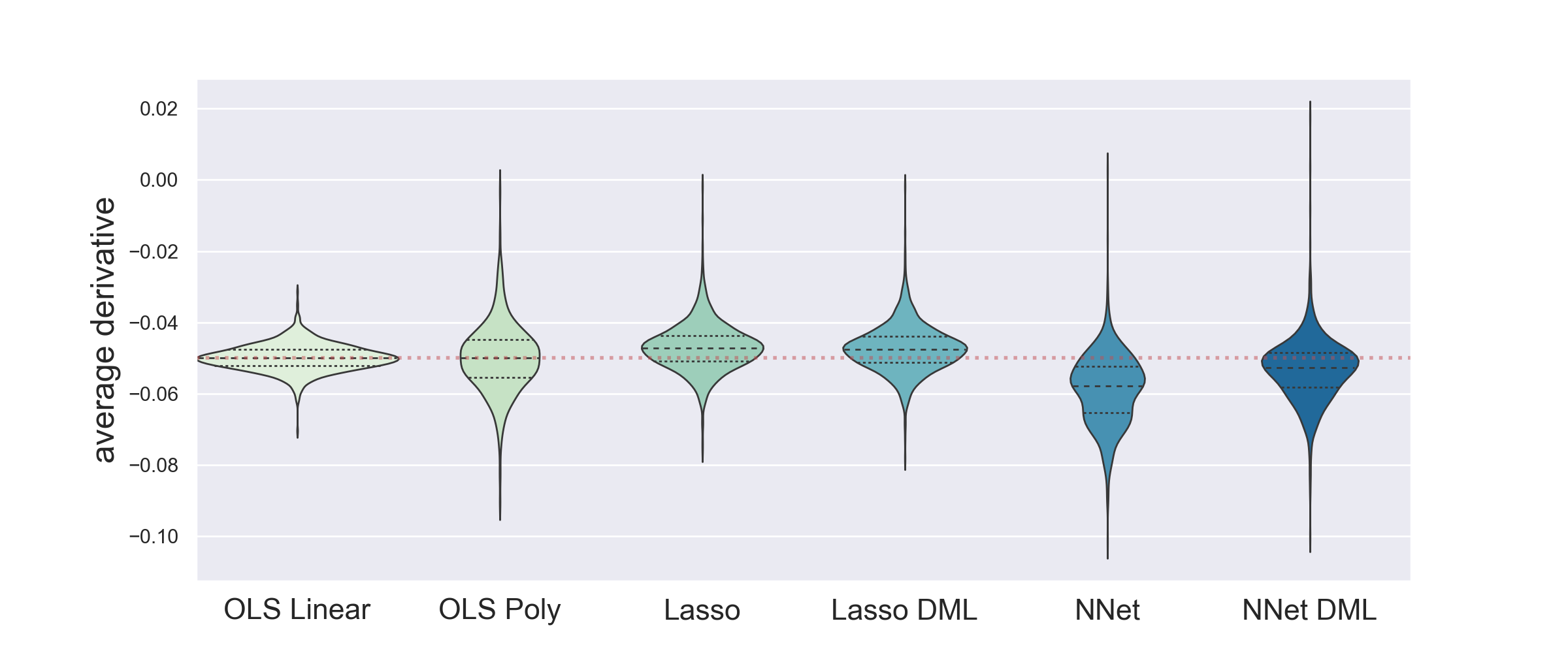}
         \caption{Yearly Linear}
         \label{fig:Yearly Linear violin full sim}
     \end{subfigure}
     % \hfill 
     % \newline 
     % \hfill 
      \begin{subfigure}[b]{0.8\textwidth}
         \centering
         \includegraphics[width=\textwidth]{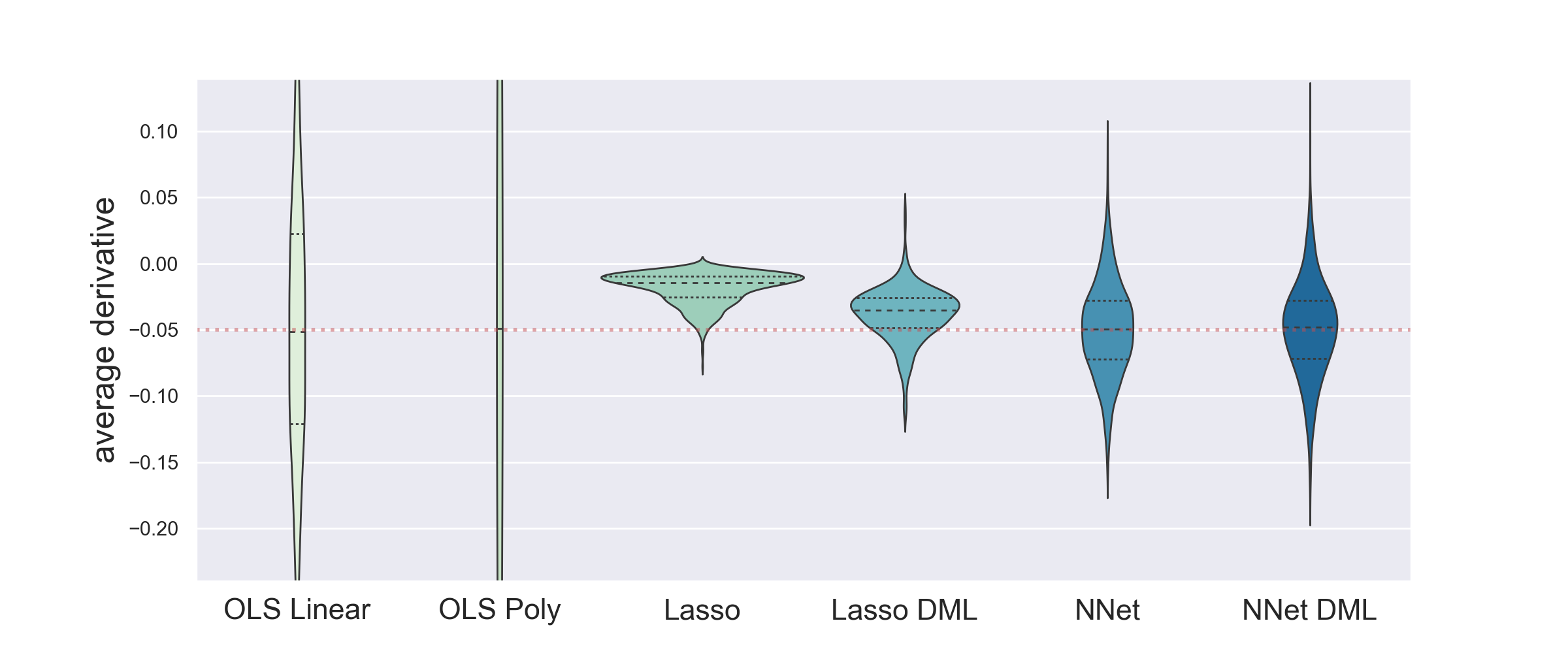}
         \caption{Yearly Flexible}
         \label{fig:Yearly Flexible violin full sim}
     \end{subfigure}
     % \hfill 
     % \newline 
     % \hfill 
      \begin{subfigure}[b]{0.8\textwidth}
         \centering
         \includegraphics[width=\textwidth]{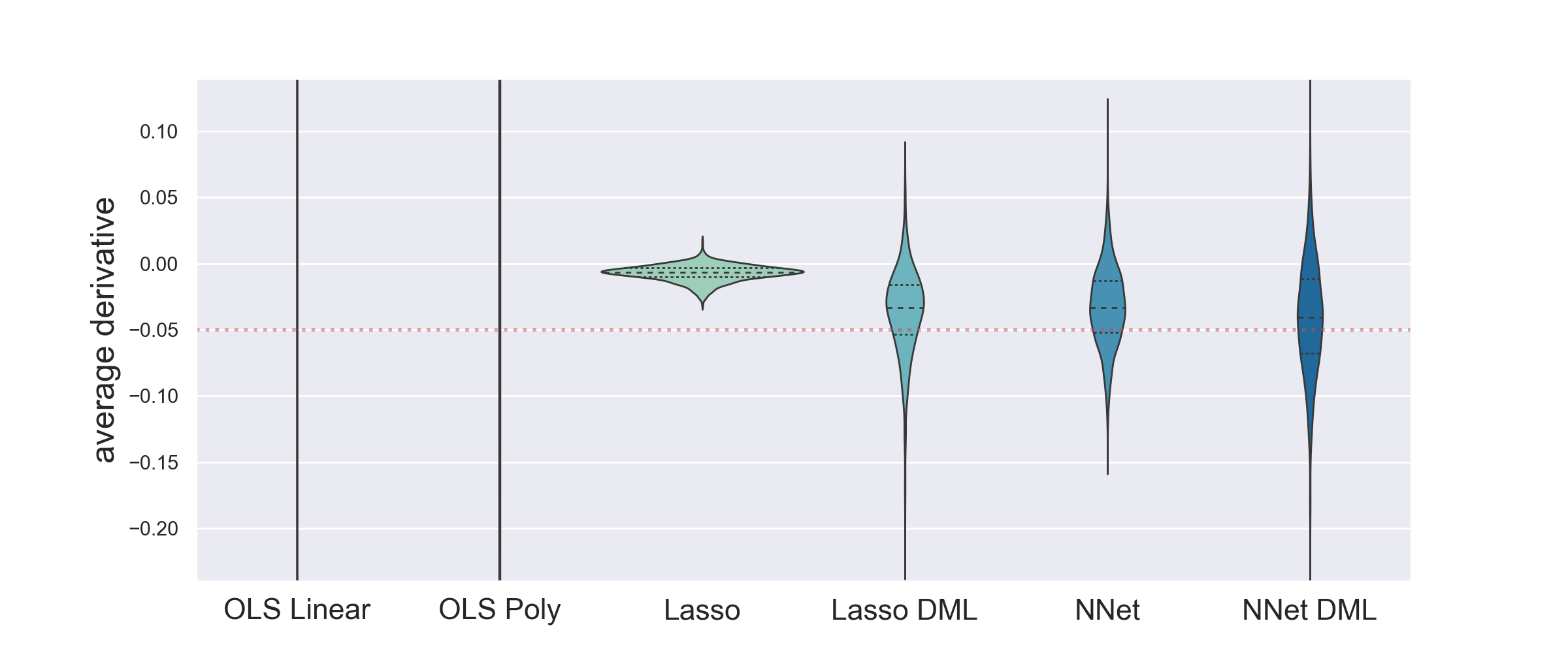}
         \caption{Monthly Flexible}
         \label{fig:monthly violin full sim}
     \end{subfigure}
     % \hfill 
        \caption[Violin plots of simulation results]{Violin plots of distribution of parameter estimates from 1,000 Monte Carlo simulations.
        The subfigures indicate which set of weather variables are used, using the notation from \Cref{fig:GDD_explainer}.
        Each subfigure has a separate violin plot for each method used. The plot shows the density of the parameter at the value on the y axis, based on the simulation trials. 
The dotted lines within each plot show the median and upper/lower quartiles of the distribution. 
The dotted horizontal line through the plots shows the true value of the parameter, -0.05. 
}
\label{fig:sim_results}
\end{figure}

I conduct 1,000 Monte Carlo simulation trials of this estimation procedure. 
In each trial, I first randomly draw a 1,000 county sample, and randomly select a 2-year panel from these counties.
I use the weather observations from this sample, guaranteeing that there are realistic correlations between weather variables. 
Then, I generate a $y$ variable according to the following function: $y_{it} = a_i + \beta_1 lower_{it} + \beta_2 higher_{it} + \beta_3 prec_{it} + \varepsilon_{it}$, where $lower$ ($higher$) is the total growing season accumulated GDD below (above) 29\textdegree C and $prec$ is the total growing season precipitation.
This functional form closely matches the parsimonious form suggested by \citet{Schlenker2009}.
Note that in this functional form, OLS Linear is correctly specified for all sets of weather variables I consider. 
I set $\beta_1 = 0.02; \beta_2 = -0.05; \beta_3 = 0.001$, and take $a_i \sim N(1,1)$ and $\varepsilon_{it} \sim N(0,1)$. 

\begin{table}[t]
\resizebox{\textwidth}{!}{
\begin{tabular}{llcccccc}
\toprule
                 & method &  OLS Linear &    OLS Poly &       Lasso &   Lasso DML &        NNet &    NNet DML \\
Weather & {} &             &             &             &             &             &             \\
\midrule
\multirow{3}{*}{Yearly Linear} & $\theta$ &     -0.0498 &     -0.0499 &      -0.047 &     -0.0473 &     -0.0588 &     -0.0533 \\
                 & \  &  (0.003756) &  (0.009589) &  (0.006619) &  (0.006594) &  (0.009762) &  (0.008802) \\
                 & MSE &       0.995 &       0.968 &        1.01 &        1.01 &        1.08 &        1.08 \\
\cline{1-8}
\multirow{3}{*}{Yearly Flexible} & $\theta$ &     -0.0507 &     -0.0469 &     -0.0184 &     -0.0383 &     -0.0498 &     -0.0498 \\
                 & \  &     (0.108) &     (0.386) &   (0.01192) &      (0.02) &   (0.03537) &   (0.03649) \\
                 & MSE &       0.954 &       0.511 &        1.03 &        1.03 &         1.1 &         1.1 \\
\cline{1-8}
\multirow{3}{*}{Monthly Flexible} & $\theta$ &        2.36 &     -0.0317 &    -0.00716 &     -0.0367 &     -0.0327 &     -0.0413 \\
                 & \  &     (58.22) &    (0.9779) &  (0.006234) &   (0.03336) &   (0.02993) &   (0.04387) \\
                 & MSE &       0.752 &    7.32e-09 &        1.08 &        1.08 &        1.13 &        1.13 \\
\bottomrule
\end{tabular}

}
\caption[Summary of Monte Carlo Simulations]{Summary of results from 1000 Monte Carlo simulation draws of the estimation procedure. 
The true value of the average derivative in all trials is -0.05; $\theta$ is the average of the estimated average derivative from all trials. 
Each column represents a separate method used for the regression function. 
Standard errors are in parentheses, and are computed from the distribution of bootstrap values.
}
\label{tab:sim results table}
\end{table}

The results of this simulation exercise are summarized visually in \Cref{fig:sim_results} and numerically in \Cref{tab:sim results table}. 
I estimate $\beta_2$, the parameter of interest, using OLS, ML, and DML for weather variables in the Yearly Linear, Yearly Flexible, and Monthly Flexible sets (as illustrated in \Cref{fig:GDD_explainer}).
For OLS, I OLS Linear and OLS Poly as described in \Cref{sec:OLS}.
For ML, I consider Lasso and NNet estimators as described in \Cref{sec:Lasso} and \Cref{sec:Neural Network}. 
For DML, I adjust each ML result with the DML procedure in \Cref{sec:Double Machine Learning}. 
I use Lasso DML and NNet DML to describe the Lasso and NNet estimates with the DML correction. 
I use the cross-folds and sample splitting procedure described in \Cref{sec:Methods}.

With Yearly Linear weather inputs, OLS performs best among all models. 
OLS Poly and OLS Linear have the lowest bias, and OLS Linear has the lowest variance. 
The machine learning models performed reasonably well, especially with the DML correction, although they have greater variance and bias than the OLS results. 
The central estimate from both DML approaches lie within 7\% of the true elasticity, and a 95\% confidence interval contains the true elasticity. 

With Yearly Flexible weather variables, ML estimates have considerably lower variance than OLS. 
The estimates using OLS Linear have a standard deviation approximately three times as large as the DML estimates, while estimates from OLS Poly have a standard deviation approximately ten times as large. 
There is substantial bias from the Lasso estimates, although the DML correction greatly reduces this bias. 
The NNet and NNet DML estimates have very little bias; central estimates from both are within 0.4\% of the true value and have less bias than OLS Linear.
The 95\% confidence interval from both DML procedures contains the true elasticity, although the 95\% confidence interval from Lasso without DML does not.

With Monthly Flexible variables, the benefits of using DML are even more dramatic. 
OLS Linear and OLS Poly have extremely high variance, and both have greater bias than the DML methods. 
The high variance is especially limiting, as confidence intervals using these approaches are uninformative. 
As \Cref{fig:monthly violin full sim} shows, the interquartile range of the bootstrap estimates is not visible on a plot whose range is 8 times the magnitude of the true target elasticity. 
OLS Linear estimates a mean value of 2.359, relative to the true value of -0.05. 
OLS Poly has a bias of .018, which is greater than bias from Lasso DML (0.012) or NNet DML (0.0087).
% The variance of DML estimates is greater than when using Yearly Linear or Yearly Flexible weather variables.

Note that while mean squared error (MSE) can help suggest a preferred model, a straightforward comparison of MSE does not select the model with lowest bias. 
MSE appears higher among ML models than OLS because the MSE reported by ML is the MSE of the model evaluated on a test set.
When MSE from an ML model is low, this indicates that ML is providing a better fit because the model extrapolates well to unseen data. 
When MSE from an OLS approach is low, this could indicate overfitting. 
OLS Poly has the lowest MSE in all trials, but is overfitting the data in the Yearly Flexible and Monthly Flexible cases. 
In the data generating process (DGP), the additive error term has variance 1. 
The MSE from OLS Poly is much lower than the true squared error from the DGP, indicating that OLS Poly is fitting the noise instead of the desired pattern in the data. 
The MSE from OLS Linear with Monthly Flexible weather terms is significantly lower than the true squared error, indicating that OLS Linear may be overfitting the data in that setting. 
This suggests that a comparison of MSE alone should not be used to select the preferred model, but the researcher can consider test-set MSE and performance in simulation trials. 

This simulation exercise shows that when a researcher wishes to measure a relationship with a high-dimensional set of weather variables, DML can provide results with lower bias and less variance than OLS.
With Yearly Flexible or Monthly Flexible weather, NNet DML has the lowest bias among all models, and the DML methods have significantly lower variance. 
OLS performs best in the Yearly Linear case, which is expected because OLS is correctly specified and the weather variation is low-dimensional. 

\FloatBarrier
\section{Results}
\label{sec:Results}
I implement the above estimation procedure to study adaptation to damaging heat exposure in U.S. corn and soy production. 
When using the Yearly Linear set of weather variables (as in \citealt{Burke2016}), I find little to no evidence of adaptation. 
However, when using a richer set of weather variation, I find evidence that a considerable share of the short-run damages from extreme heat are offset with greater exposure to those temperatures. 
To help explain this discrepancy, I visualize OLS results with Yearly Flexible temperature variation to show that the simple linear model may not accurately describe the long-run role of extreme heat. 

I run 500 bootstrap trials of the procedure described in \Cref{sec:Methods} to measure the elasticity of crop yields with respect to extreme heat. 
As described in \Cref{sec:Data}, I study corn and soy production from 1990-2019 in counties east of the 100 \textdegree W meridian.
% I weight each county observation by the acres of corn planted, to capture the average effect per acre of corn instead of the average effect over counties; 
I use a panel of the full data to capture short-run variation, and capture long-run variation by comparing average yield and weather from 1990-1999 to 2010-2019.
In each bootstrap trial, I draw a random subsamples of 80\% of counties.
This resampling scheme addresses the intertemporal correlation as suggested by \citet{kapetanios2008bootstrap}, while avoiding the risk of own-sample bias from machine learning methods by including the same observations in train and test data.

I estimate the elasticity using the three sets of weather variation described in \Cref{fig:GDD_explainer} and all estimation approaches described in \Cref{sec:Methods}.
Each method uses the within transformation to remove individual fixed effects, and includes additive yearly fixed effects. 
For estimation methods that use a set of basis functions, I use polynomial expansions of all temperature and precipitation variables, and interactions of the polynomial expansions of each precipitation variable with the polynomial expansions of each temperature variable.
I use third-order polynomial expansions for Yearly Linear and Yearly Flexible weather variables, and second-order polynomial expansions for Monthly Flexible. 
The machine learning models use a 5-fold cross validation procedure. 

I only report results from OLS Linear, Lasso DML, and NNet DML in this section. 
The simulation results showed that the debiasing procedures are effective at reducing bias from naive machine learning methods, and that estimates using OLS Poly can have extremely high variance. 
Results from the machine learning models without the debiasing procedure and from OLS with polynomial basis functions are included in \ref{sec:Appendix Results}.
Generally, estimates using machine learning without bias correction are similar to their debiased counterparts, and estimates using OLS Poly have unacceptably high variance when using richer sets of weather variation. 

\begin{figure}[t]
     \centering
     % \hfill 
     \begin{subfigure}[b]{0.305\textwidth}
         \centering
         \includegraphics[width=\textwidth]{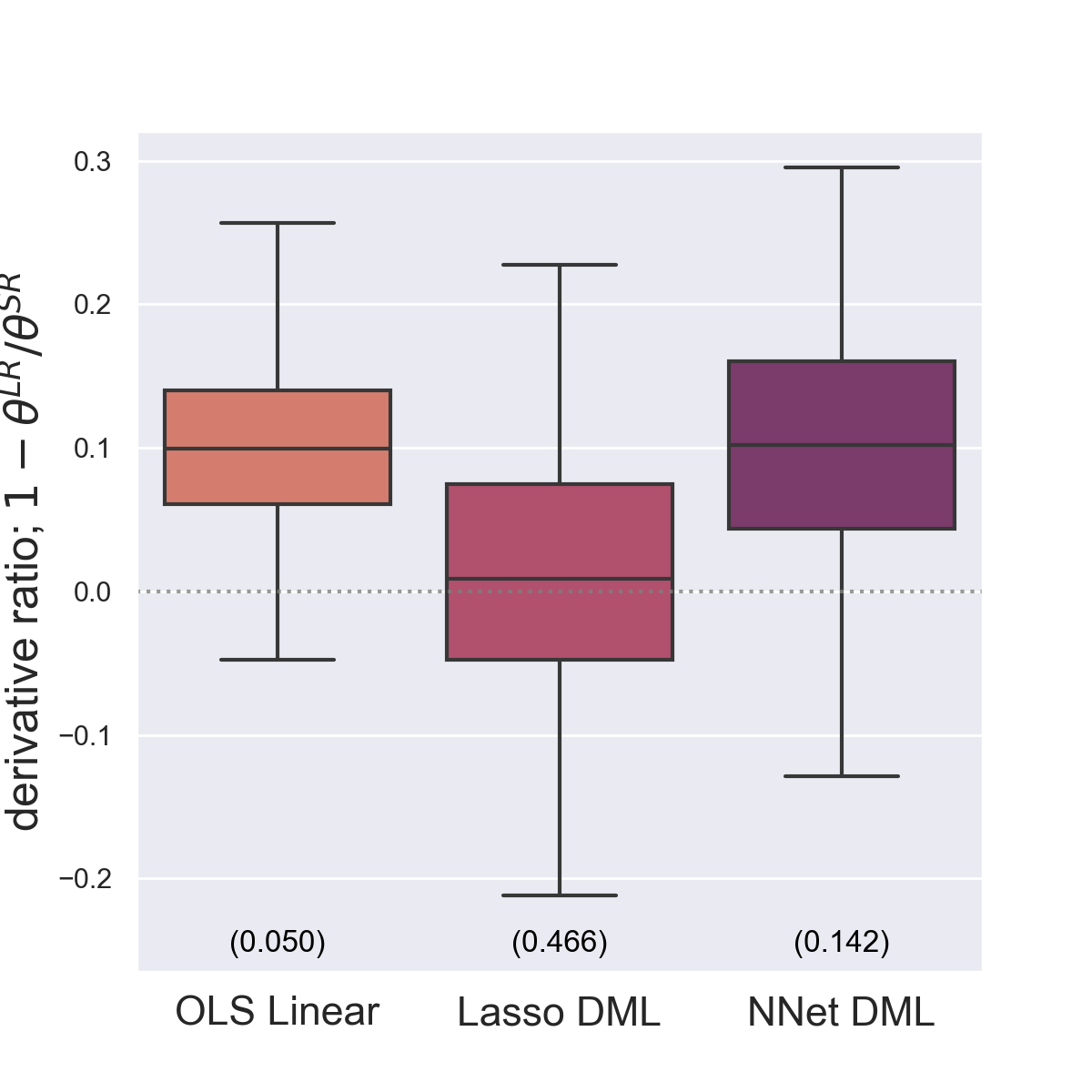}
         \caption{Corn, Yearly Linear}
         \label{fig:corn Yearly Linear box sim}
     \end{subfigure}
      \begin{subfigure}[b]{0.305\textwidth}
         \centering
         \includegraphics[width=\textwidth]{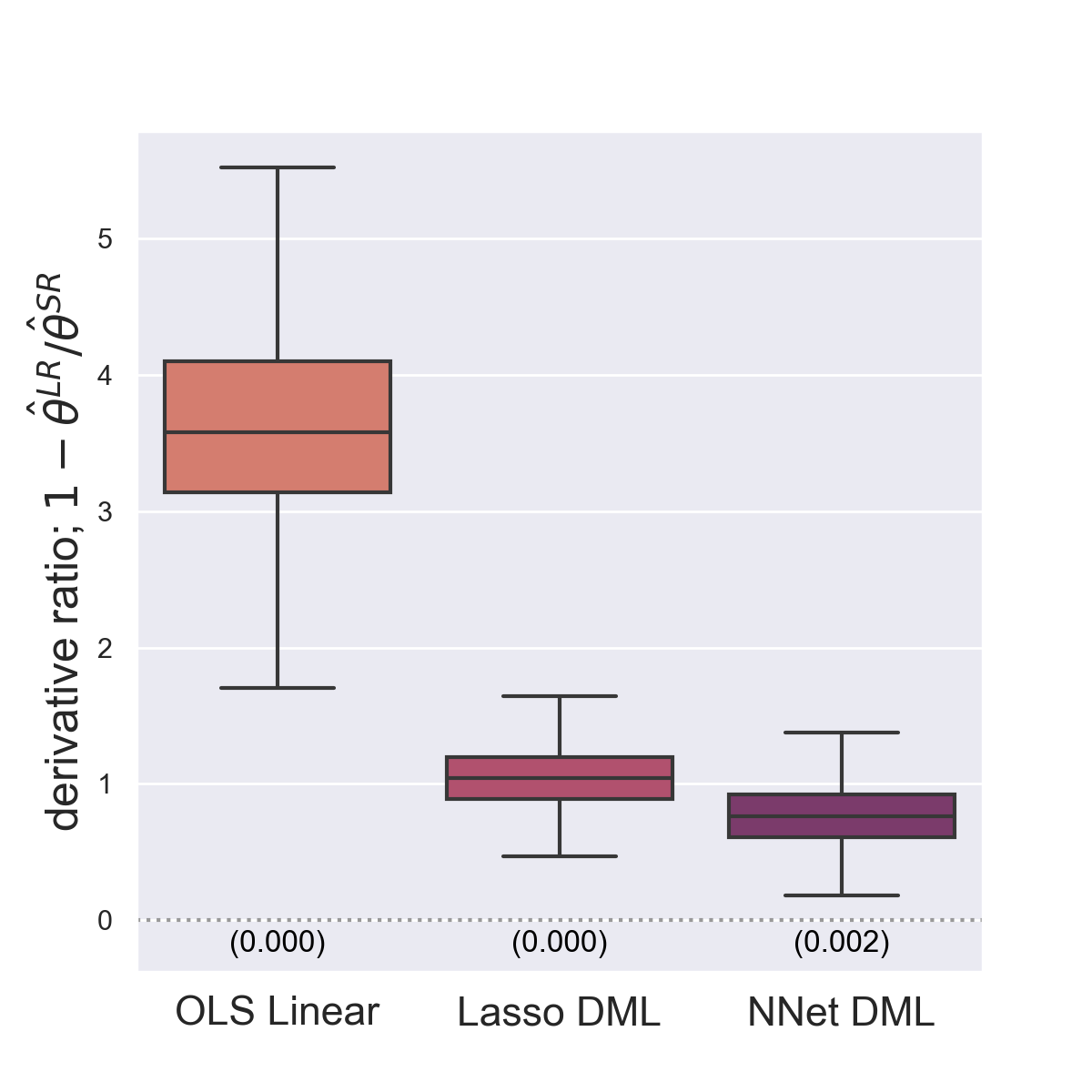}
         \caption{Corn, Yearly Flexible}
         \label{fig:corn Yearly Flexible box sim}
     \end{subfigure}
      \begin{subfigure}[b]{0.305\textwidth}
         \centering
         \includegraphics[width=\textwidth]{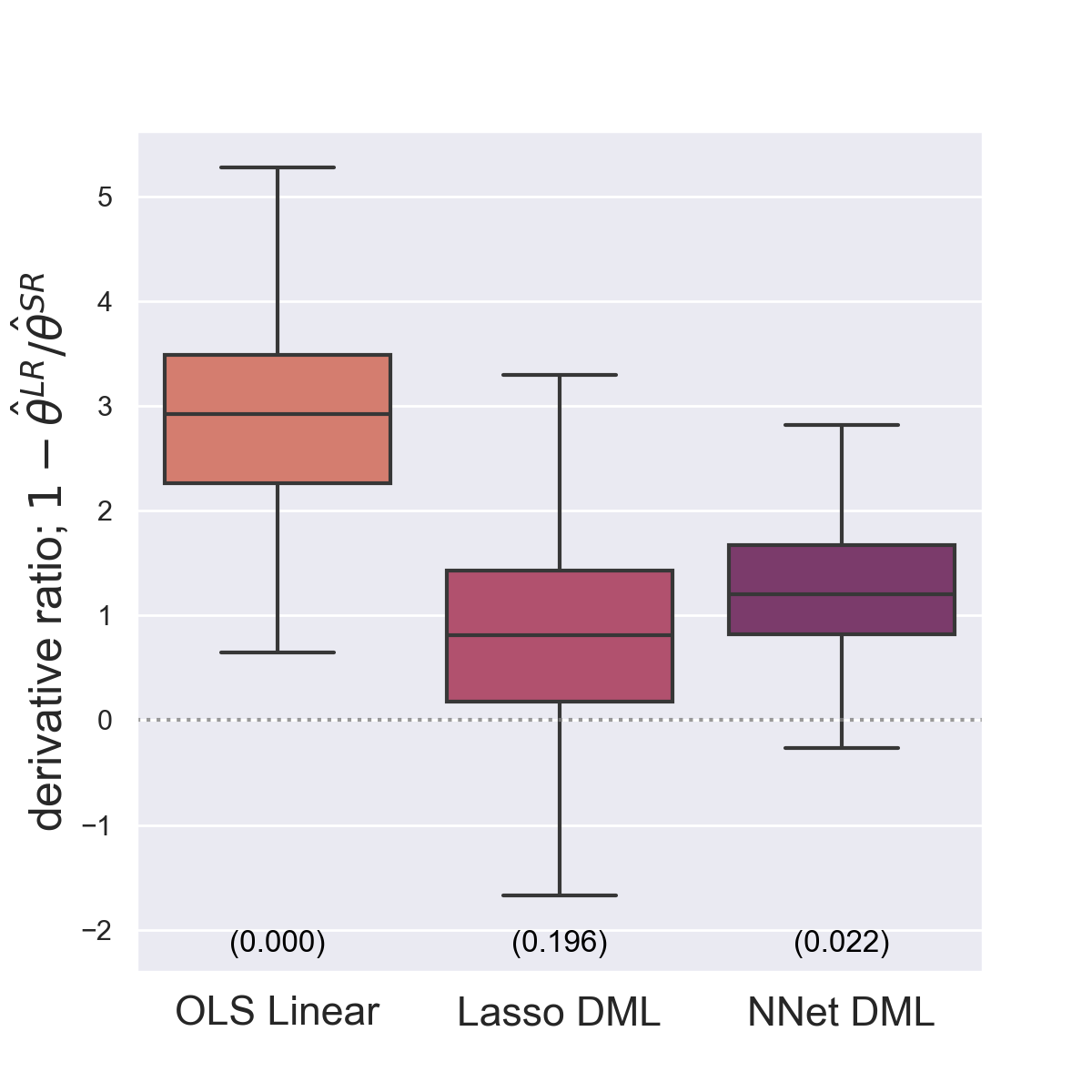}
         \caption{Corn, Monthly Flexible}
         \label{fig:corn monthly box sim}
     \end{subfigure}
     \begin{subfigure}[b]{0.305\textwidth}
         \centering
         \includegraphics[width=\textwidth]{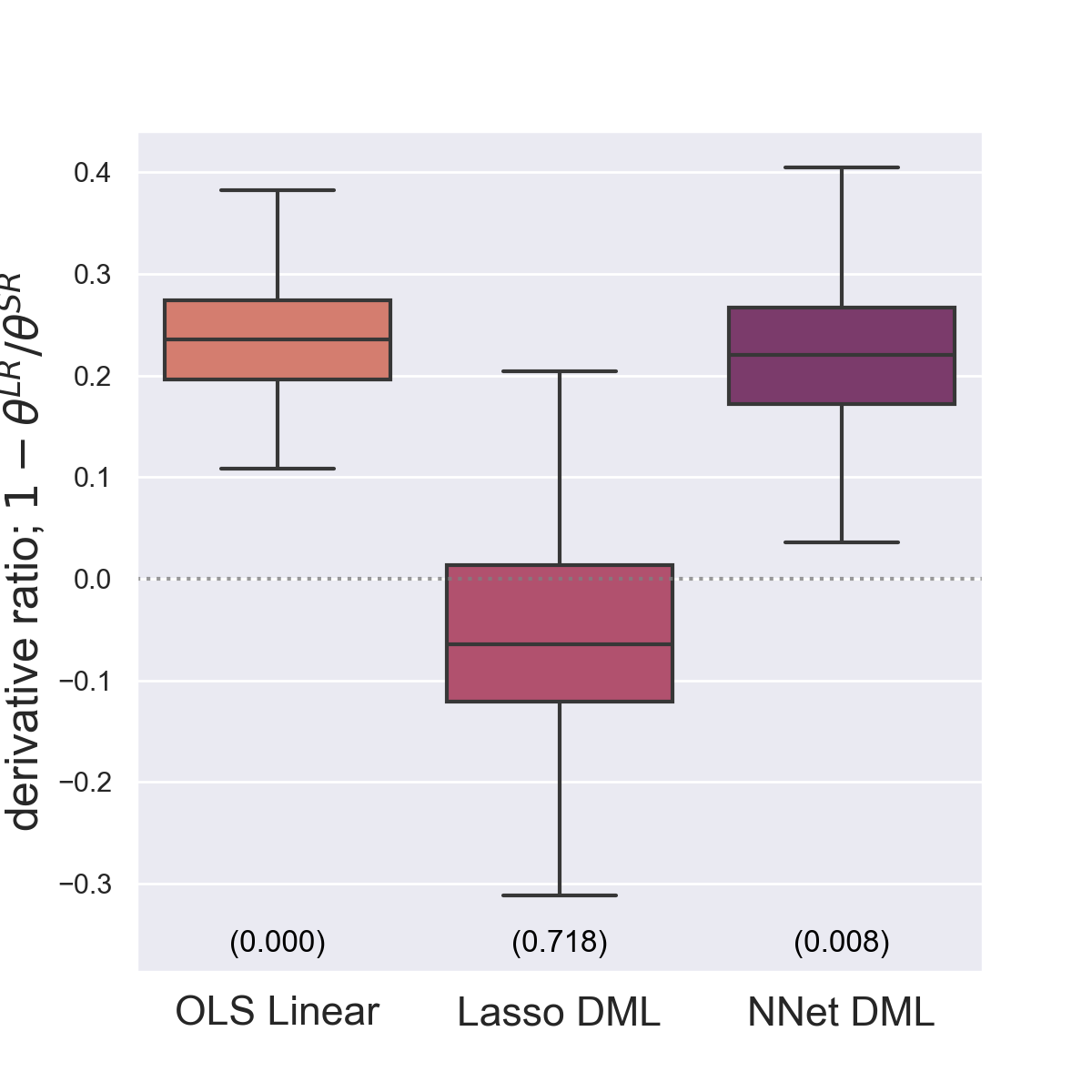}
         \caption{Soy, Yearly Linear}
         \label{fig:soy Yearly Linear box sim}
     \end{subfigure}
      \begin{subfigure}[b]{0.305\textwidth}
         \centering
         \includegraphics[width=\textwidth]{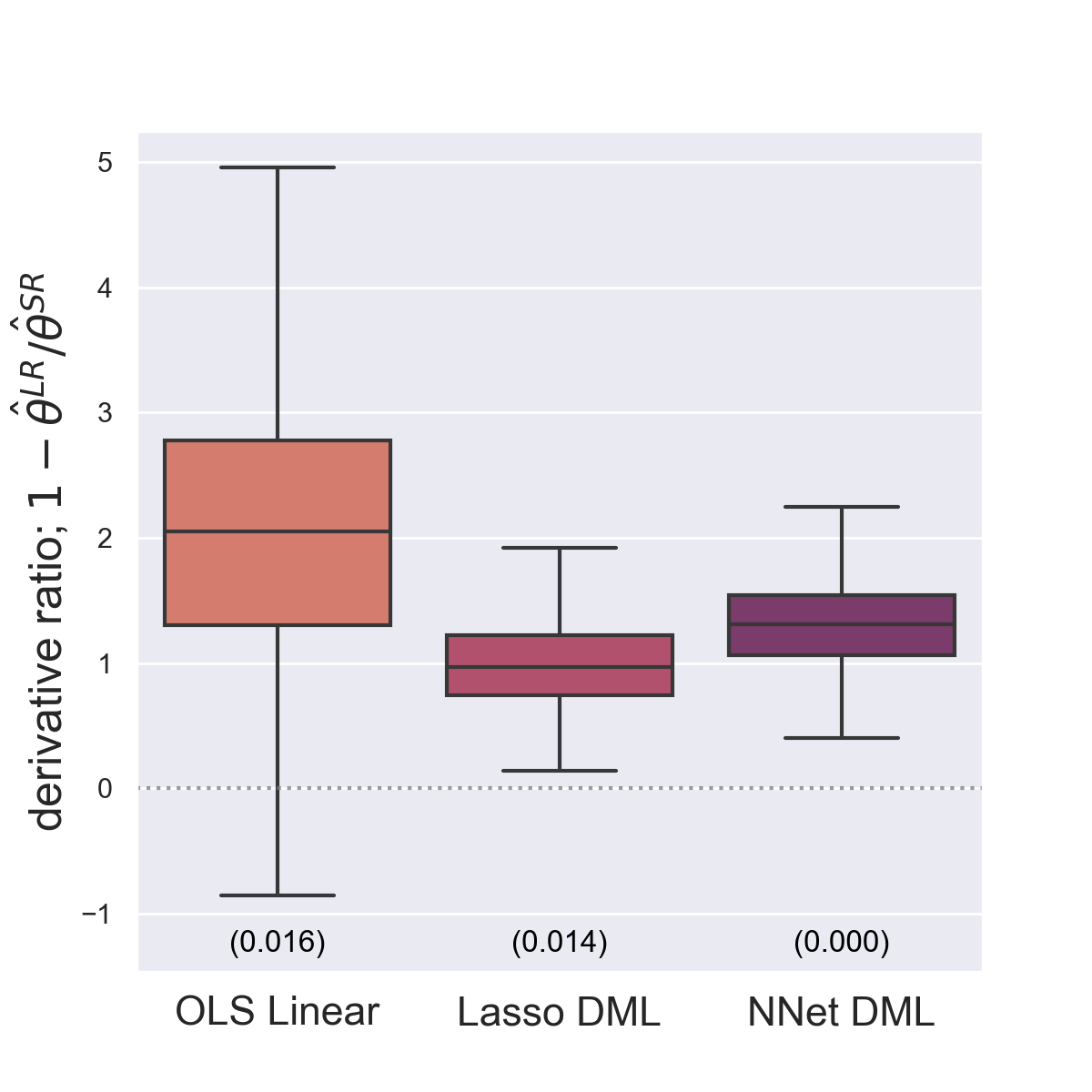}
         \caption{Soy, Yearly Flexible}
         \label{fig:soy Yearly Flexible box sim}
     \end{subfigure}
      \begin{subfigure}[b]{0.305\textwidth}
         \centering
         \includegraphics[width=\textwidth]{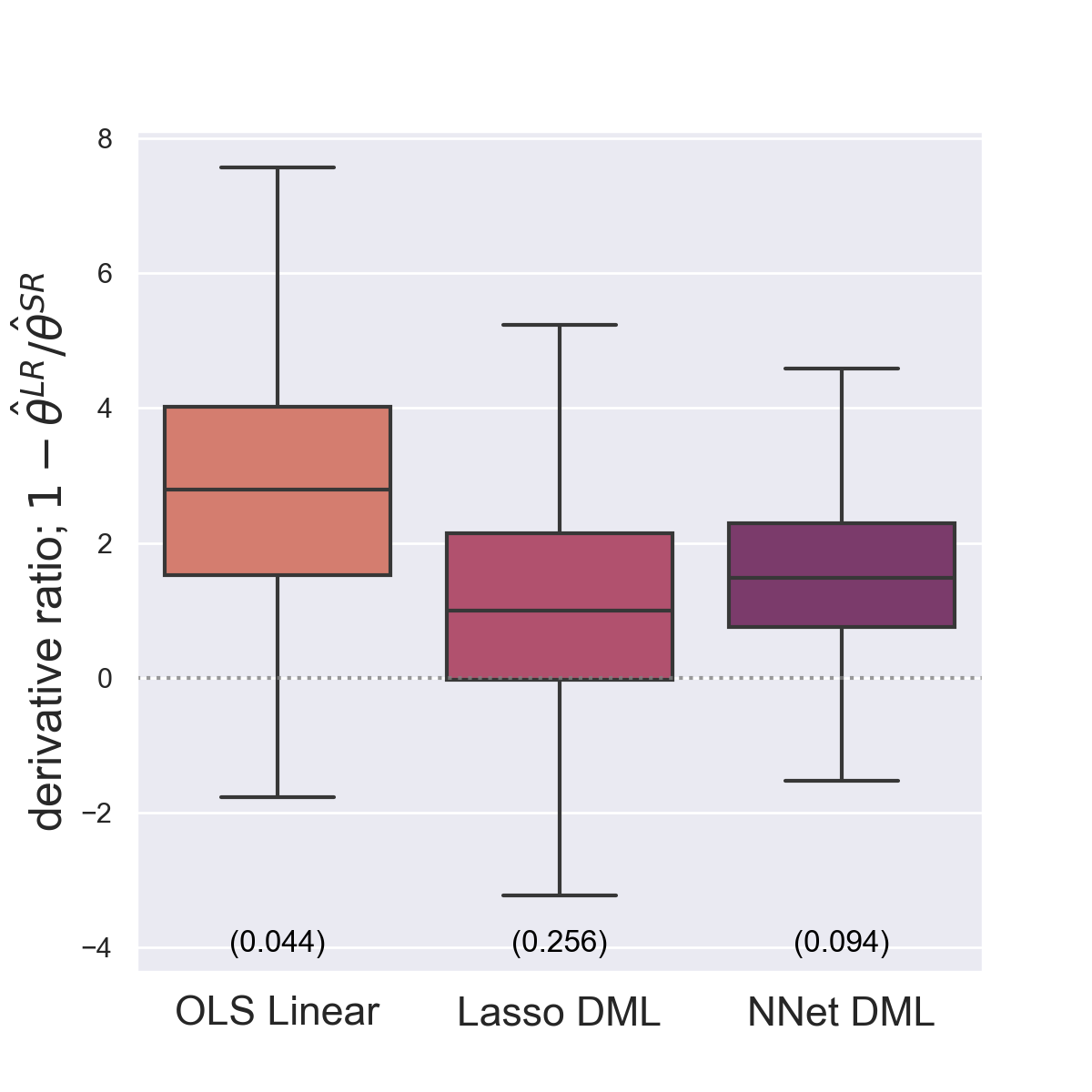}
         \caption{Soy, Monthly Flexible}
         \label{fig:Soy monthly box sim}
     \end{subfigure}
     % \hfill 
        \caption[Box plots, main estimation]{Box plots of 500 bootstrap samples of ratio $1 - \hat{\theta}^{LR}/\hat{\theta}^{SR}$, from both corn and soy production. 
        The subfigures indicate which set of weather variables are used, using the notation from \Cref{fig:GDD_explainer}.
        The number in parentheses at the bottom of each box plot is the $p$ value for the one-sided test that the ratio is greater than 0. 
        Each subfigure has a separate box plot for each method used. The line inside each box is the median, the edges of the box are the upper and lower quartile of the distribution, and the whiskers extending from each box are 1.5 times the interquartile range. 
}
\label{fig:ag_results}
\end{figure}

\begin{table}[t]
\resizebox{\textwidth}{!}{
\begin{tabular}{lllccc|ccc}
\toprule
    &                  & {} & \multicolumn{3}{c}{Long-Run} & \multicolumn{3}{c}{Short-Run} \\
    &                  & method &        OLS Linear &        Lasso DML &          NNet DML &        OLS Linear &         Lasso DML &          NNet DML \\
Crop & Weather & {} &                   &                  &                   &                   &                   &                   \\
\midrule
\multirow{12}{*}{Corn} & \multirow{4}{*}{Yearly Linear} & $\theta$ &  -0.00481$^{***}$ &   -0.007$^{***}$ &  -0.00597$^{***}$ &  -0.00535$^{***}$ &  -0.00709$^{***}$ &  -0.00663$^{***}$ \\
    &                  & \  &        (0.000326) &       (0.000568) &        (0.000464) &        (5.68e-05) &        (8.05e-05) &        (0.000337) \\
    &                  & MSE &           0.00652 &          0.00648 &           0.00673 &            0.0387 &            0.0363 &            0.0362 \\
    &                  & N &              2850 &             2850 &              2850 &             38633 &             38633 &             38633 \\
\cline{2-9}
    & \multirow{4}{*}{Yearly Flexible} & $\theta$ &    0.0322$^{***}$ &         0.000429 &          -0.00249 &   -0.0125$^{***}$ &   -0.0106$^{***}$ &   -0.0104$^{***}$ \\
    &                  & \  &         (0.00826) &        (0.00234) &         (0.00247) &         (0.00106) &         (0.00115) &        (0.000603) \\
    &                  & MSE &           0.00531 &          0.00582 &           0.00603 &            0.0376 &            0.0343 &            0.0352 \\
    &                  & N &              2850 &             2850 &              2850 &             38633 &             38633 &             38633 \\
\cline{2-9}
    & \multirow{4}{*}{Monthly Flexible} & $\theta$ &      0.0213$^{*}$ &        -0.000793 &           0.00148 &   -0.0114$^{***}$ &  -0.00505$^{***}$ &  -0.00674$^{***}$ \\
    &                  & \  &         (0.00874) &        (0.00533) &         (0.00404) &         (0.00101) &        (0.000612) &        (0.000718) \\
    &                  & MSE &           0.00263 &          0.00544 &           0.00461 &            0.0334 &            0.0303 &            0.0306 \\
    &                  & N &              2850 &             2850 &              2850 &             38633 &             38633 &             38633 \\
\cline{1-9}
\cline{2-9}
\multirow{12}{*}{Soy} & \multirow{4}{*}{Yearly Linear} & $\theta$ &  -0.00409$^{***}$ &  -0.0055$^{***}$ &  -0.00409$^{***}$ &  -0.00537$^{***}$ &  -0.00525$^{***}$ &  -0.00525$^{***}$ \\
    &                  & \  &        (0.000307) &       (0.000574) &        (0.000331) &         (5.2e-05) &        (8.07e-05) &        (0.000252) \\
    &                  & MSE &           0.00579 &          0.00562 &           0.00578 &            0.0308 &             0.029 &            0.0291 \\
    &                  & N &              2422 &             2422 &              2422 &             33799 &             33799 &             33799 \\
\cline{2-9}
    & \multirow{4}{*}{Yearly Flexible} & $\theta$ &           0.00805 &        -7.48e-05 &           0.00239 &  -0.00788$^{***}$ &  -0.00698$^{***}$ &  -0.00764$^{***}$ \\
    &                  & \  &          (0.0079) &        (0.00276) &          (0.0026) &         (0.00109) &        (0.000843) &        (0.000648) \\
    &                  & MSE &           0.00432 &           0.0056 &           0.00519 &              0.03 &            0.0276 &            0.0282 \\
    &                  & N &              2422 &             2422 &              2422 &             33799 &             33799 &             33799 \\
\cline{2-9}
    & \multirow{4}{*}{Monthly Flexible} & $\theta$ &           0.00955 &         0.000192 &           0.00247 &  -0.00565$^{***}$ &  -0.00362$^{***}$ &  -0.00465$^{***}$ \\
    &                  & \  &         (0.00914) &         (0.0069) &         (0.00572) &         (0.00104) &        (0.000717) &        (0.000611) \\
    &                  & MSE &           0.00204 &          0.00451 &           0.00421 &            0.0253 &            0.0236 &            0.0234 \\
    &                  & N &              2422 &             2422 &              2422 &             33799 &             33799 &             33799 \\
\bottomrule
\end{tabular}

}
\caption[Summary of Bootstrap Trials]{Summary of results from 500 bootstrap draws of the estimation procedure. $\theta$ is the mean of the average derivative over bootstrap trials. Standard errors are in parentheses, and are computed from the distribution of bootstrap values. N is the average number of samples per simulation trial. Stars indicate significance at the $p=0.05 (^{*}), 0.01 (^{**}), \text{ and } 0.001 (^{***})$ levels, based on a Z-score from the mean and standard error of bootstrap trials.}
\label{Tab: adaptation results }
\end{table}

Reassuringly, the results using Yearly Linear weather variation are similar to those of \citet{Burke2016}.
\Cref{Tab: adaptation results } summarizes the results from the bootstrap trials.
Long-run and short-run datasets both find that additional extreme heat decreases crop yields, with comparable magnitudes for both corn and soy cultivation.
The magnitude of these elasticities are economically significant -- an estimate of -0.005 implies that crop yields decline by 0.5\% for each additional day crops are exposed to temperatures above 29\textdegree C.
The findings are similar to the results reported by \citet{Burke2016}. 
They find that the elasticity of corn yield ranges from $-0.0037$ to $-0.0062$. 

As in \citet{Burke2016}, with Yearly Linear weather variation I find little to no evidence that damages from the short run are offset in the long run. 
\Cref{fig:ag_results} shows the results of taking bootstrap samples of the ratio $1 - \hat{\theta}^{LR}/\hat{\theta}^{SR}$. 
As described in \Cref{sec:adaptation}, $\hat{\theta}^{SR} \ (\hat{\theta}^{LR}) $ is the estimated elasticity with short-run (long-run) variation. 
For corn cultivation, I fail to reject the null hypothesis that this ratio is different from 0 for any estimation method.
I test at the $p = 0.05$ level, with a Bonferroni correction to account for taking three hypothesis tests.  
For soy cultivation, I find mixed results: for both OLS Linear and NNet DML I reject the null hypothesis at this level, while with Lasso DML I fail to reject the null hypothesis.
For these soy estimates, the 95\% confidence interval is $[0.1278, 0.3488]$ using OLS Linear and $[0.07153, 0.3649]$ using NNet DML.
Confidence intervals are computed using the sample mean and standard deviation among bootstrap trials. 

Using Yearly Flexible weather variables, I find significant evidence of adaptation. 
The panel with short-run variation finds statistically and economically significant damages from a marginal increase in extreme heat exposure. 
Central estimates range from $-0.01057$ to $-0.01248$ for corn, and $-0.006983$ to $-0.00788$ for soy. 
However, I fail to reject the null hypothesis that the long-run elasticity is different from zero, for both crops using all specifications.
\Cref{fig:ag_results} confirms this finding.
For both crops and using all estimation methods, I reject the null hypothesis that the degree of short-run damages that are offset in the long run is equal to zero. 

Using this set of weather variables, my preferred estimator is NNet DML. 
Simulation results showed the this estimator performed best with Yearly Flexible weather variables. 
The MSE also suggests that the NNet Double is performing well in this case. 
The MSE of NNet DML decreases for long-run regressions for both corn and soy. 
As the MSE of NNet DML is the test-set MSE, an improvement relative to the MSE from the Yearly Linear weather variables reflects am improvement in modeling the true functional form instead of overfitting. 
% The MSE of OLS Linear also decreases substantially for long-run regressions for both corn and soy, . 
Using this preferred estimate, I find a 95\% confidence interval that the ratio is within $[0.3023, 1.222]$ for corn and $[0.6331, 2.001]$ for soy. 

The results from using the Monthly Flexible set of weather variables are similar to those using Yearly Flexible weather variables, although with greater variance. 
This greater variance is not surprising, as the simulation results demonstrated that DML and OLS have higher bias and variance using this set of weather variables. 
With short-run variation using this set of weather variables, there is an economically and statistically significant decline in yields associated with a marginal increase in heat exposure. 
With long-run variation, there is no evidence of significant declines in yields from a marginal increase in heat exposure. 
Reassuringly, this evidence is consistent with the result using Yearly Flexible weather variation. 
Given the greater variance of these estimates, I fail to reject the null hypothesis that the ratio of these elasticities is different from zero except using DML methods. Using OLS Linear, I find that a marginal increase in heat exposure significantly increases crop yields. 

These results show that, when the temperature-crop yield relationship is modeled flexibly, there are not significant average declines in yields associated with a marginal change in long-run damaging heat exposure.
I do not estimate a significant decline from long-run exposure to temperatures above 29\textdegree C because other variation is better able to explain differences in average long-run yields.
This is not the case when using short-run variation, where flexible modeling confirms that a marginal increase in exposure to temperatures above 29\textdegree C is significantly associated with a decline in crop yields. 
This finding is novel, and suggests significant adaptation to damaging heat exposure.

\begin{figure}[t]
     \centering
     % \hfill 
      \begin{subfigure}[b]{0.4\textwidth}
         \centering
         \includegraphics[width=\textwidth]{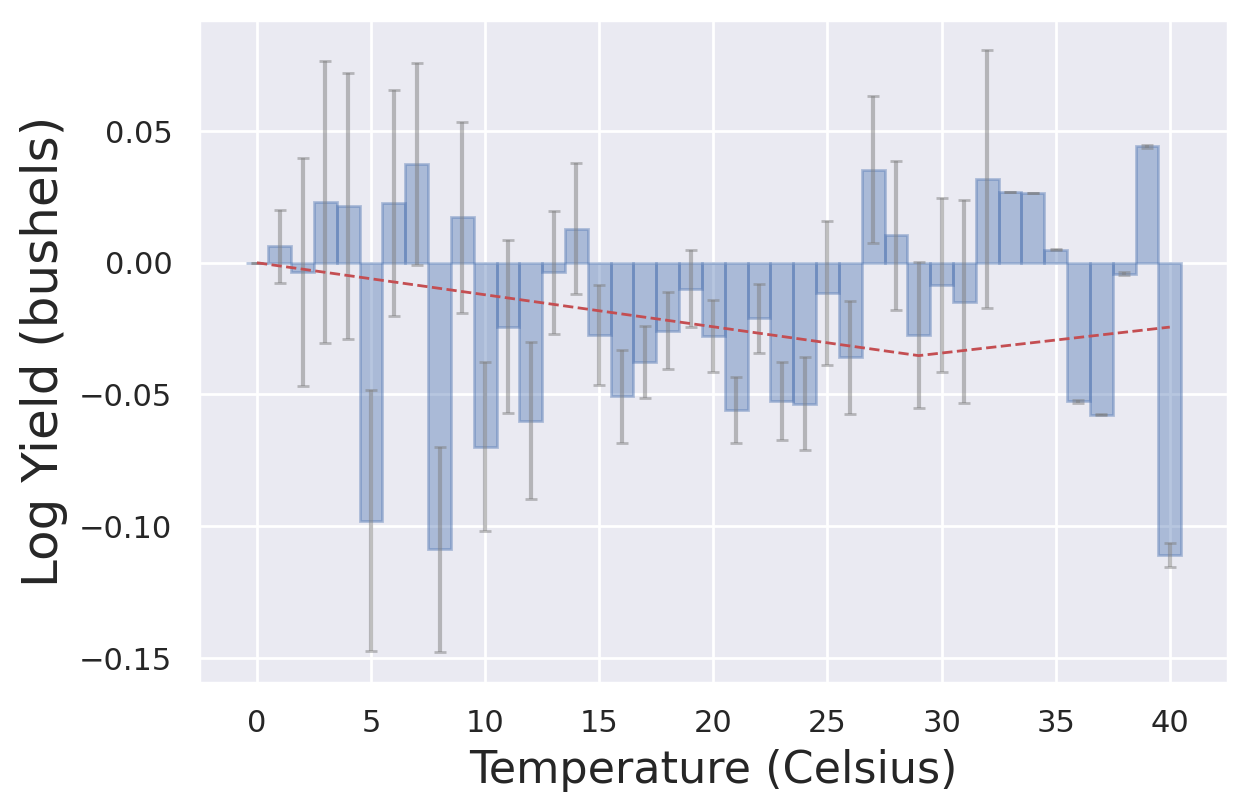}
         \caption{Corn, Long-Run Variation}
         \label{fig:corn long diff nnet}
     \end{subfigure}
     \begin{subfigure}[b]{0.4\textwidth}
         \centering
         \includegraphics[width=\textwidth]{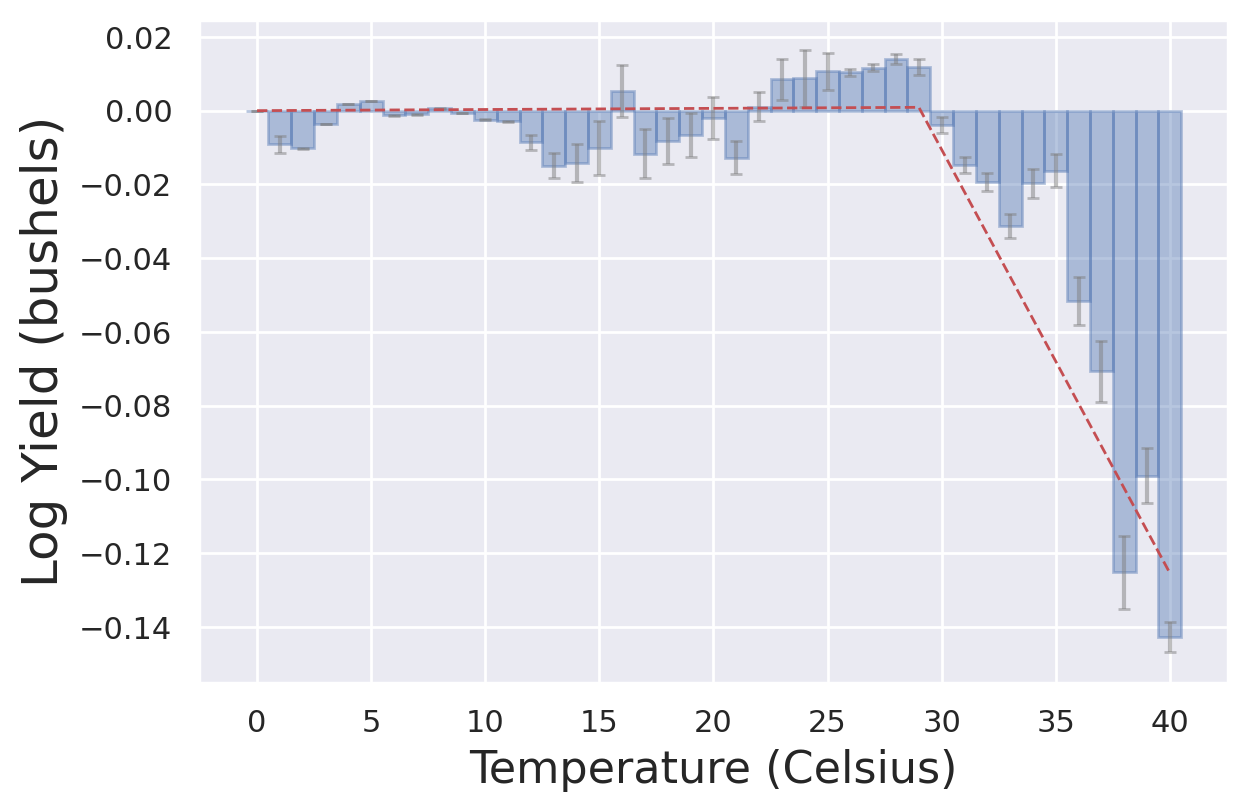}
         \caption{Corn, Short-Run Variation }
         \label{fig:corn panel nnet}
     \end{subfigure}
      \begin{subfigure}[b]{0.4\textwidth}
         \centering
         \includegraphics[width=\textwidth]{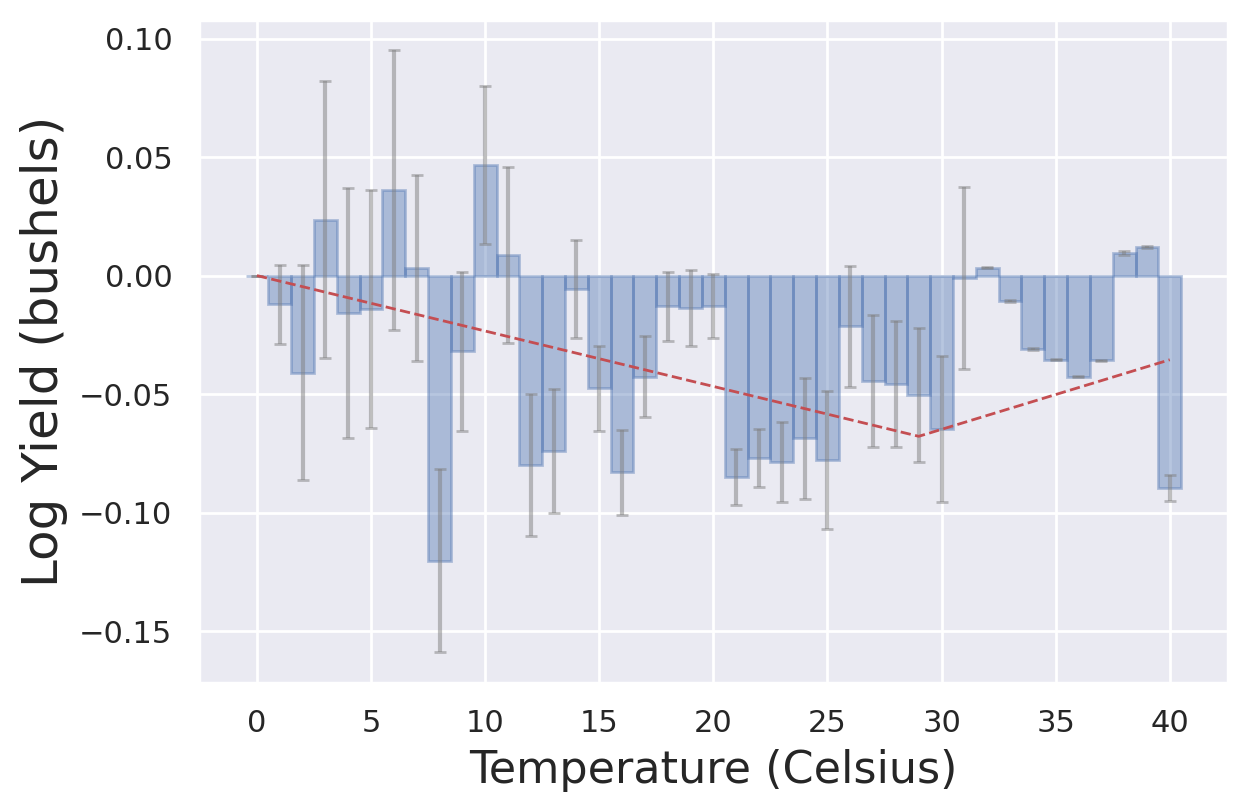}
         \caption{Soy, Long-Run Variation}
         \label{fig:soy long diff nnet}
     \end{subfigure}
     \begin{subfigure}[b]{0.4\textwidth}
         \centering
         \includegraphics[width=\textwidth]{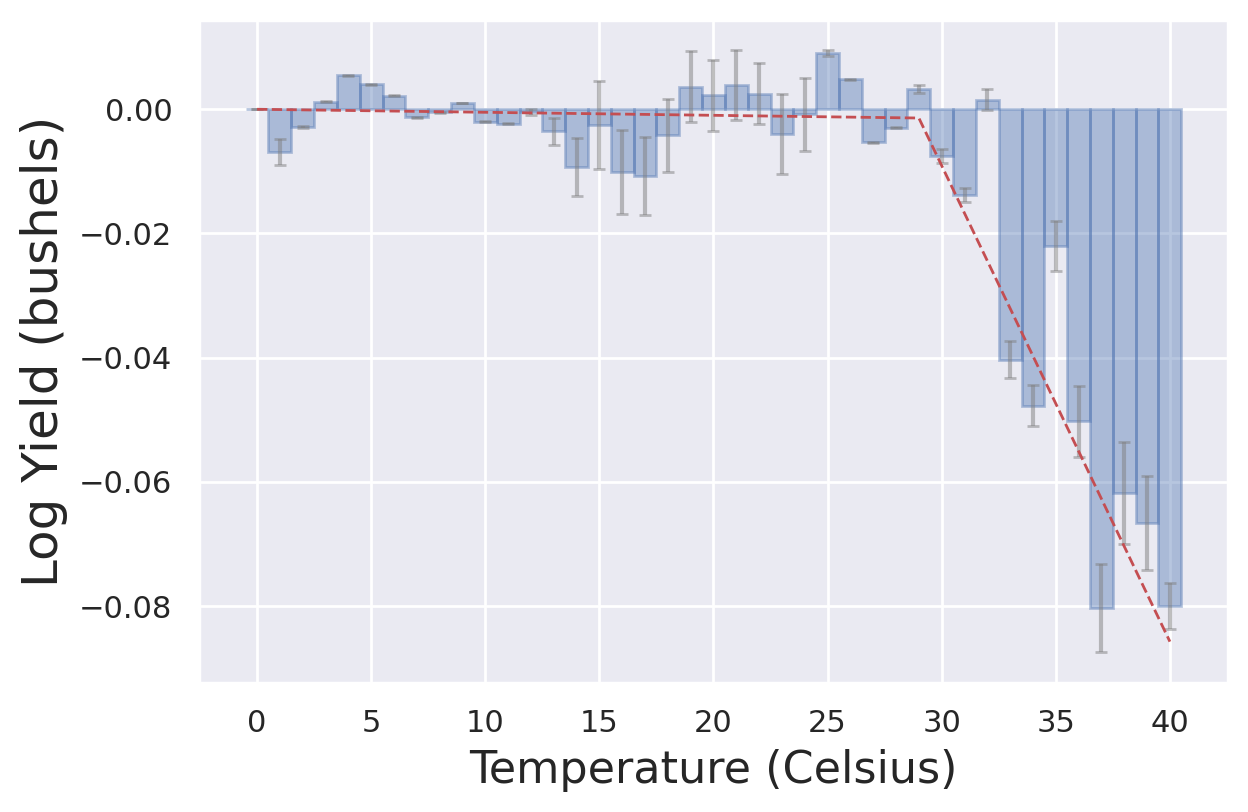}
         \caption{Soy, Short-Run Variation }
         \label{fig:soy panel nnet}
     \end{subfigure}
     % \hfill 
        \caption[NNet DML Estimates for Yearly Flexible]{Comparison of relationship between temperature and yields with short-run and long-run variation. 
        Each plot summarizes the coefficient estimates from results with the Yearly Flexible set of weather variables, using NNet DML. 
        The bar plot shows the expected change in yields from exposure to a single degree day at the given temperature level. 
        Error bars show one standard deviation above/below each estimated coefficient. 
        The line shows the best piecewise linear fit to the bar plot, with one piece from 0-29\textdegree C and another from 29-40 \textdegree C. 
        % Each observation is weighted by acres of crop planted. 
}
\label{fig:NNet Estimates for Yearly Flexible}
\end{figure}

To better explain this result, I examine the estimates of temperature coefficients from each weather bin of the Yearly Flexible weather variables. 
I use my NNet DML procedure to estimate these plots. \Cref{sec: alternate weather bin plots} includes results using alternate estimation procedures. 
Due to the computational cost of estimating each average derivative using the DML procedure, I do not replicate the DML procedure for each of these temperature coefficients; the results are similar. 
% Given the results from \Cref{Tab: adaptation results }, I expect central estimates from the DML procedure to be similar, but variance in estimates to be smaller. 
\Cref{fig:NNet Estimates for Yearly Flexible} visualizes the results of these regressions. 
\Cref{fig:corn panel nnet} and \Cref{fig:soy panel nnet} show the results of this estimate using short-run variation. 
These results are similar to the results from \citet{Schlenker2009}, who noted that the piecewise linear fit (the dotted line in these figures) explains most of the variation from the flexible modeling approach (the bar chart in these figures). 
Here, the piecewise linear model seems to be a good fit: before around 29\textdegree C, additional heat exposure is generally associated with an increase in yields. After this point, additional heat exposure is associated with a decline in yields. 
% Like \citet{Burke2016}, I find that 29\textdegree C appears to be threshold for damaging heat exposure. 
% This flexible modeling also confirms that 29\textdegree C is a reasonable threshold for when beneficial heat exposure becomes damaging heat exposure. 
% This seems to be largely true given \Cref{Tab: adaptation results} - when using panel data, there is little improvement in mean squared error from using more flexible sources of variation or different estimation techniques. 

The results using long-run variation do not share this pattern. 
\Cref{fig:corn long diff nnet} and \Cref{fig:soy long diff nnet} show the results of this estimate using long-run variation. 
The coefficients do not demonstrate a clear piecewise linear pattern. 
This suggests that the piecewise linear model relating temperature exposure and crop yield may not be appropriate in long-run models. 
In \ref{sec: alternate weather bin plots}, I replicate this figure for different time periods, starting from 1950-1979. These figures confirm that the piecewise linear model is a better fit when using short-run variation than when using long run variation.
While the results from the Yearly Linear analysis show that long-run damaging heat exposure is correlated with a decline in crop yields, these results show that other temperature variation is better able to explain the long-run changes in crop yields. 
This shows the importance of flexible modeling to understanding the role of a marginal increase in damaging heat exposure.

% Another important element in my conclusion is the definition of marginal increase in heat exposure above 29\textdegree C. 
% The density plots in each panel of \Cref{fig:OLS Estimates for Yearly Flexible} show that density of GDD rapidly declines as the temperature increases above 29\textdegree C. 
% I assume that a marginal increase in this damaging heat exposure will be additional to the observed distribution of GDD.
% This places less weight on the coefficients for higher temperature bins. 
% An alternate assumption that places more weight on higher temperature bins may come to a different conclusion about the impacts of marginal increase in long-run heat exposure, although it is unclear why additional heat exposure would not follow the empirical distribution. 

\section{Discussion}
\label{sec:Discussion}

I introduced a DML procedure that can estimate average directional derivatives in high-dimensional settings, and demonstrated the benefits of this procedure over OLS estimation in a simulation exercise. 
The results show that this procedure can estimate average directional derivatives with lower bias and variance than OLS. 
The simulation results also show that my method is less reliable than having access to a true, parsimonious model of the underlying function.
In settings where the researcher has high-dimensional weather variation and does not have a strong prior about the true functional form, this DML approach can be used to estimate elasticities and the degree of adaptation to a weather feature. 

Applying this estimator to panel of U.S. corn and soy yields with a rich set of temperature variation, I conclude that there has been significant adaptation to damaging heat exposure.
Using this flexible method, a panel of short-run damages finds evidence that a marginal increase in heat exposure above 29\textdegree C is damaging for both corn and soy yields. 
However, by constructing a dataset of average changes to capture long-run variation, I cannot reject the null hypothesis that a marginal increase in long-run damaging heat exposure is unrelated to long-run crop yields.
OLS estimates using a Yearly Flexible set of weather variables support this finding.
I demonstrate that long-run exposure to temperatures above 29\textdegree C is not clearly associated with yield declines, as is the case with short-run exposures.
This implies that there has been significant adaptation to climate change in this setting, as the short-run damages are significantly offset in the long run. 

This approach does not offer insight into what form this adaptation may take. 
Farmers have many possible adaptation mechanisms, such as investing in irrigation, purchasing improved seeds, or adjusting planting times.
It is important to learn which mechanisms may be responsible for offsetting damages in the long run, in order to study the long-term effectiveness of these mechanisms or the ability to use them in other agricultural contexts.
More research is needed to understand how these damages are offset.  

The apparent contradiction between this result and prior literature suggests that while there is adaptation to a marginal increase in damaging heat exposure, there is limited adaptation to some other damaging feature of climate change.
Prior literature, such as \citet{Burke2016}, did not find evidence of substantial adaptation to damaging heat exposure. 
My analysis shows that while a long-run increase in damaging heat exposure is correlated with a decline in crop yields, variation in heat exposure at other temperature levels is better able to explain this pattern.
This is evidence of omitted variable bias in long-run estimates using only beneficial and damaging heat exposure. 
While prior results indicate limited adaptation to some change in the temperature distribution, my detailed analysis finds that there has been substantial adaptation to damaging heat exposure. 

\FloatBarrier
\newpage 
\singlespacing
% \bibliography{/Users/maxv/Documents/bib/thesis_ref.bib}
\printbibliography

\newpage 
\doublespacing
\appendix 

\renewcommand{\thesubsection}{Appendix \Alph{section}.\arabic{subsection}}
\renewcommand{\thesection}{Appendix \Alph{section}}

\FloatBarrier

\section{Machine Learning Estimation Details}
\label{sec:Appendix Double Machine Learning}

This appendix additional details on the ML estimation procedures. 

\subsection{Lasso}
\label{sec:Appendix Lasso}
I vary the Lasso hyperparameter (the regularization penalty) on a grid of 15 values, evenly distributed (in log space) from $10^{-10}$ to $10^0$. 
The cross folds validation procedure generally selects a hyperparameter from the interior of this grid, suggesting that the range is appropriate. 

Recall the form of the optimization problem: 
\begin{equation}
\hat{\beta}_\ell = \argmin_{\beta} \sum_{i \in \mathcal{I}_\ell} \sum_{t = 1}^{T} w_i (\ddot{y}_{it} - \ddot{b}(X_{it})\beta)^2 + \lambda |\beta|_1
\end{equation}

As this is a (weakly) convex optimization problem, I solve for $\hat{\beta}_\ell$ using the optimization package CVXPY \citep{diamond2016cvxpy} with the optimizer Mosek \citep{mosek}. 
In the optimization package implementation, the optimizer can occasionally fail to find a unique optimal solution. 
In these cases, I introduce an $\ell2$ of $10^{-20}$ to find a unique solution. 
Using the optimization package ensures that the Lasso procedure converges, although this procedure is more memory intensive than coordinate descent and will raise an error if it fails to converge.
I find that the model fails to converge in roughly 5\% of simulation trials.

In the main specification, I include yearly fixed effects terms but do not apply a regularization to these terms. 
I accomplish this by selectively applying the regularization factor to terms in $\beta$ not in those yearly fixed effects terms. 
The additional fixed effects are retuned before evaluating the estimator on the test set; this does not impact the estimate of the gradient, but does improve the mean squared error of the estimator. 

I use the optimization package for two reasons. 
When the solver does not return an error, the solution is guaranteed to be optimal, unlike in iterative methods such as stochastic gradient descent or coordinate descent. 
% Additionally, it is possible to customize the loss function to weight observations and selectively impose the penalty term.
Some packages contain this functionality for iterative methods, but I am not aware of an implementation in Python. 

\subsection{Neural Network}
\label{sec:Appendix Neural Network}

I use a relatively simple network configuration, and vary the width of the neural network (NNet) in each simulation trial. 
Three key researcher degrees of freedom when using a NNet are the depth (the number of layers), the width (the number of nodes per hidden layer), and the activation function (the nonlinear function applied to the outputs of each layer).
I use a network with one hidden layer, the minimal depth for which a NNet can approximate an arbitrary function \citep{Hornik1990,Park1991}. 
In each trial run, I use cross-folds validation to select the width of the network.  
To ensure that the network is differentiable, I use the Continuously Differentiable Exponential Linear Unit \citep{Barron2017} as the activation function in the network. 

I select the hyperparameter from a grid space of 2 to 256, evenly spaced in log terms. 
The cross folds procedure generally selects a width from the interior of this set, suggesting that the range is appropriate. 
To train the network, I use batch normalization, the Adam optimizer \citep{Kingma2014}, a learning rate of $0.01$, and $1000$ epochs of training. 
I retune the parametric components (the county and yearly fixed effects terms) before evaluating on the test set.
This does not impact the estimate of the gradient, but does improve the mean squared error of the estimator. 

Simulation results show that this simple configuration performs well, although it is possible that a more complex network configuration would perform better. 
Given the computational cost of exploring a wide range of potential network configurations, I leave such exploration for future work. 

\subsection{Automatic Double Machine Learning}
\label{sec:Appendix Auto DML}
The theoretical basis for the double machine learning (DML) procedure comes from  the Riesz representation theorem.
This theorem states that, for the linear functional\footnote{A functional is a scalar summary of a function. The functional in this setting is the directional derivative.} $m$, there exists a function $\alpha_0$ such that for any function $h$:
\begin{equation}
\E[m(h, X_{i})] = \E[h(X_{i})\alpha_0(X_i)]
\end{equation}
Because this relationship holds regardless of function $h$, one can set known functions for which the target $m(h, X_{i})$ is known and estimate $\hat{\alpha}$ from the empirical distribution of $X$.
Alternate procedures involve the researcher solving for the functional form of $\alpha$ given the functional $m$, and often estimating densities or derivatives of densities. 
I use this procedure because it does not require the researcher specifying the form of $\alpha$, but instead learns $\hat{\alpha}$ from data. 

I follow \citet{chernozhukov2022automatic} to estimate $\hat{\alpha}$ using Lasso and the basis functions used to estimate the regression function. 
The goal is to find the estimator that minimizes the mean squared error between $\hat{\alpha}$ and $\alpha_0$, where $\alpha_0$ is the true Riesz representer:
 \begin{equation*}
     \hat{\alpha} = \argmin_\alpha \E[(\alpha_0(X_{i,t}) - \alpha(X_{i,t}))^2]
 \end{equation*}
Using the Lasso functional form, I have $\hat{\alpha} = \ddot{b}(X_{i,t}) \hat{\rho}$.
I find $\hat{\rho}$ by solving the following regularized problem: 
\begin{equation*}
\hat{\rho} = \argmin_\rho \E\left[ (\alpha_0(X_{i,t}) - \ddot{b}(X_{i,t})\rho)^2\right] + \kappa |\rho|_1
\end{equation*}
Expanding the polynomial and applying the definition of the Riesz representer, the expression simplifies:
\begin{equation*}
\hat{\rho} = \argmin_\rho -2\E[ m(\ddot{b},X_{i,t}) ]\rho + \rho' \E[\ddot{b}(X_{i,t})' \ddot{b}(X_{i,t})]\rho +  \kappa |\rho|_1
\end{equation*}
\citet{chernozhukov2022automatic} include an additional iterative procedure to scale each component's regularization term by the inverse of its variance; I implement their suggested procedure without modification. 
% To incorporate weights into this estimation, I use weighted averages when constructing $\E[ m(\ddot{b},X_{i,t}) ]$ and $\E[\ddot{b}(X_{i,t})' \ddot{b}(X_{i,t})]$. I also multiply each observation in the cross folds procedure by its weight. 

As this is a (weakly) convex optimization problem, I use the optimization package Mosek \citep{mosek} to find the optimal value of parameter vector $\hat{\rho}$ for a given value of regularization $\kappa$.  
\citet{klosin2022} present simulation results comparing this optimization method to the iterative approach from \citet{chernozhukov2022automatic}; they find that the optimization approach results in lower mean squared error in estimating the true average derivative. 

I select the value of the hyperparameter $\kappa$ by minimizing the above loss function through the cross-folds procedure described in \Cref{sec:ML methods}.
That is: 
\begin{equation*}
\mathcal{L}_\alpha(\hat{\alpha}_{\ell}, X_{i,t}; \kappa) = -2m(\ddot{b},X_{i,t}) \hat{\rho}_{\ell} + \hat{\rho}_{\ell}' \ddot{b}(X_{i,t})' \ddot{b}(X_{i,t})\hat{\rho}_{\ell}
\end{equation*}

I search the hyperparameter over a grid of values suggested by \citet{chernozhukov2022automatic}:
$\kappa = c (N - N_{\ell})^{-1/2} \Phi^{-1} (1 - .05 / p) $, for $c \in \{5/4, 1, 3/4, 5/8, 9/16, 1/2 \}$, where $\Phi^{-1}$ is the inverse of the standard normal density function and $N_{\ell}$ is the size of the set of observations in fold $\ell$.

\FloatBarrier
\section{Additional Results}
\label{sec:Appendix Results}

\begin{figure}[p]
     \centering
     % \hfill 
     \begin{subfigure}[b]{0.8\textwidth}
         \centering
         \includegraphics[width=\textwidth]{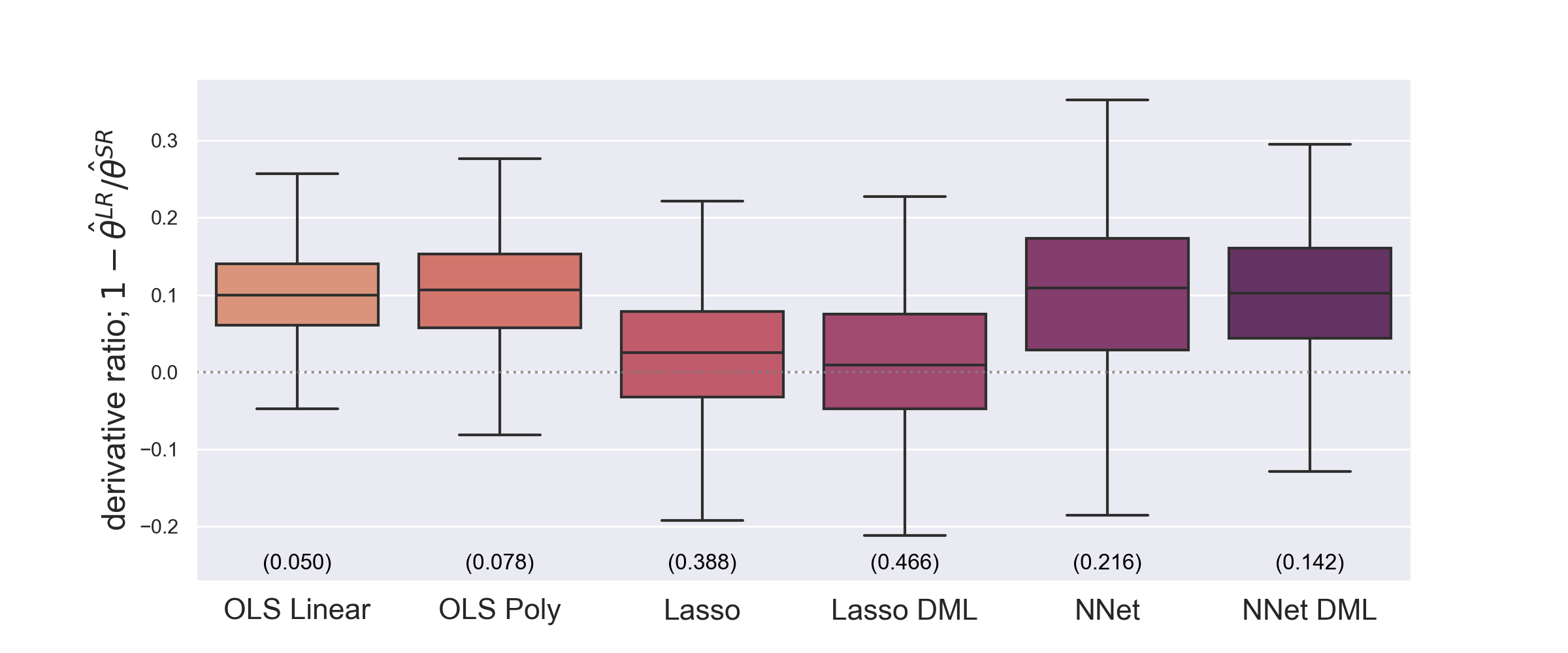}
         \caption{Corn, Yearly Linear}
         \label{fig:corn Yearly Linear box full sim}
     \end{subfigure}
     % \hfill 
     % \newline 
     % \hfill 
      \begin{subfigure}[b]{0.8\textwidth}
         \centering
         \includegraphics[width=\textwidth]{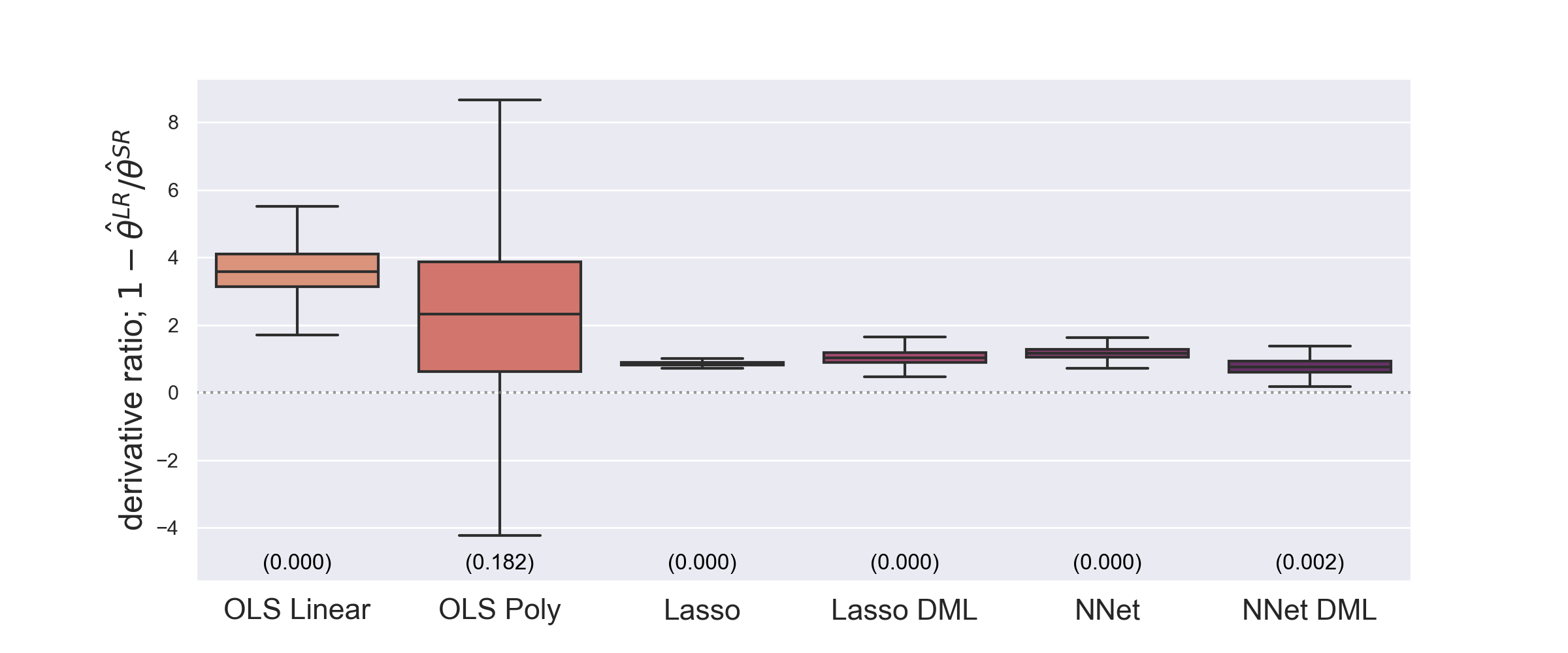}
         \caption{Corn, Yearly Flexible}
         \label{fig:corn Yearly Flexible box full sim}
     \end{subfigure}

      \begin{subfigure}[b]{0.8\textwidth}
         \centering
         \includegraphics[width=\textwidth]{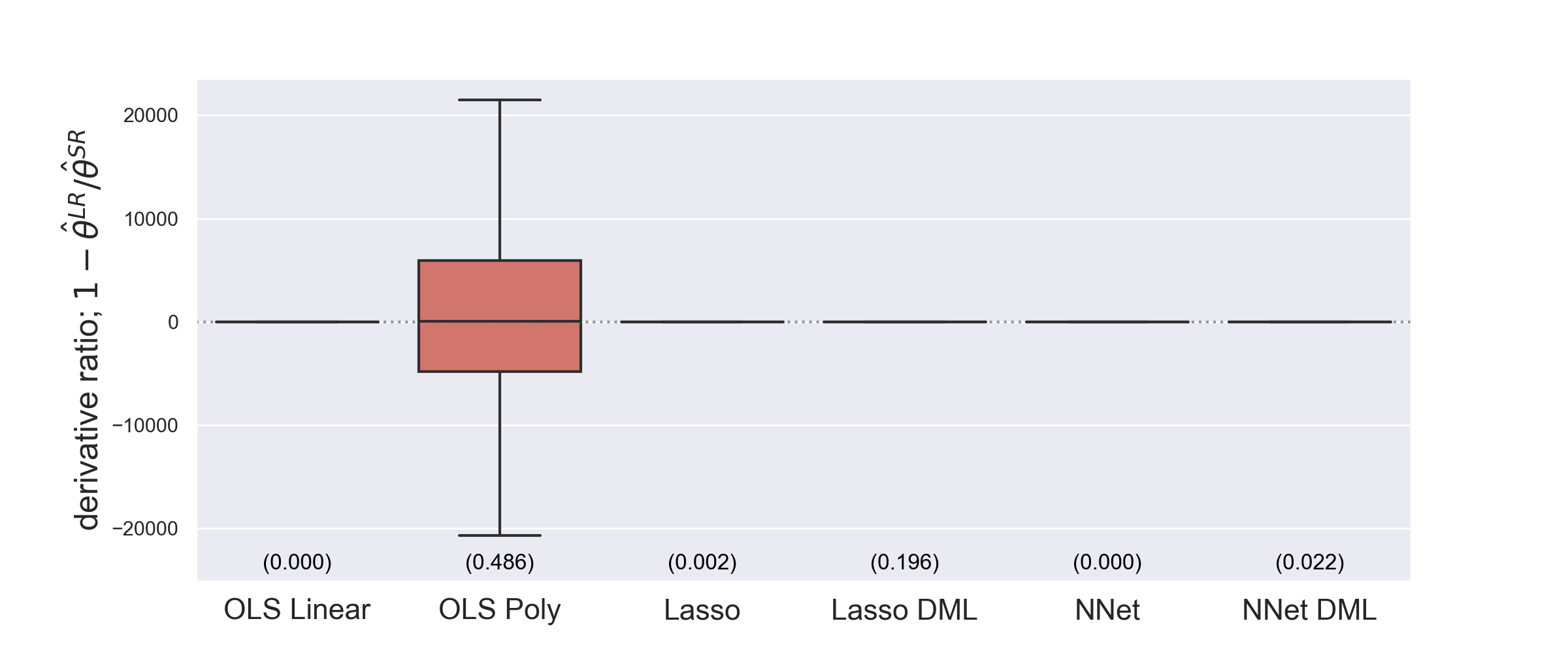}
         \caption{Corn, Monthly Flexible}
         \label{fig:corn monthly box full sim}
     \end{subfigure}
     % \hfill 
        \caption[Appendix: Box Plots of Simulation Results, Corn]{Box plots of 500 bootstrap samples of ratio $1 - \hat{\theta}^{LR}/\hat{\theta}^{SR}$, from corn production. 
        The subfigures indicate which set of weather variables are used, using the notation from \Cref{fig:GDD_explainer}.
        The number in parentheses at the bottom is the $p$ value for the one-sided test that the ratio is greater than 0. 
        Each subfigure has a separate box plot for each method used. The line inside each box is the median, the edges of the box are the upper and lower quartile of the distribution, and the whiskers extending from each box are 1.5 times the interquartile range. 
}
\label{fig:corn box plots}
\end{figure}

\begin{figure}[p]
     \centering
     % \hfill 
     \begin{subfigure}[b]{0.8\textwidth}
         \centering
         \includegraphics[width=\textwidth]{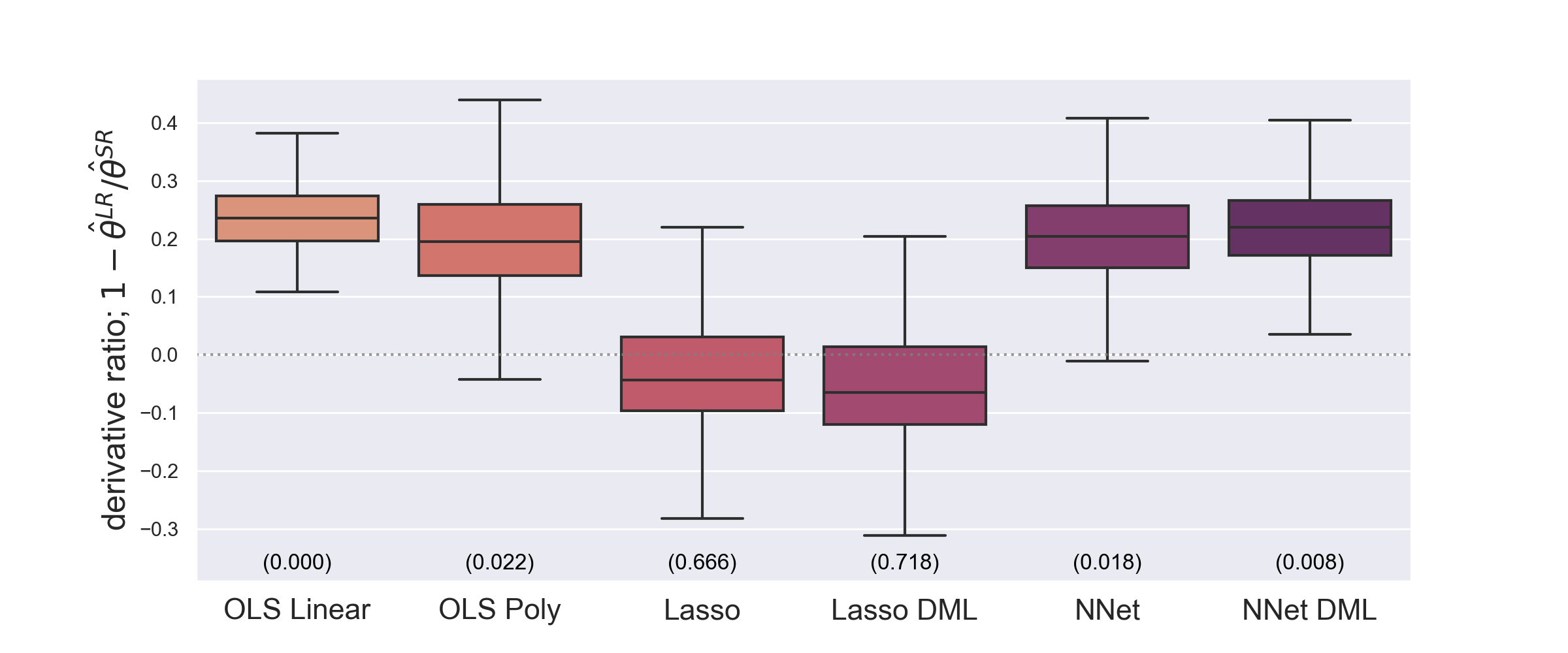}
         \caption{Soy, Yearly Linear}
         \label{fig:soy Yearly Linear box full sim}
     \end{subfigure}
     % \hfill 
     % \newline 
     % \hfill 
      \begin{subfigure}[b]{0.8\textwidth}
         \centering
         \includegraphics[width=\textwidth]{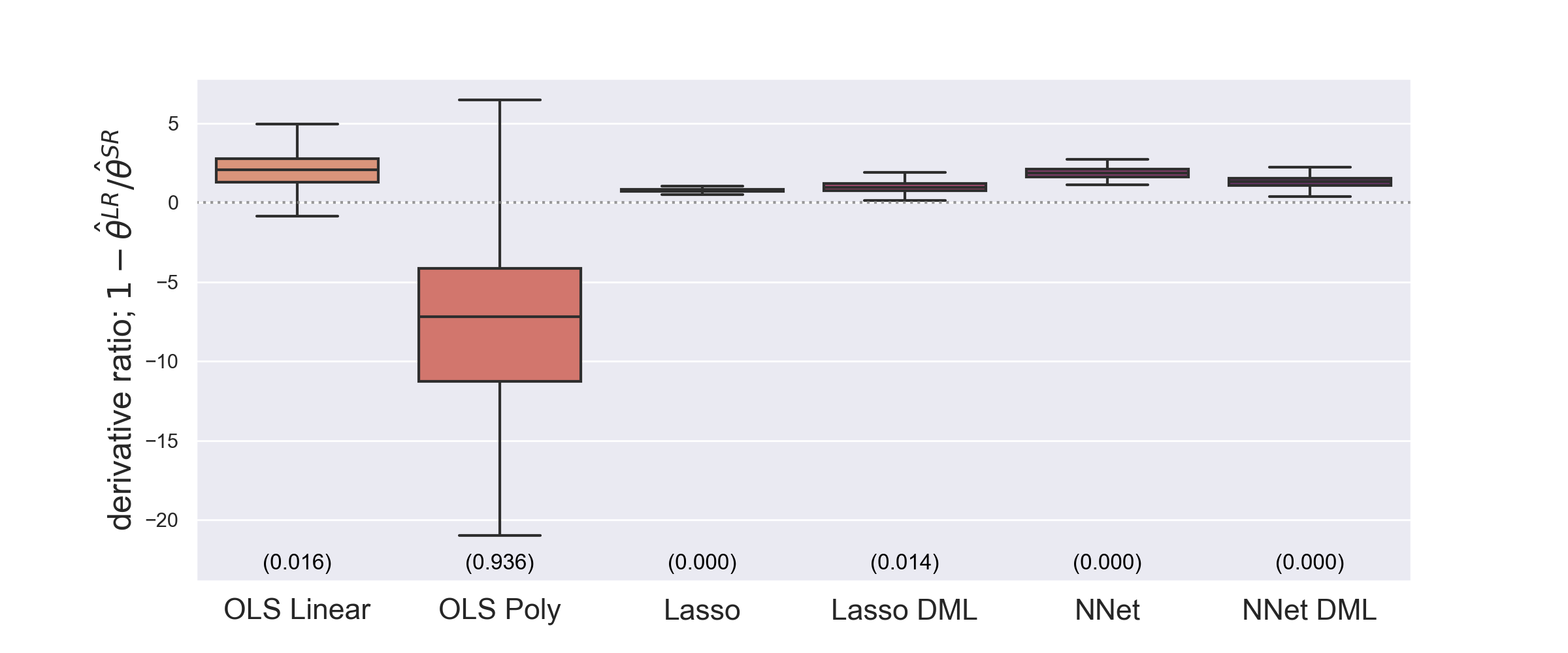}
         \caption{Soy, Yearly Flexible}
         \label{fig:soy Yearly Flexible box full sim}
     \end{subfigure}

     % \hfill 
     % \newline 
     % % \hfill 
      \begin{subfigure}[b]{0.8\textwidth}
         \centering
         \includegraphics[width=\textwidth]{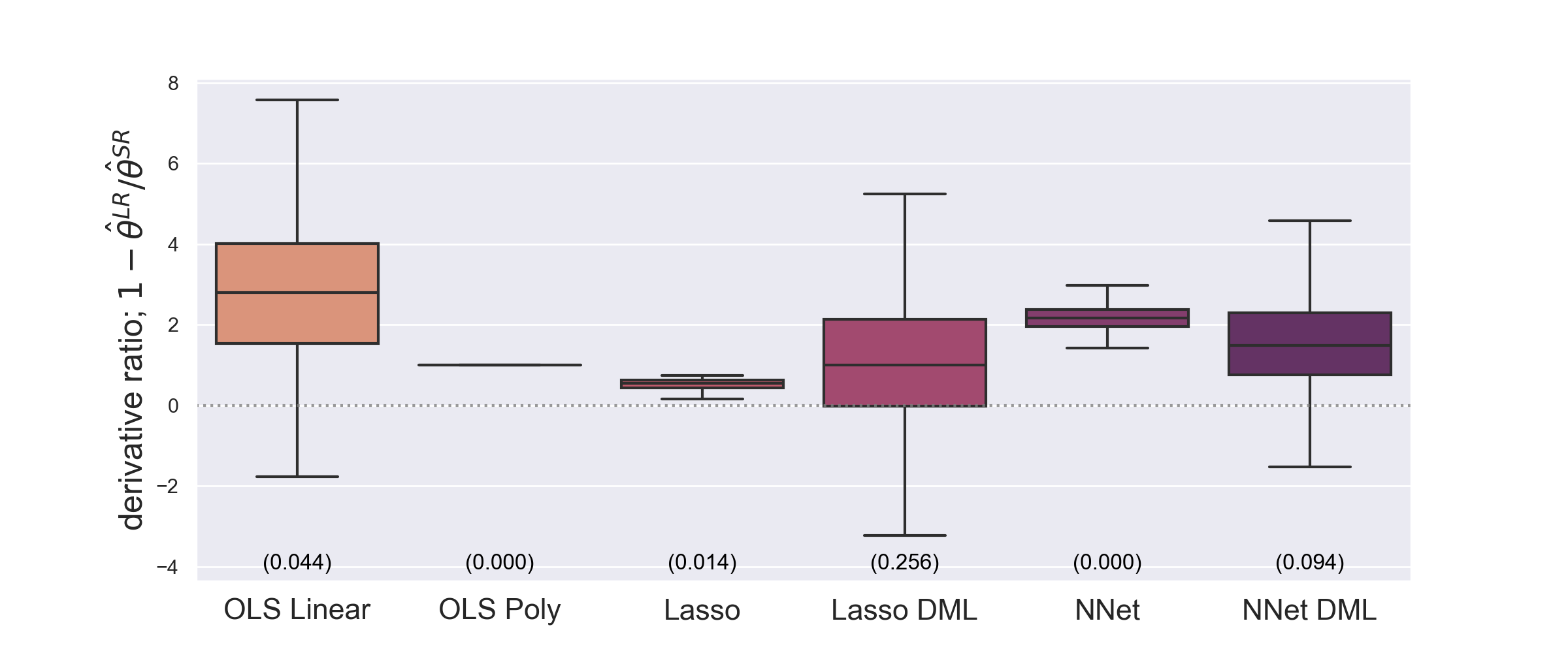}
         \caption{Soy, Monthly Flexible}
         \label{fig:soy monthly box full sim}
     \end{subfigure}
     % \hfill 
        \caption[Appendix: Box Plots of Simulation Results, Soy]{Box plots of 500 bootstrap samples of ratio $1 - \hat{\theta}^{LR}/\hat{\theta}^{SR}$, from soy production. 
        The subfigures indicate which set of weather variables are used, using the notation from \Cref{fig:GDD_explainer}.
        The number in parentheses at the bottom is the $p$ value for the one-sided test that the ratio is greater than 0. 
        Each subfigure has a separate box plot for each method used. The line inside each box is the median, the edges of the box are the upper and lower quartile of the distribution, and the whiskers extending from each box are 1.5 times the interquartile range. 
}
\label{fig:soy box plots}
\end{figure}

This appendix includes additional results from the bootstrap trials described in \Cref{sec:Results}.
\Cref{Tab: corn results } shows the full results from the estimation of corn yields, and \Cref{Tab: soy results } shows the full results from the estimation of soy yields. 
As expected, the OLS Poly results have higher variance than other methods, particularly in higher-dimensional settings. 
The results from machine learning approaches without bias correction have very low variance, but are likely biased relative to the debiased estimates.
After applying the bias correction, the variance typically increases.
\Cref{fig:corn box plots} show the bootstrapped values of the ratio $1 - \hat{\theta}^{LR}/\hat{\theta}^{SR}$ from all models for corn yields, and \Cref{fig:soy box plots} show the bootstrapped values of this ratio from soy yields. 
Note that the variance from the OLS Poly estimates dominates. These standard errors are large and in charge, particularly in \Cref{fig:corn monthly box full sim} and \Cref{fig:soy monthly box full sim}.

\begin{table}[p]
\footnotesize
\resizebox{\textwidth}{!}{
\begin{tabular}{lllcccccc}
\toprule
          &                  & method &        OLS Linear &          OLS Poly &             Lasso &         Lasso DML &              NNet &          NNet DML \\
Variation & Weather & {} &                   &                   &                   &                   &                   &                   \\
\midrule
\multirow{12}{*}{Long-Run} & \multirow{4}{*}{Yearly Linear} & $\theta$ &  -0.00481$^{***}$ &   -0.0063$^{***}$ &  -0.00692$^{***}$ &    -0.007$^{***}$ &  -0.00583$^{***}$ &  -0.00597$^{***}$ \\
          &                  & \  &        (0.000326) &         (0.00052) &        (0.000547) &        (0.000568) &        (0.000485) &        (0.000464) \\
          &                  & MSE &           0.00652 &           0.00584 &           0.00648 &           0.00648 &           0.00673 &           0.00673 \\
          &                  & N &              2850 &              2850 &              2850 &              2850 &              2850 &              2850 \\
\cline{2-9}
          & \multirow{4}{*}{Yearly Flexible} & $\theta$ &    0.0322$^{***}$ &             0.014 &          -0.00135 &          0.000429 &           0.00147 &          -0.00249 \\
          &                  & \  &         (0.00826) &          (0.0294) &        (0.000882) &         (0.00234) &         (0.00155) &         (0.00247) \\
          &                  & MSE &           0.00531 &           0.00159 &           0.00582 &           0.00582 &           0.00603 &           0.00603 \\
          &                  & N &              2850 &              2850 &              2850 &              2850 &              2850 &              2850 \\
\cline{2-9}
          & \multirow{4}{*}{Monthly Flexible} & $\theta$ &      0.0213$^{*}$ &         -5.83e+04 &  -0.00184$^{***}$ &         -0.000793 &    0.00356$^{**}$ &           0.00148 \\
          &                  & \  &         (0.00874) &        (1.25e+06) &         (0.00042) &         (0.00533) &         (0.00126) &         (0.00404) \\
          &                  & MSE &           0.00263 &          3.72e-06 &           0.00544 &           0.00544 &           0.00461 &           0.00461 \\
          &                  & N &              2850 &              2850 &              2850 &              2850 &              2850 &              2850 \\
\cline{1-9}
\cline{2-9}
\multirow{12}{*}{Short-Run} & \multirow{4}{*}{Yearly Linear} & $\theta$ &  -0.00535$^{***}$ &  -0.00704$^{***}$ &  -0.00708$^{***}$ &  -0.00709$^{***}$ &  -0.00645$^{***}$ &  -0.00663$^{***}$ \\
          &                  & \  &        (5.68e-05) &        (7.64e-05) &        (7.96e-05) &        (8.05e-05) &        (0.000595) &        (0.000337) \\
          &                  & MSE &            0.0387 &            0.0363 &            0.0363 &            0.0363 &            0.0362 &            0.0362 \\
          &                  & N &             38633 &             38633 &             38633 &             38633 &             38633 &             38633 \\
\cline{2-9}
          & \multirow{4}{*}{Yearly Flexible} & $\theta$ &   -0.0125$^{***}$ &   -0.0115$^{***}$ &   -0.0106$^{***}$ &   -0.0106$^{***}$ &  -0.00845$^{***}$ &   -0.0104$^{***}$ \\
          &                  & \  &         (0.00106) &        (0.000988) &         (0.00112) &         (0.00115) &         (0.00055) &        (0.000603) \\
          &                  & MSE &            0.0376 &            0.0316 &            0.0343 &            0.0343 &            0.0352 &            0.0352 \\
          &                  & N &             38633 &             38633 &             38633 &             38633 &             38633 &             38633 \\
\cline{2-9}
          & \multirow{4}{*}{Monthly Flexible} & $\theta$ &   -0.0114$^{***}$ &               -14 &   -0.0038$^{***}$ &  -0.00505$^{***}$ &  -0.00562$^{***}$ &  -0.00674$^{***}$ \\
          &                  & \  &         (0.00101) &             (491) &        (0.000589) &        (0.000612) &        (0.000677) &        (0.000718) \\
          &                  & MSE &            0.0334 &            0.0258 &            0.0303 &            0.0303 &            0.0306 &            0.0306 \\
          &                  & N &             38633 &             38633 &             38633 &             38633 &             38633 &             38633 \\
\bottomrule
\end{tabular}

}
\caption[Appendix: Summary of Corn Bootstrap Results]{Summary of results from 500 bootstrap draws of the estimation procedure, for corn production.  $\theta$ is the mean of the average derivative over bootstrap trials. Standard errors are in parentheses, and are computed from the distribution of bootstrap values. N is the average number of samples per simulation trial. Stars indicate significance at the $p=0.05 (^{*}), 0.01 (^{**}), \text{ and } 0.001 (^{***})$ levels, based on a Z-score from the mean and standard error of bootstrap trials.}
\label{Tab: corn results }
\end{table}

\begin{table}[p]
\resizebox{\textwidth}{!}{
\begin{tabular}{lllcccccc}
\toprule
          &                  & method &        OLS Linear &          OLS Poly &             Lasso &         Lasso DML &              NNet &          NNet DML \\
Variation & Weather & {} &                   &                   &                   &                   &                   &                   \\
\midrule
\multirow{12}{*}{Long-Run} & \multirow{4}{*}{Yearly Linear} & $\theta$ &  -0.00409$^{***}$ &   -0.0042$^{***}$ &  -0.00538$^{***}$ &   -0.0055$^{***}$ &  -0.00404$^{***}$ &  -0.00409$^{***}$ \\
          &                  & \  &        (0.000307) &        (0.000485) &        (0.000585) &        (0.000574) &        (0.000374) &        (0.000331) \\
          &                  & MSE &           0.00579 &           0.00479 &           0.00562 &           0.00562 &           0.00578 &           0.00578 \\
          &                  & N &              2422 &              2422 &              2422 &              2422 &              2422 &              2422 \\
\cline{2-9}
          & \multirow{4}{*}{Yearly Flexible} & $\theta$ &           0.00805 &           -0.0554 &    -0.00169$^{*}$ &         -7.48e-05 &    0.00559$^{**}$ &           0.00239 \\
          &                  & \  &          (0.0079) &          (0.0328) &        (0.000677) &         (0.00276) &         (0.00204) &          (0.0026) \\
          &                  & MSE &           0.00432 &          0.000905 &            0.0056 &            0.0056 &           0.00519 &           0.00519 \\
          &                  & N &              2422 &              2422 &              2422 &              2422 &              2422 &              2422 \\
\cline{2-9}
          & \multirow{4}{*}{Monthly Flexible} & $\theta$ &           0.00955 &           0.00374 &  -0.00182$^{***}$ &          0.000192 &   0.00518$^{***}$ &           0.00247 \\
          &                  & \  &         (0.00914) &          (0.0779) &        (0.000299) &          (0.0069) &         (0.00128) &         (0.00572) \\
          &                  & MSE &           0.00204 &          2.74e-23 &           0.00451 &           0.00451 &           0.00421 &           0.00421 \\
          &                  & N &              2422 &              2422 &              2422 &              2422 &              2422 &              2422 \\
\cline{1-9}
\cline{2-9}
\multirow{12}{*}{Short-Run} & \multirow{4}{*}{Yearly Linear} & $\theta$ &  -0.00537$^{***}$ &  -0.00522$^{***}$ &  -0.00525$^{***}$ &  -0.00525$^{***}$ &  -0.00509$^{***}$ &  -0.00525$^{***}$ \\
          &                  & \  &         (5.2e-05) &        (7.43e-05) &        (8.01e-05) &        (8.07e-05) &        (0.000321) &        (0.000252) \\
          &                  & MSE &            0.0308 &             0.029 &             0.029 &             0.029 &            0.0291 &            0.0291 \\
          &                  & N &             33799 &             33799 &             33799 &             33799 &             33799 &             33799 \\
\cline{2-9}
          & \multirow{4}{*}{Yearly Flexible} & $\theta$ &  -0.00788$^{***}$ &  -0.00651$^{***}$ &  -0.00686$^{***}$ &  -0.00698$^{***}$ &  -0.00649$^{***}$ &  -0.00764$^{***}$ \\
          &                  & \  &         (0.00109) &         (0.00102) &        (0.000884) &        (0.000843) &        (0.000877) &        (0.000648) \\
          &                  & MSE &              0.03 &            0.0258 &            0.0276 &            0.0276 &            0.0282 &            0.0282 \\
          &                  & N &             33799 &             33799 &             33799 &             33799 &             33799 &             33799 \\
\cline{2-9}
          & \multirow{4}{*}{Monthly Flexible} & $\theta$ &  -0.00565$^{***}$ &             -38.3 &  -0.00396$^{***}$ &  -0.00362$^{***}$ &  -0.00444$^{***}$ &  -0.00465$^{***}$ \\
          &                  & \  &         (0.00104) &         (3.4e+03) &        (0.000839) &        (0.000717) &        (0.000544) &        (0.000611) \\
          &                  & MSE &            0.0253 &            0.0193 &            0.0236 &            0.0236 &            0.0234 &            0.0234 \\
          &                  & N &             33799 &             33799 &             33799 &             33799 &             33799 &             33799 \\
\bottomrule
\end{tabular}

}
\caption[Appendix: Summary of Soy Bootstrap Results]{Summary of results from 500 bootstrap draws of the estimation procedure, for soy production.  $\theta$ is the mean of the average derivative over bootstrap trials. Standard errors are in parentheses, and are computed from the distribution of bootstrap values. N is the average number of samples per simulation trial. Stars indicate significance at the $p=0.05 (^{*}), 0.01 (^{**}), \text{ and } 0.001 (^{***})$ levels, based on a Z-score from the mean and standard error of bootstrap trials.}
\label{Tab: soy results }
\end{table}

\newpage
\FloatBarrier
\section{Short-Run and Long-Run Estimates of Yearly Coefficients}
\label{sec: alternate weather bin plots}
I provide additional results to show the difference between short-run and long-run estimates of yearly coefficient bins. 
These results confirm that the functional form of long-run damages does not follow the well-established pattern of short-run damages. I demonstrate that this conclusions is robust to different modeling choices and time periods. 

\FloatBarrier
\subsection{Comparisons using alternate estimators}
The following plots show results from the comparison between short-run and long-run estimates for each yearly coefficient bin. All plots use the sample from the main estimation, corn and soy yields from 1990-2019. 
Each plot summarizes the coefficient estimates from results with the Yearly Flexible set of weather variables, using either Lasso DML, OLS Poly, or OLS Linear. 
The bar plot shows the expected change in yields from exposure to a single degree day at the given temperature level. 
Error bars show one standard deviation above/below each estimated coefficient. 
The line shows the best piecewise linear fit to the bar plot, with one piece from 0-29\textdegree C and another from 29-40 \textdegree C. 

\begin{figure}[ht]
     \centering
     % \hfill 
      \begin{subfigure}[b]{0.4\textwidth}
         \centering
         \includegraphics[width=\textwidth]{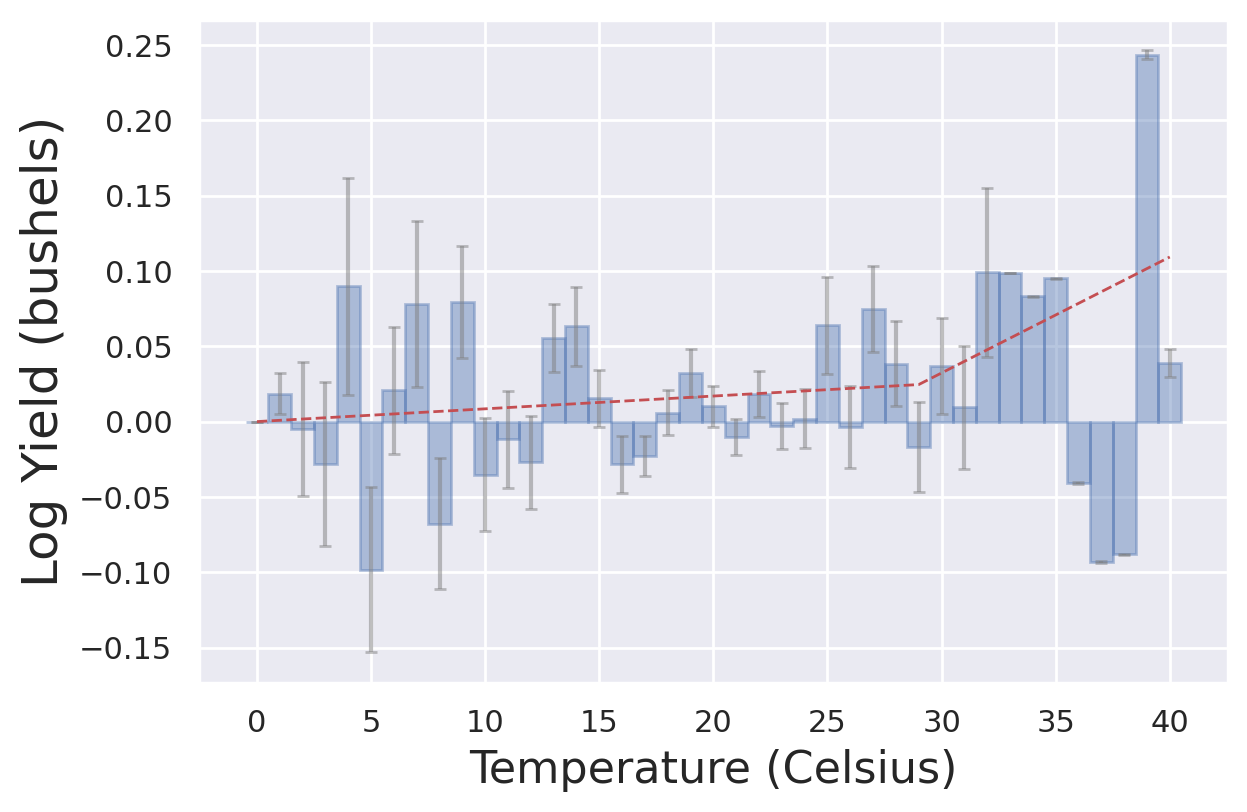}
         \caption{Corn, Long-Run Variation, Lasso DML}
     \end{subfigure}
     \begin{subfigure}[b]{0.4\textwidth}
         \centering
         \includegraphics[width=\textwidth]{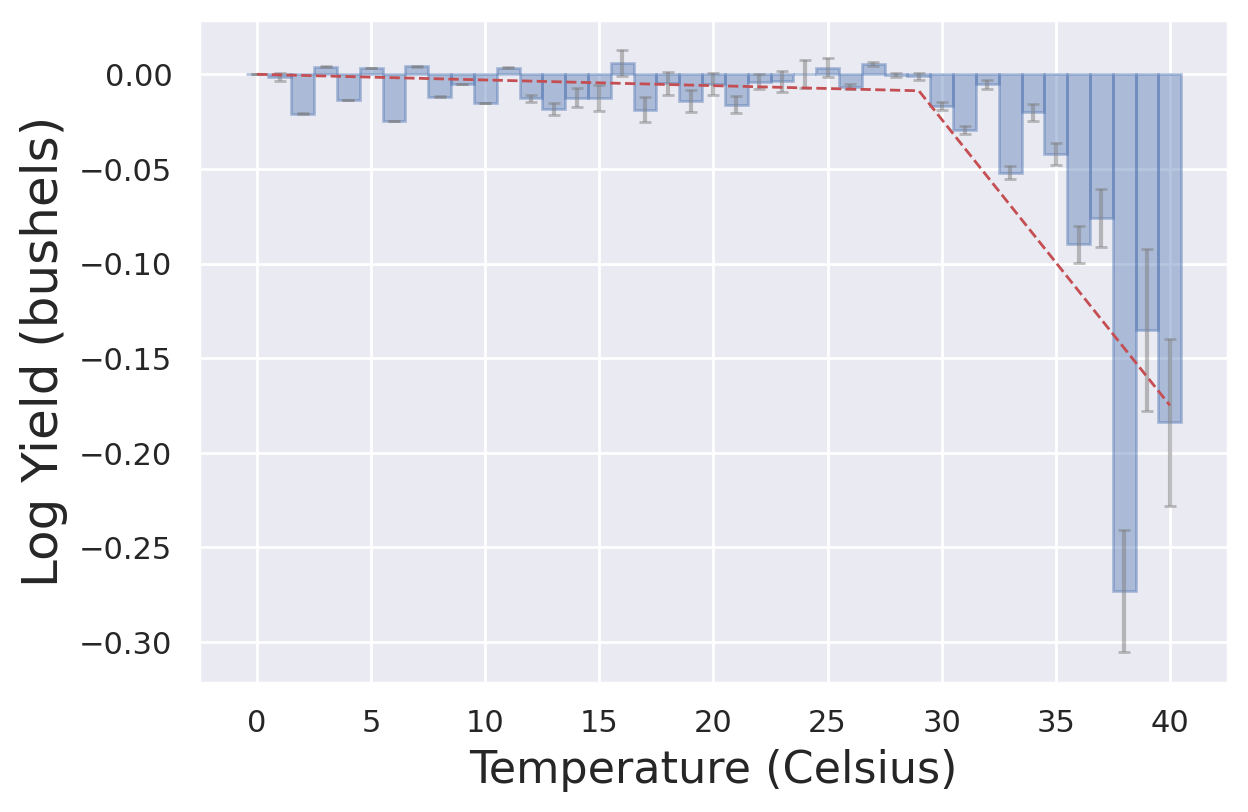}
         \caption{Corn, Short-Run Variation, Lasso  DML}
     \end{subfigure}
      \begin{subfigure}[b]{0.4\textwidth}
         \centering
         \includegraphics[width=\textwidth]{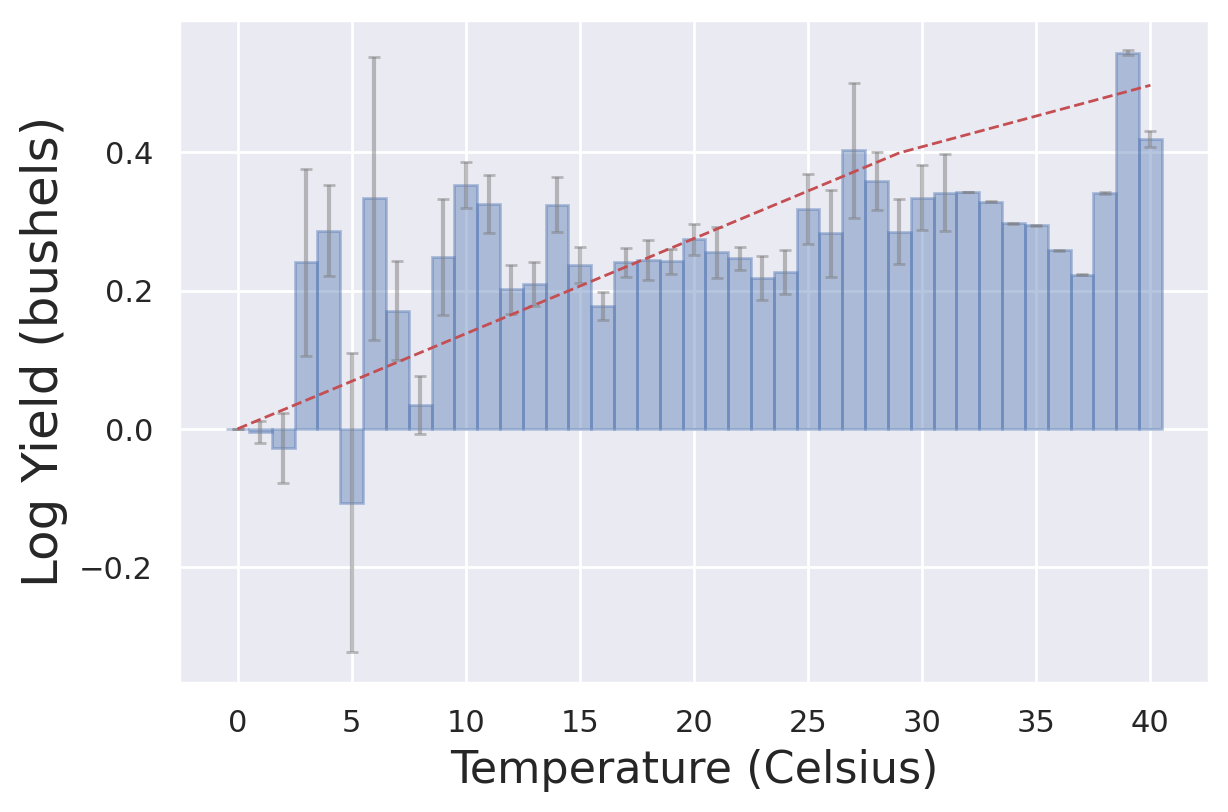}
         \caption{Soy, Long-Run Variation, Lasso DML}
     \end{subfigure}
     \begin{subfigure}[b]{0.4\textwidth}
         \centering
         \includegraphics[width=\textwidth]{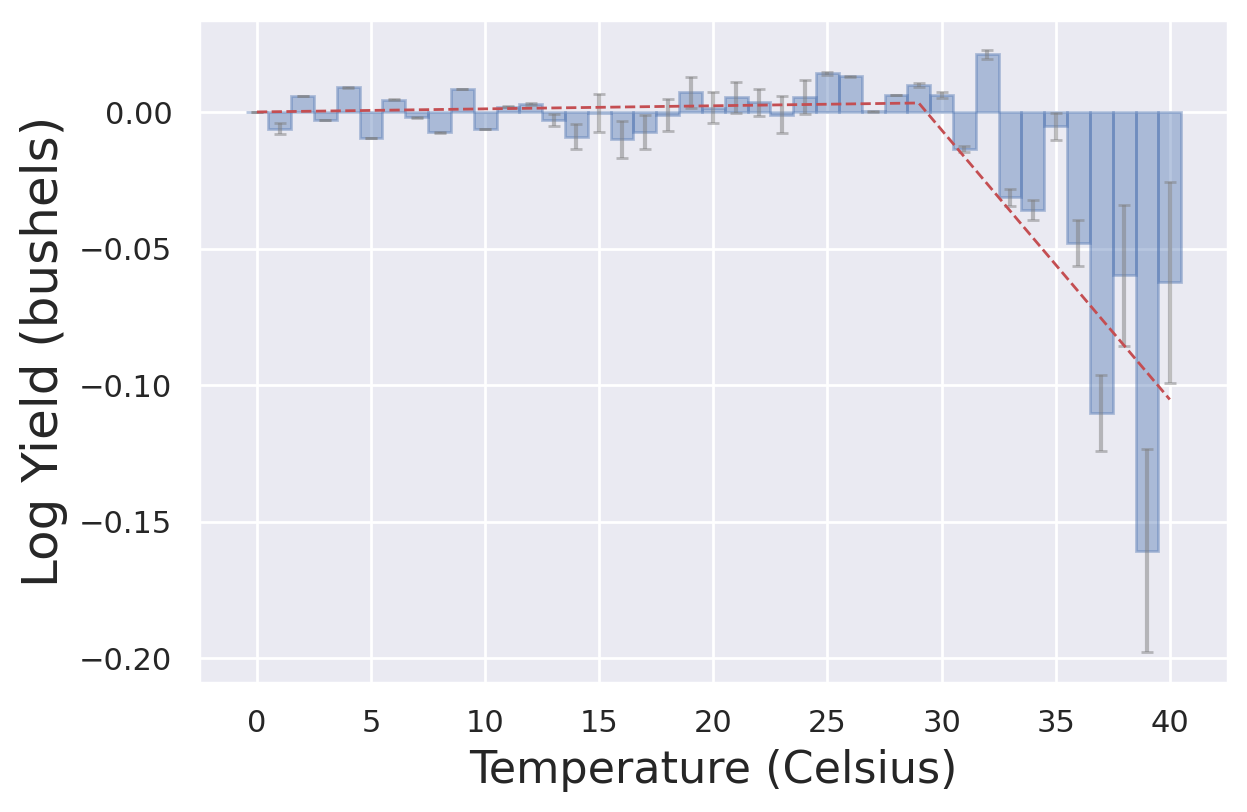}
         \caption{Soy, Short-Run Variation, Lasso DML}
     \end{subfigure}
     % \hfill 
        \caption[Lasso DML Estimates for Yearly Flexible]{Comparison of relationship between temperature and yields with short-run and long-run variation, using Lasso DML. 
        % Each observation is weighted by acres of crop planted. 
}
\end{figure}

\begin{figure}[ht]
     \centering
     % \hfill 
      \begin{subfigure}[b]{0.4\textwidth}
         \centering
         \includegraphics[width=\textwidth]{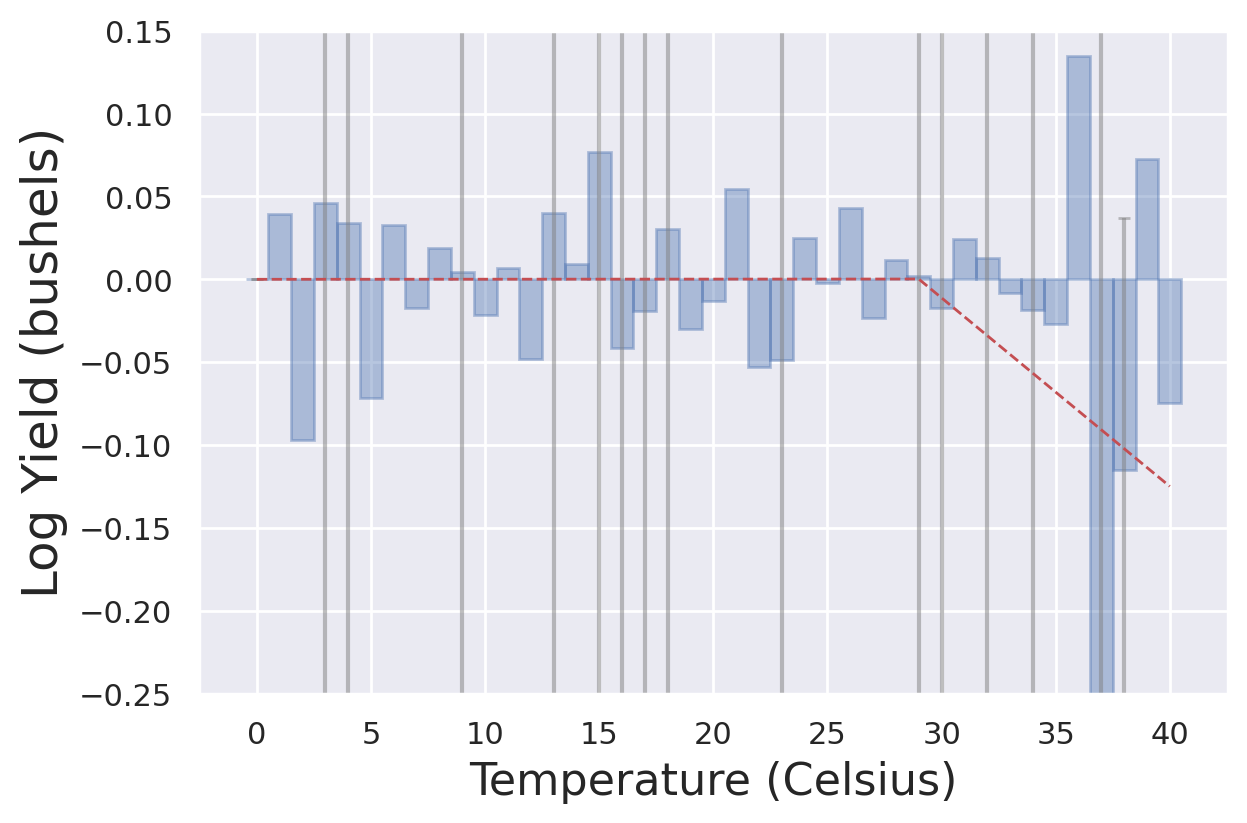}
         \caption{Corn, Long-Run Variation, OLS Poly}
     \end{subfigure}
     \begin{subfigure}[b]{0.4\textwidth}
         \centering
         \includegraphics[width=\textwidth]{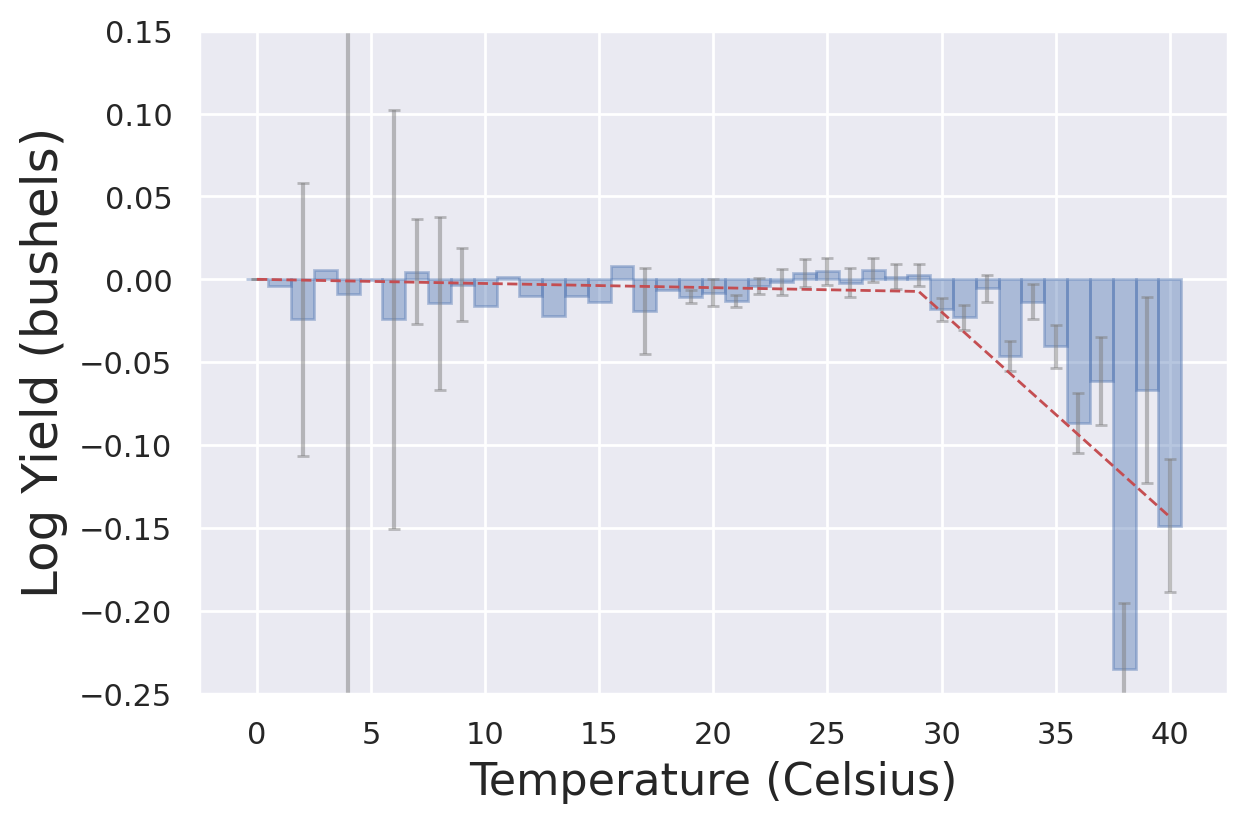}
         \caption{Corn, Short-Run Variation, Lasso  DML}
     \end{subfigure}
      \begin{subfigure}[b]{0.4\textwidth}
         \centering
         \includegraphics[width=\textwidth]{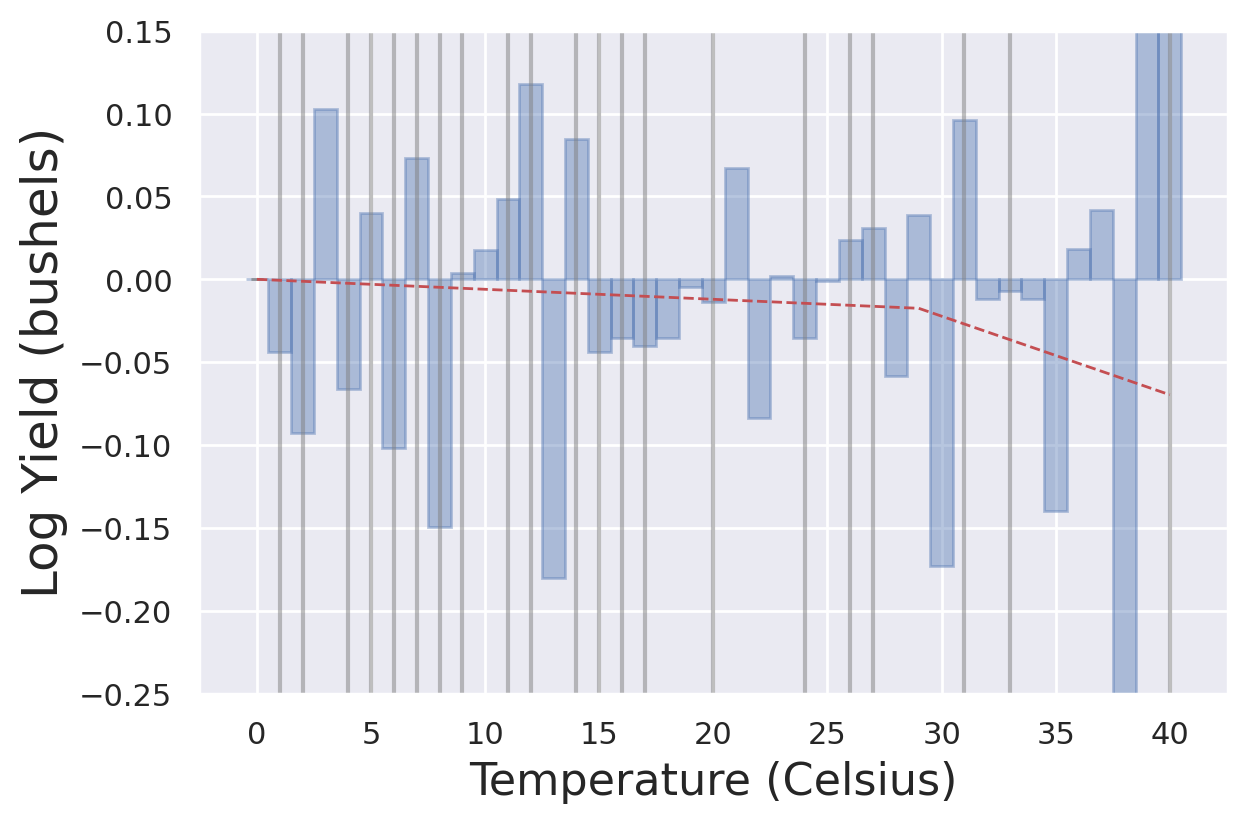}
         \caption{Soy, Long-Run Variation, OLS Poly}
     \end{subfigure}
     \begin{subfigure}[b]{0.4\textwidth}
         \centering
         \includegraphics[width=\textwidth]{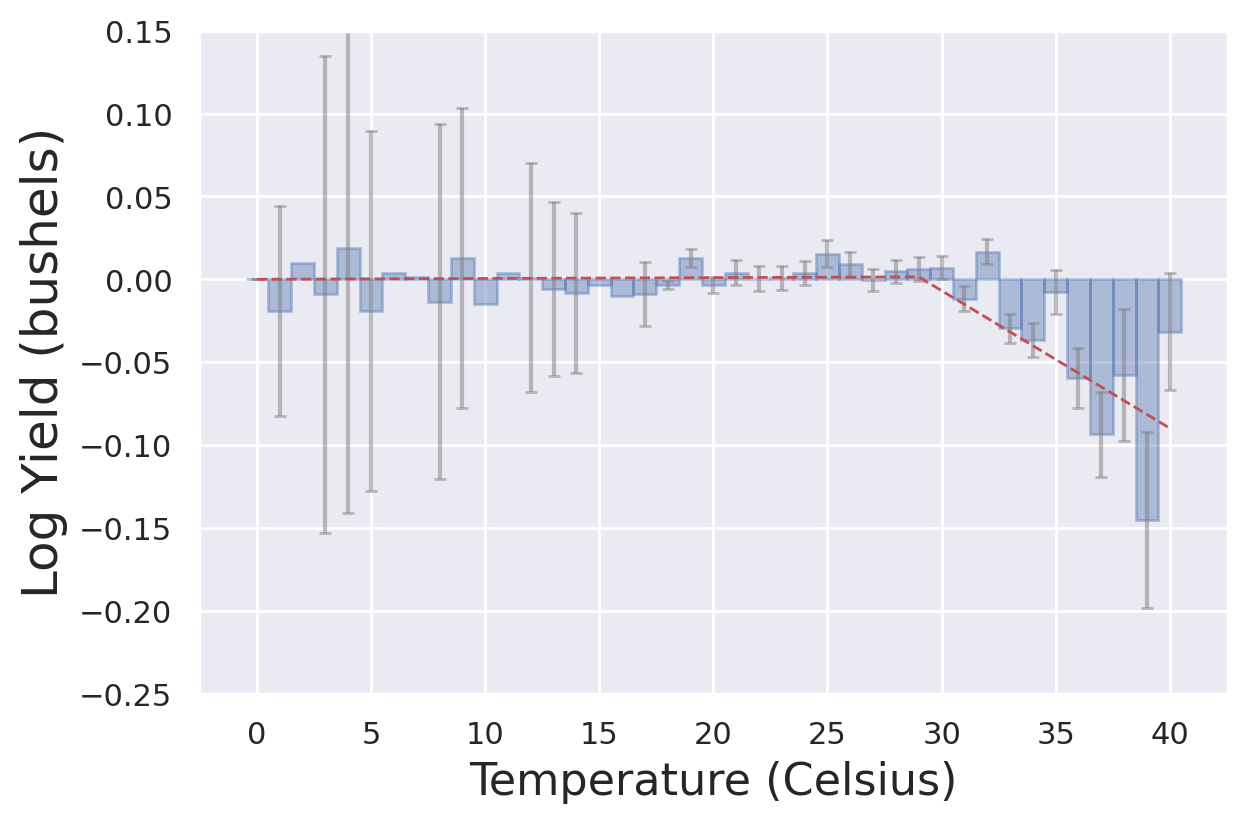}
         \caption{Soy, Short-Run Variation, OLS Poly}
     \end{subfigure}
     % \hfill 
        \caption[OLS Poly Estimates for Yearly Flexible]{Comparison of relationship between temperature and yields with short-run and long-run variation, using OLS Poly. 
        % Each observation is weighted by acres of crop planted. 
}
\end{figure}

\begin{figure}[ht]
     \centering
     % \hfill 
      \begin{subfigure}[b]{0.4\textwidth}
         \centering
         \includegraphics[width=\textwidth]{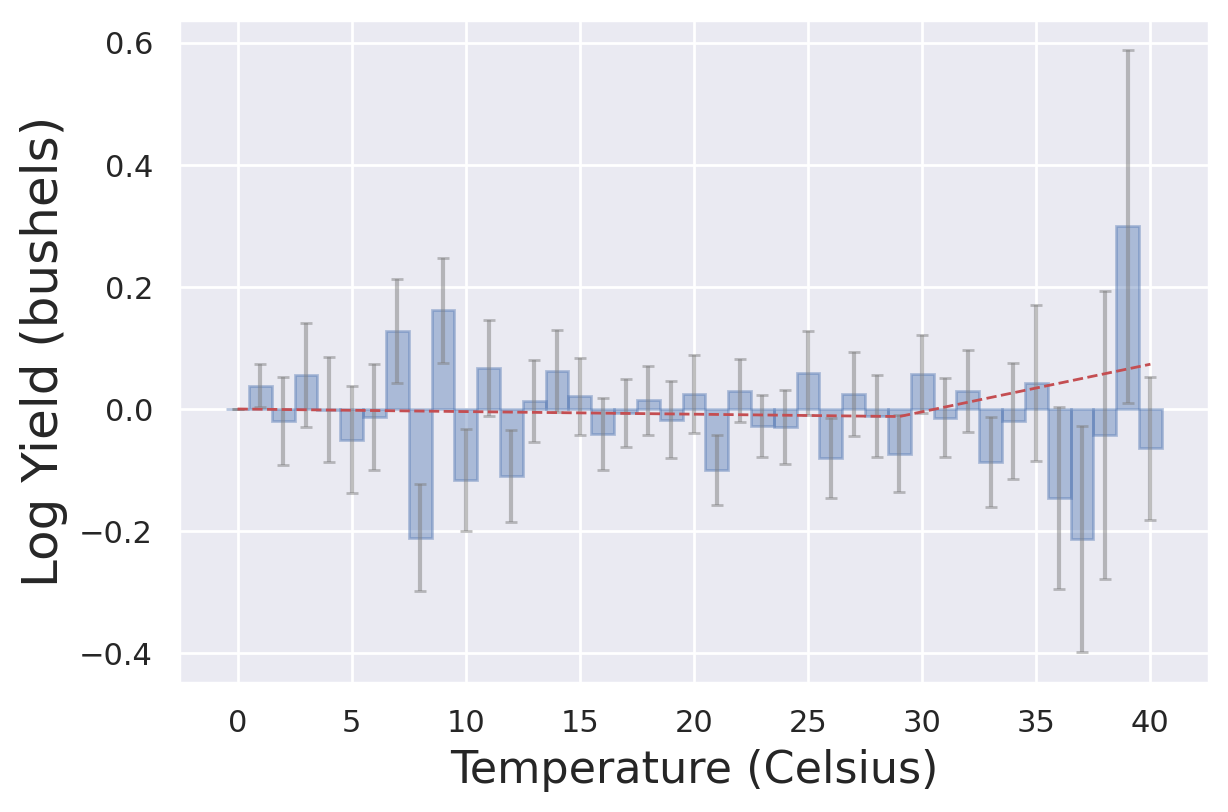}
         \caption{Corn, Long-Run Variation, OLS Linear}
     \end{subfigure}
     \begin{subfigure}[b]{0.4\textwidth}
         \centering
         \includegraphics[width=\textwidth]{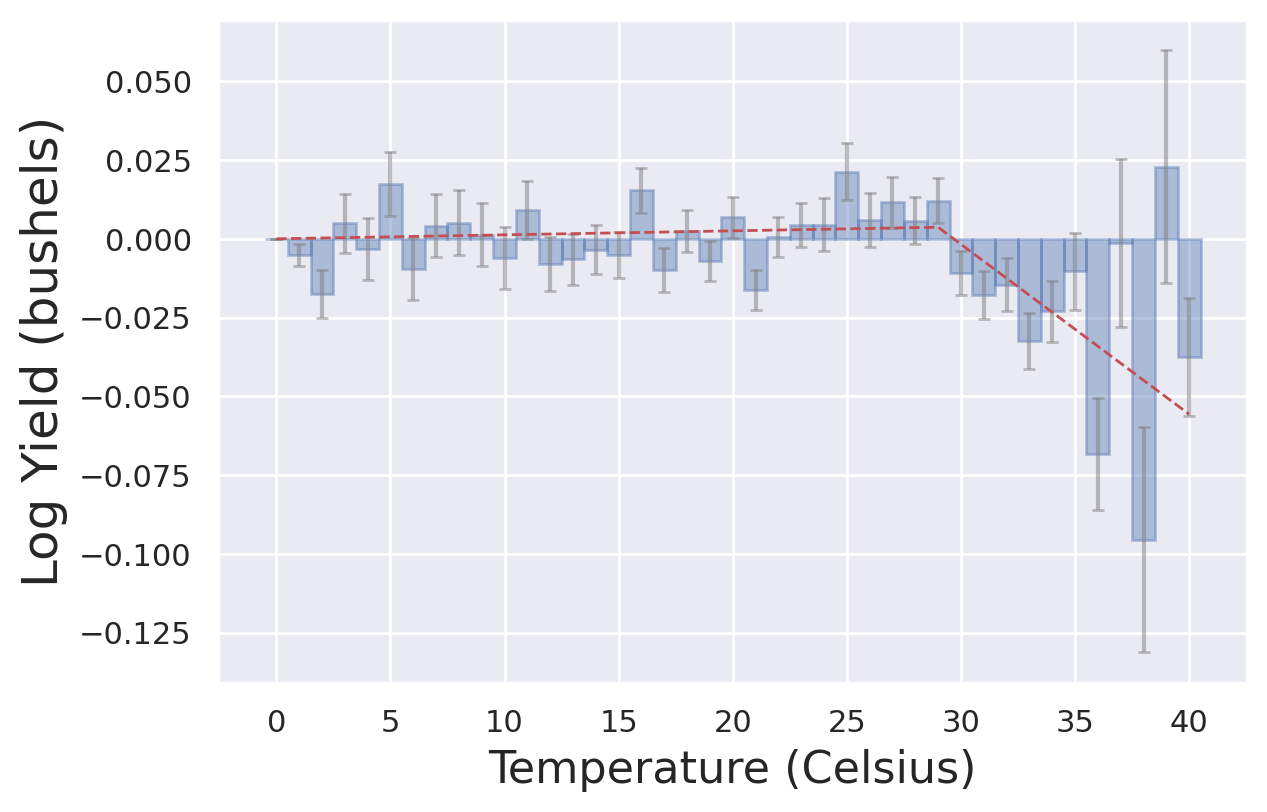}
         \caption{Corn, Short-Run Variation, Lasso  DML}
     \end{subfigure}
      \begin{subfigure}[b]{0.4\textwidth}
         \centering
         \includegraphics[width=\textwidth]{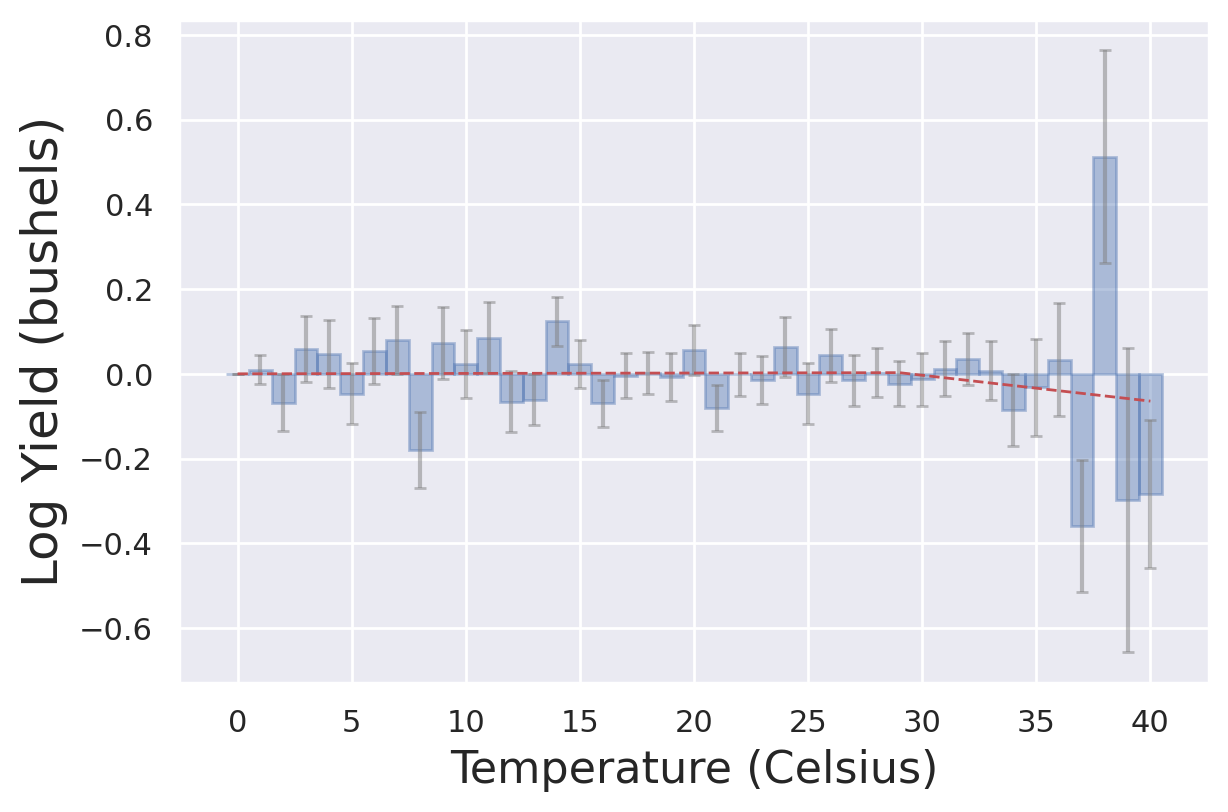}
         \caption{Soy, Long-Run Variation, OLS Linear}
     \end{subfigure}
     \begin{subfigure}[b]{0.4\textwidth}
         \centering
         \includegraphics[width=\textwidth]{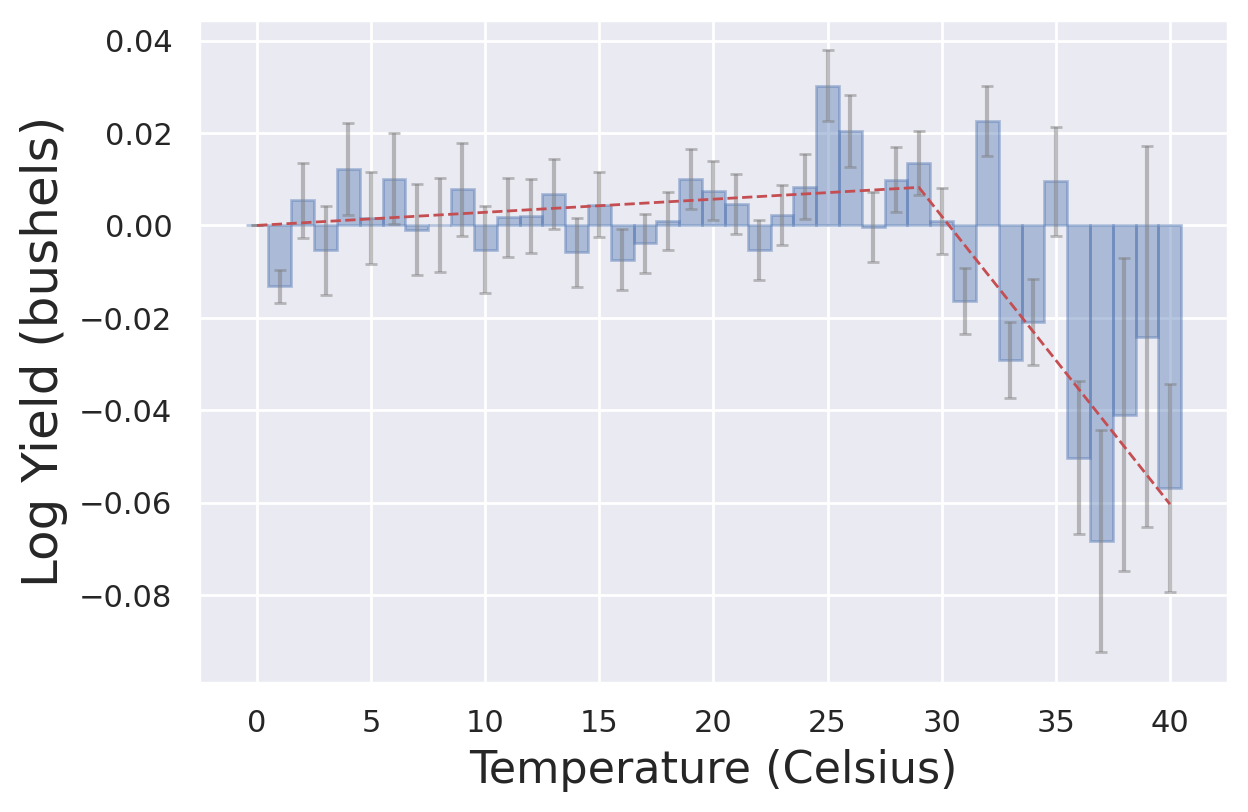}
         \caption{Soy, Short-Run Variation, OLS Linear}
     \end{subfigure}
     % \hfill 
        \caption[OLS Linear Estimates for Yearly Flexible]{Comparison of relationship between temperature and yields with short-run and long-run variation, using OLS Linear. 
        % Each observation is weighted by acres of crop planted. 
}
\end{figure}

\FloatBarrier
\subsection{Comparisons Over Time }
In this section, I provide OLS estimates of Yearly Flexible weather variables using short-run and long-run variation from different historical periods. 
In each figure, I compare short-run estimates using a panel dataset all years in the 30-year range, and long-run estimates comparing average weather and crop yield values from the first and last decades in that range. 

Each figure contains four separate plots. 
Each plot summarizes the coefficient estimates from OLS results with the Yearly Flexible set of weather variables. 
The bar plot shows the expected change in yields from exposure to a single degree day at the given temperature level.
Error bars show one standard deviation above/below each estimated coefficient.  
The line shows the best piecewise linear fit to the bar plot, with one piece from 0-29\textdegree C and another from 29-40 \textdegree C. 
Regression is taken after the within transformation, and also includes yearly fixed effects terms and precipitation; these terms are omitted from the figures. 

\begin{figure}[ht]
     \centering
     % \hfill 
      \begin{subfigure}[b]{0.4\textwidth}
         \centering
         \includegraphics[width=\textwidth]{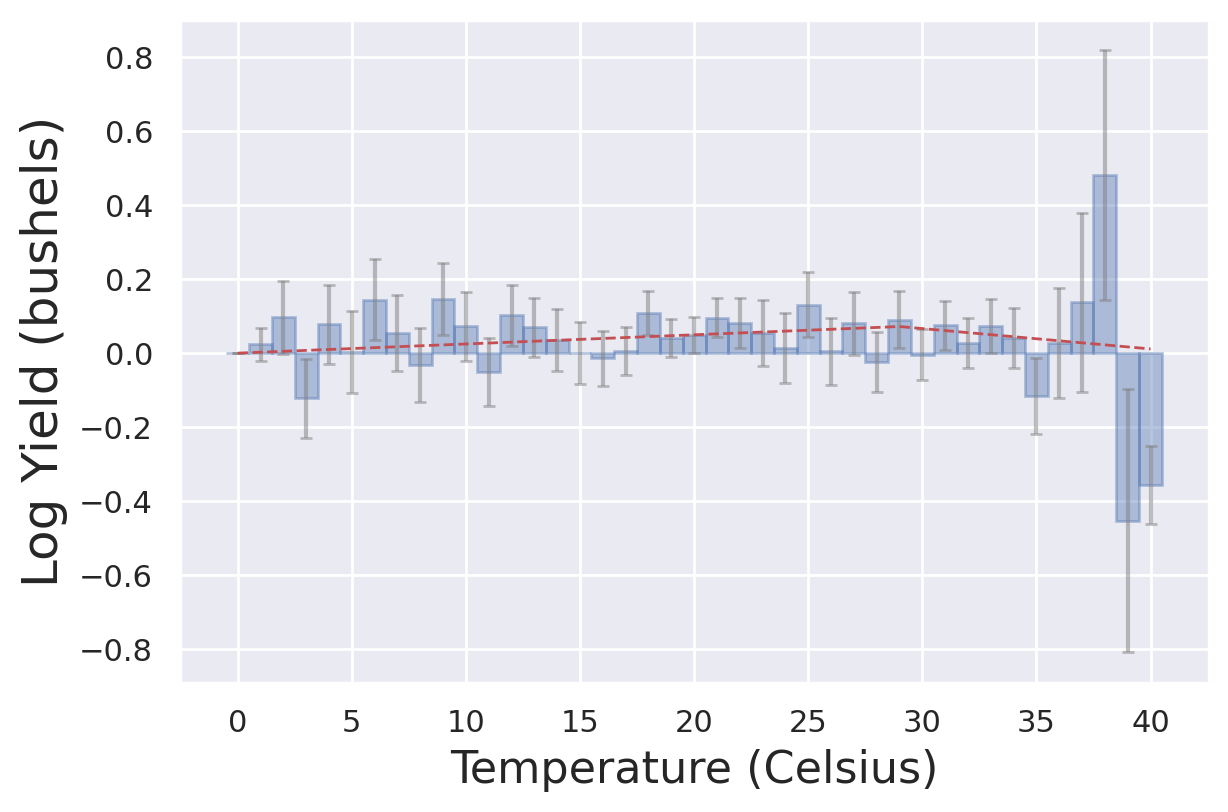}
         \caption{Corn, Long-Run Variation}
     \end{subfigure}
     \begin{subfigure}[b]{0.4\textwidth}
         \centering
         \includegraphics[width=\textwidth]{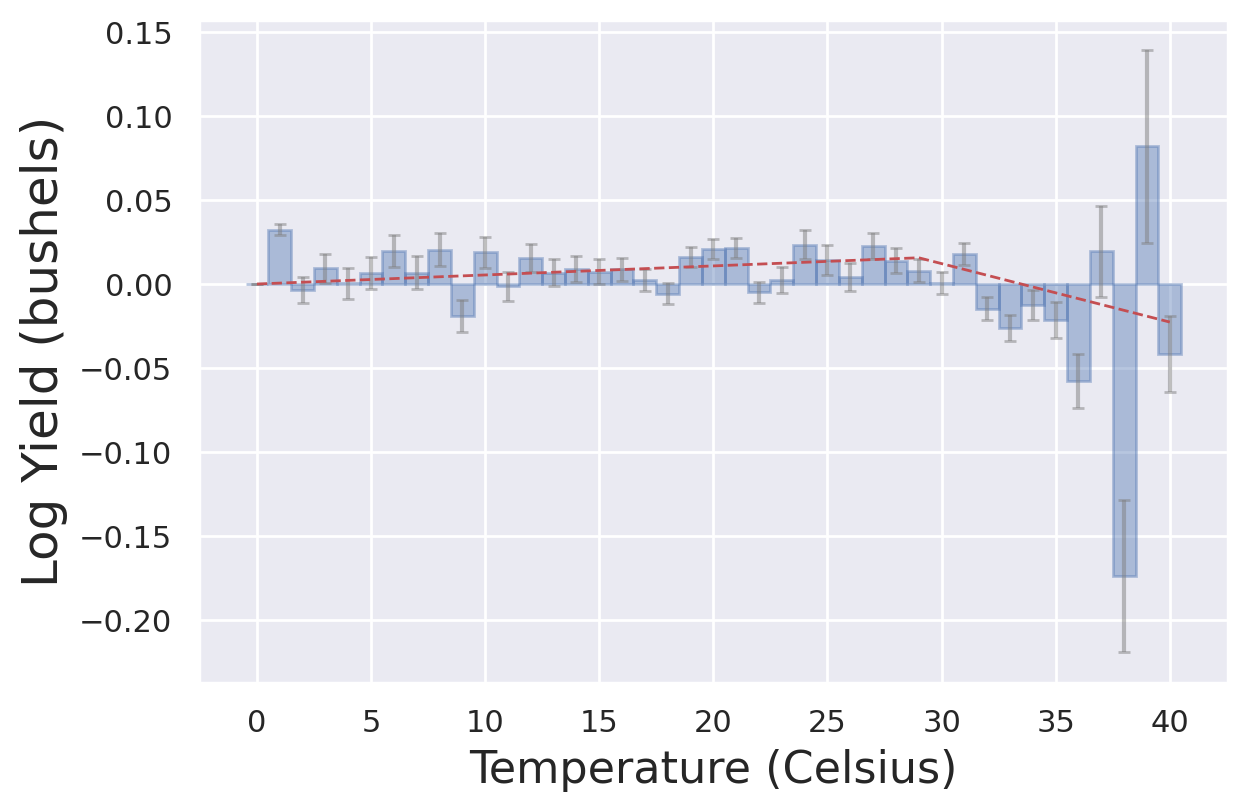}
         \caption{Corn, Short-Run Variation }
     \end{subfigure}
      \begin{subfigure}[b]{0.4\textwidth}
         \centering
         \includegraphics[width=\textwidth]{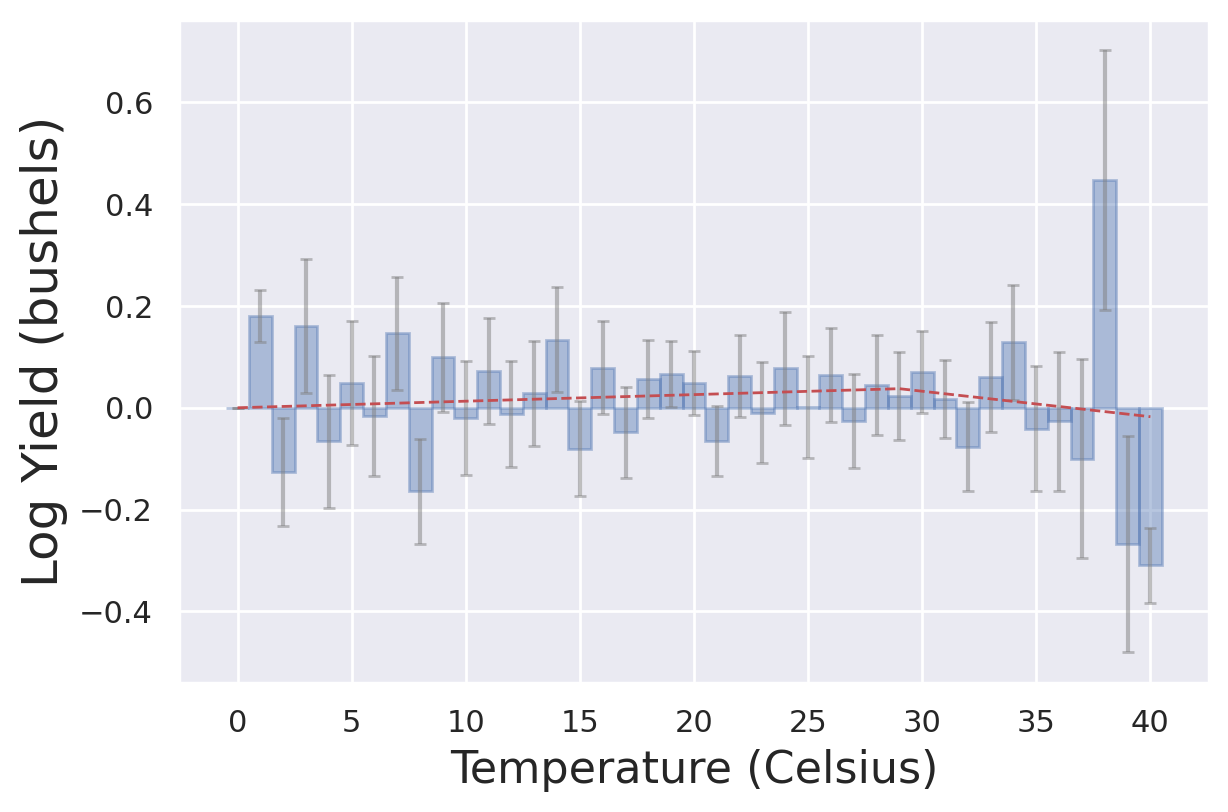}
         \caption{Soy, Long-Run Variation}
     \end{subfigure}
     \begin{subfigure}[b]{0.4\textwidth}
         \centering
         \includegraphics[width=\textwidth]{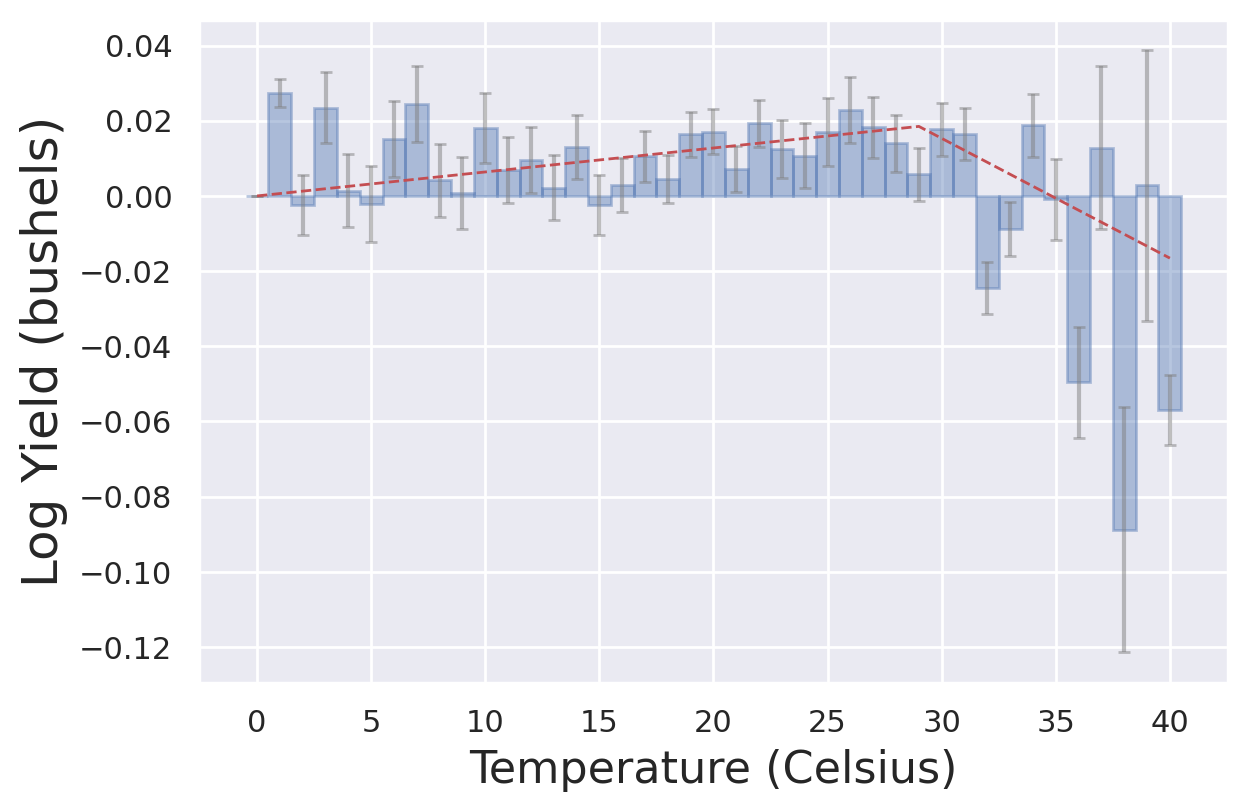}
         \caption{Soy, Short-Run Variation }
     \end{subfigure}
     % \hfill 
        \caption[OLS Estimates for Yearly Flexible, starting in 1950]{Comparison of relationship between temperature and yields with short-run and long-run variation, from 1950-1979. 
}
% \label{fig:OLS Estimates for Yearly Flexible}
\end{figure}

\begin{figure}[ht]
     \centering
     % \hfill 
      \begin{subfigure}[b]{0.4\textwidth}
         \centering
         \includegraphics[width=\textwidth]{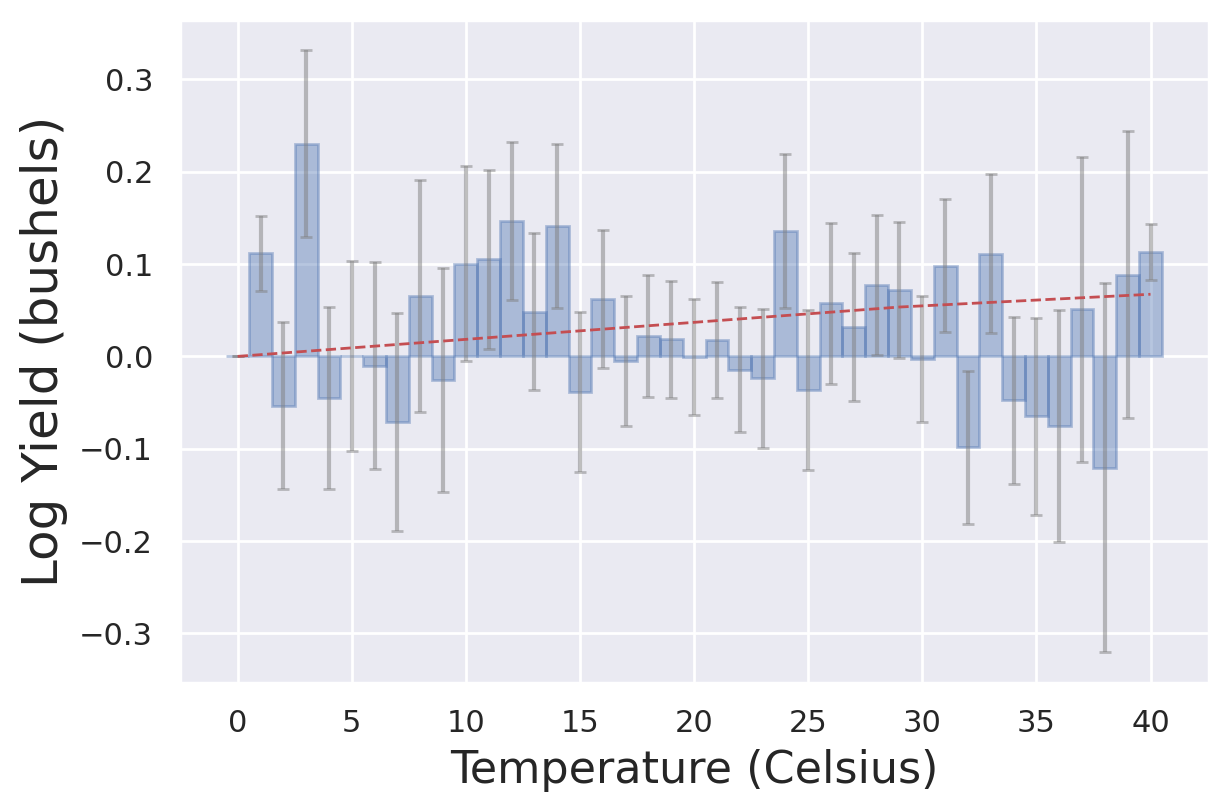}
         \caption{Corn, Long-Run Variation}
     \end{subfigure}
     \begin{subfigure}[b]{0.4\textwidth}
         \centering
         \includegraphics[width=\textwidth]{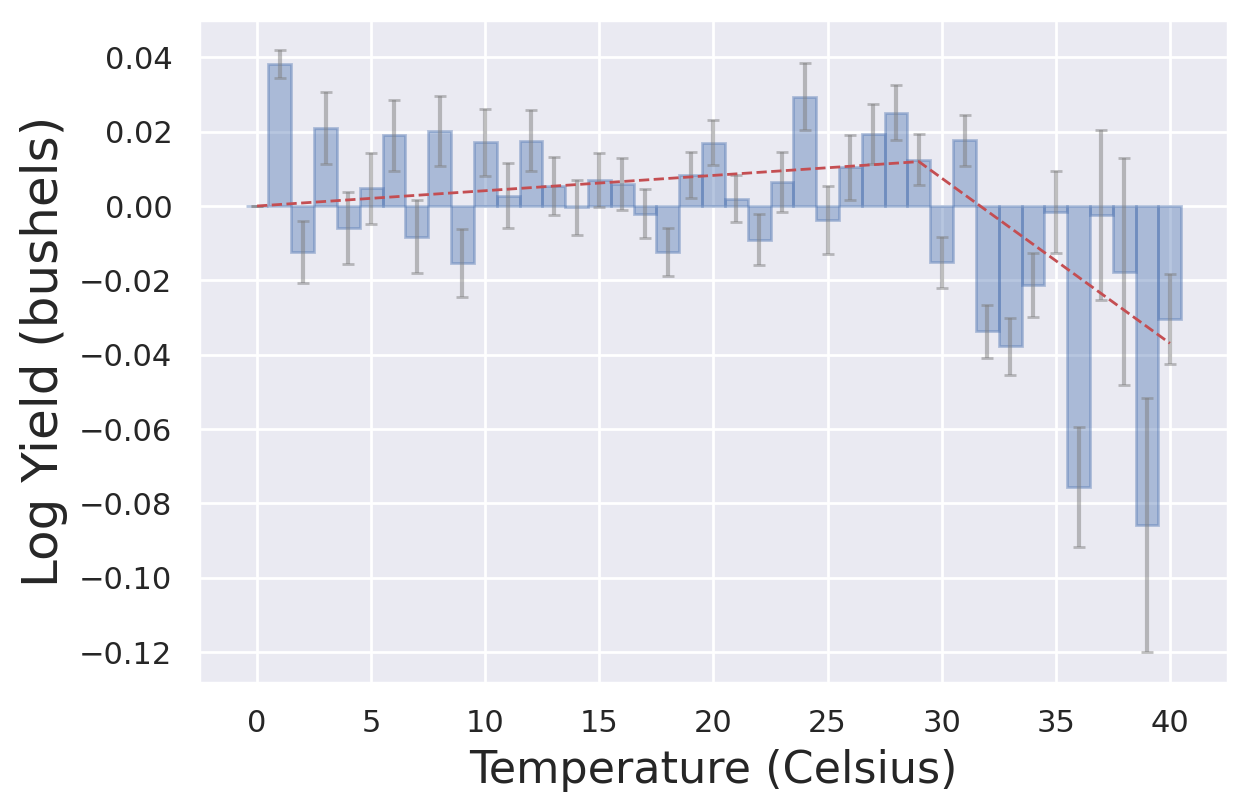}
         \caption{Corn, Short-Run Variation }
     \end{subfigure}
      \begin{subfigure}[b]{0.4\textwidth}
         \centering
         \includegraphics[width=\textwidth]{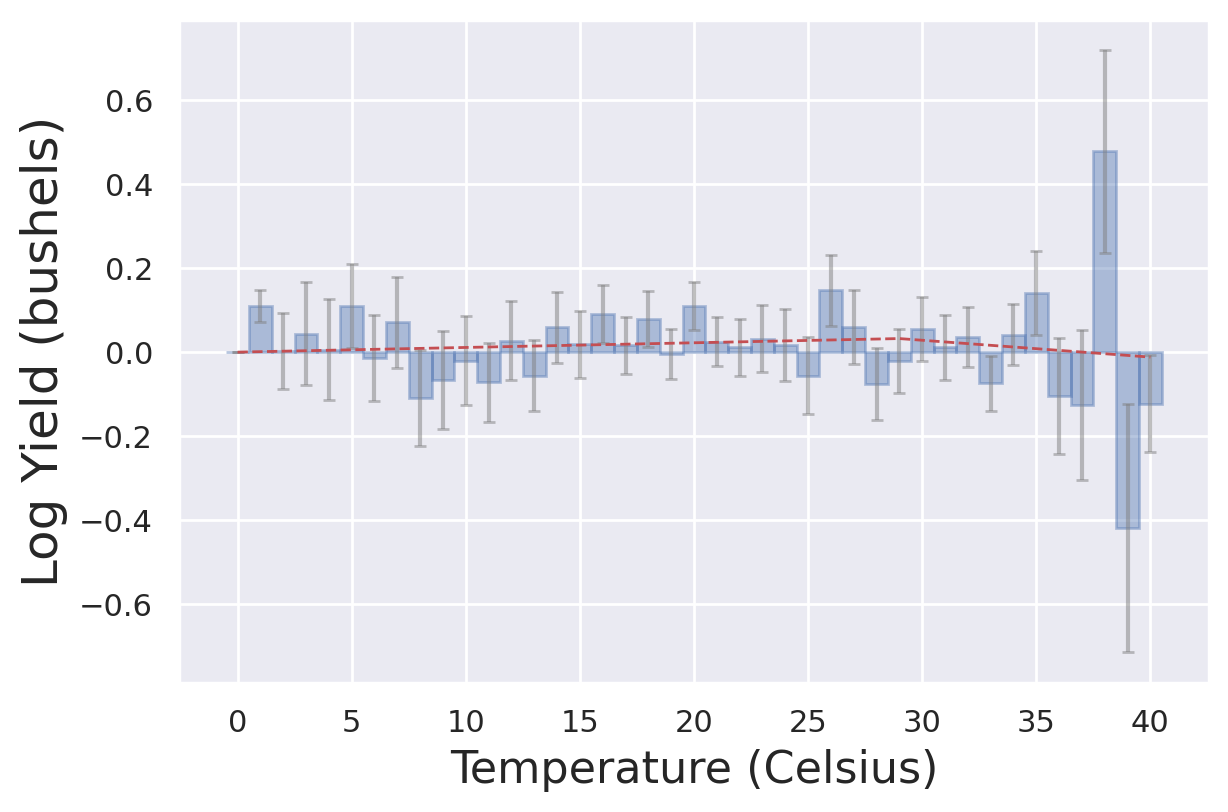}
         \caption{Soy, Long-Run Variation}
     \end{subfigure}
     \begin{subfigure}[b]{0.4\textwidth}
         \centering
         \includegraphics[width=\textwidth]{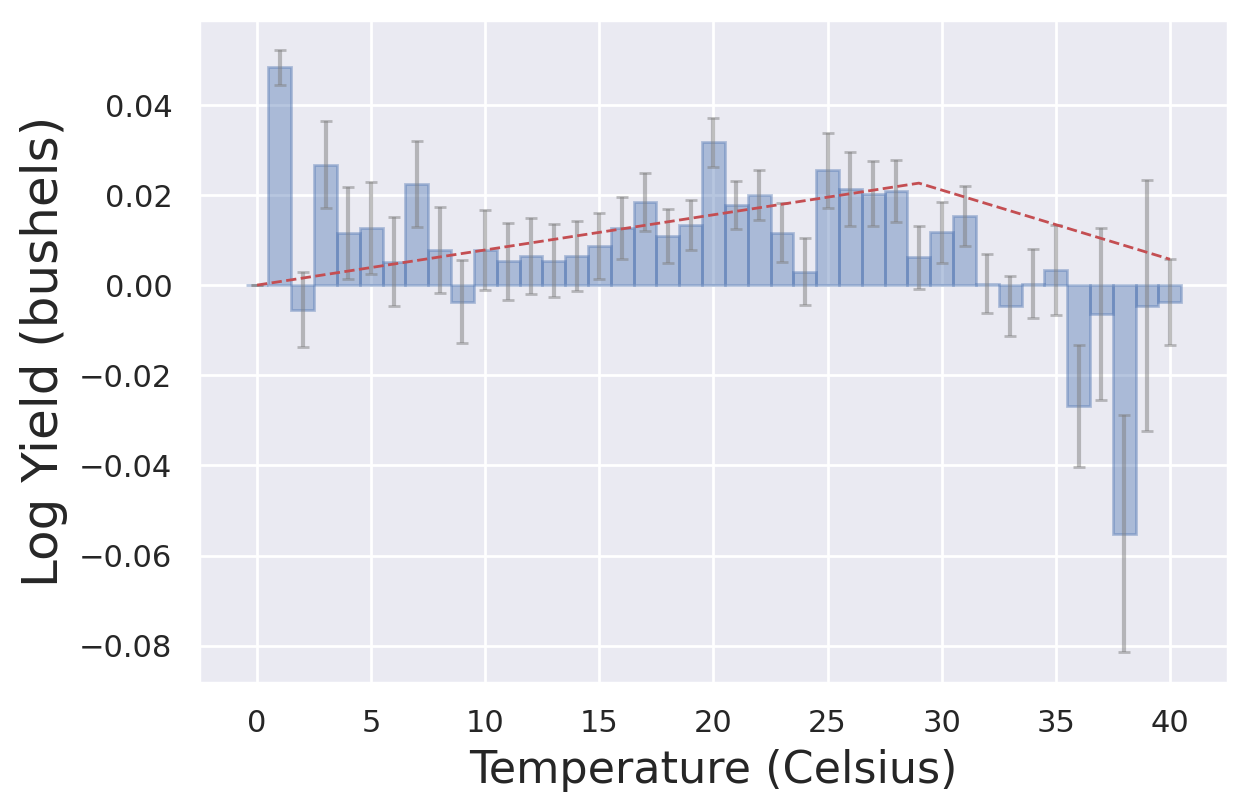}
         \caption{Soy, Short-Run Variation }
     \end{subfigure}
     % \hfill 
        \caption[OLS Estimates for Yearly Flexible, starting in 1960]{Comparison of relationship between temperature and yields with short-run and long-run variation, from 1960-1989. 
}
% \label{fig:OLS Estimates for Yearly Flexible}
\end{figure}

\begin{figure}[ht]
     \centering
     % \hfill 
      \begin{subfigure}[b]{0.4\textwidth}
         \centering
         \includegraphics[width=\textwidth]{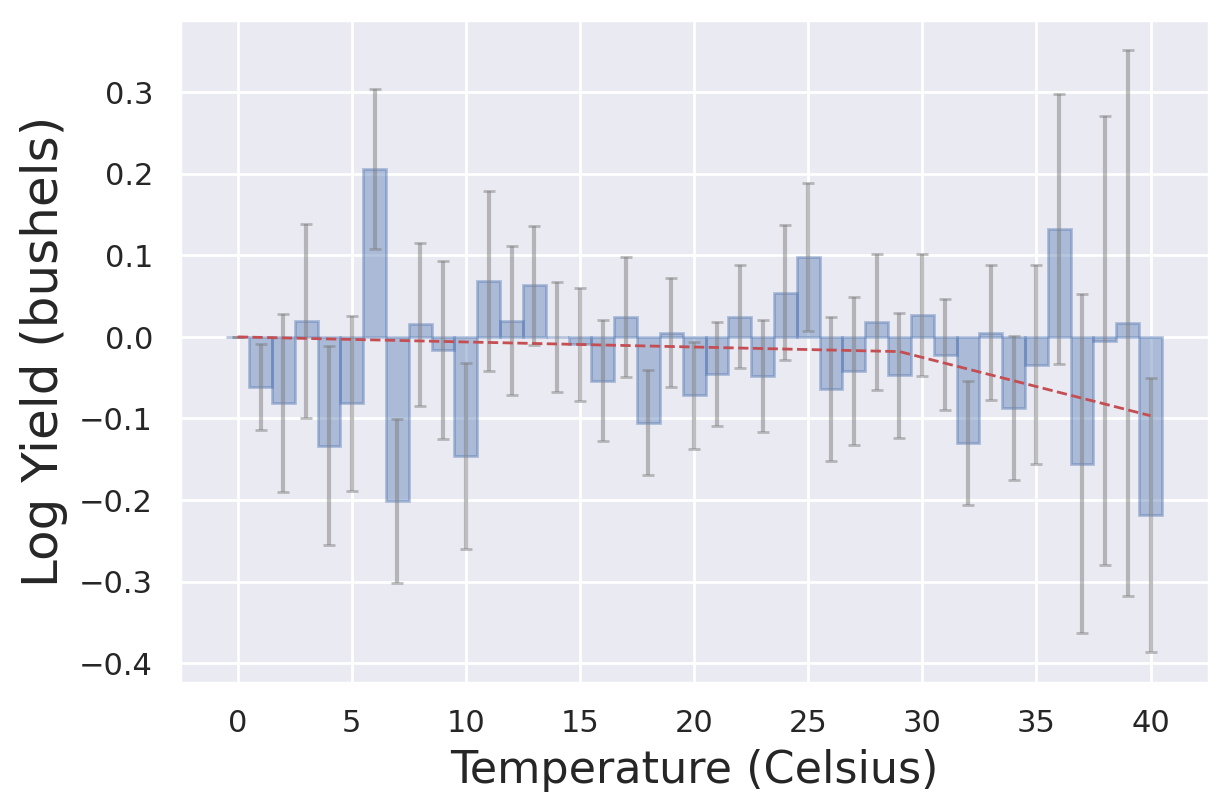}
         \caption{Corn, Long-Run Variation}
     \end{subfigure}
     \begin{subfigure}[b]{0.4\textwidth}
         \centering
         \includegraphics[width=\textwidth]{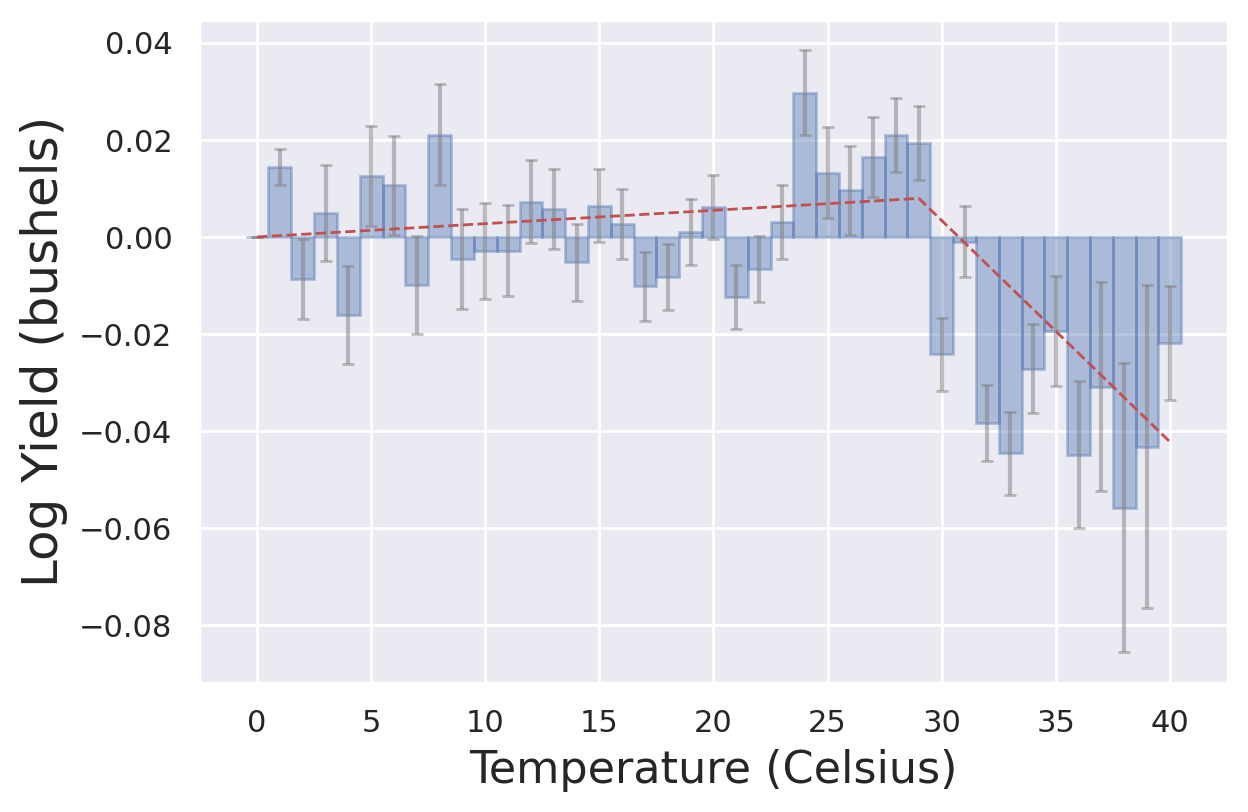}
         \caption{Corn, Short-Run Variation }
     \end{subfigure}
      \begin{subfigure}[b]{0.4\textwidth}
         \centering
         \includegraphics[width=\textwidth]{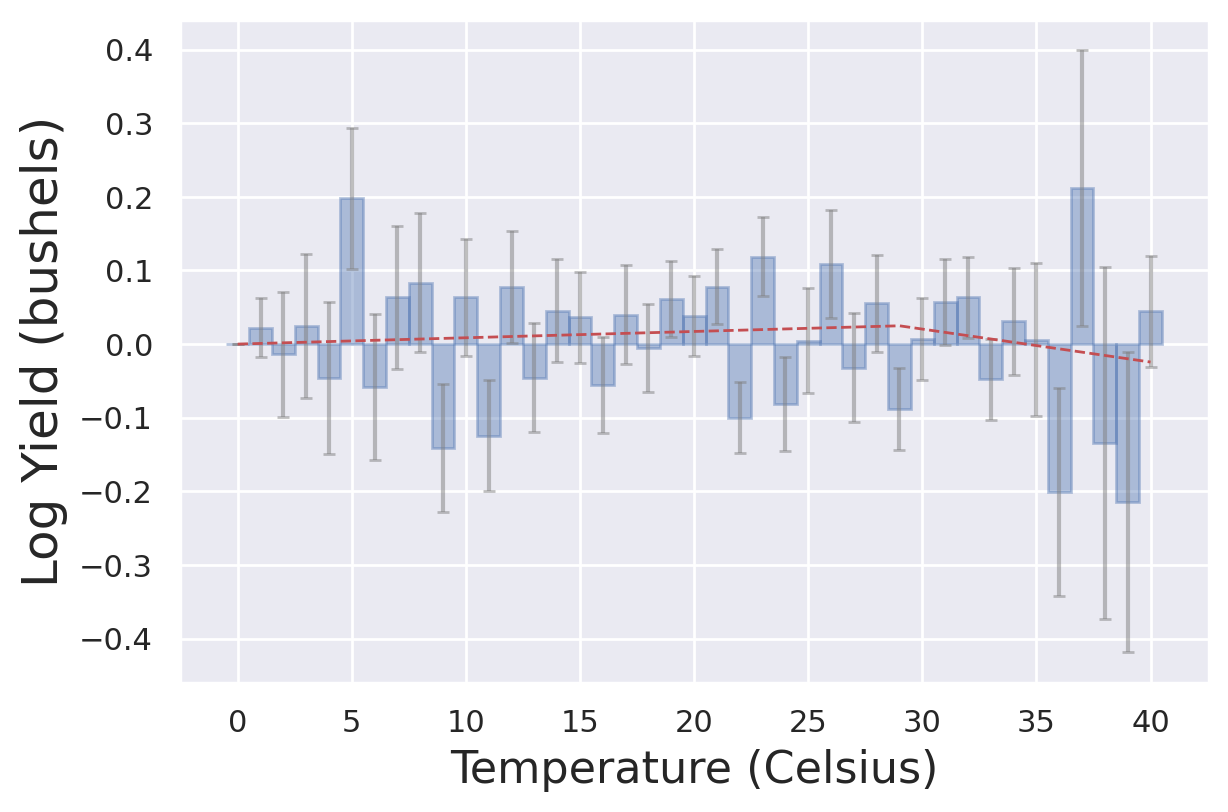}
         \caption{Soy, Long-Run Variation}
     \end{subfigure}
     \begin{subfigure}[b]{0.4\textwidth}
         \centering
         \includegraphics[width=\textwidth]{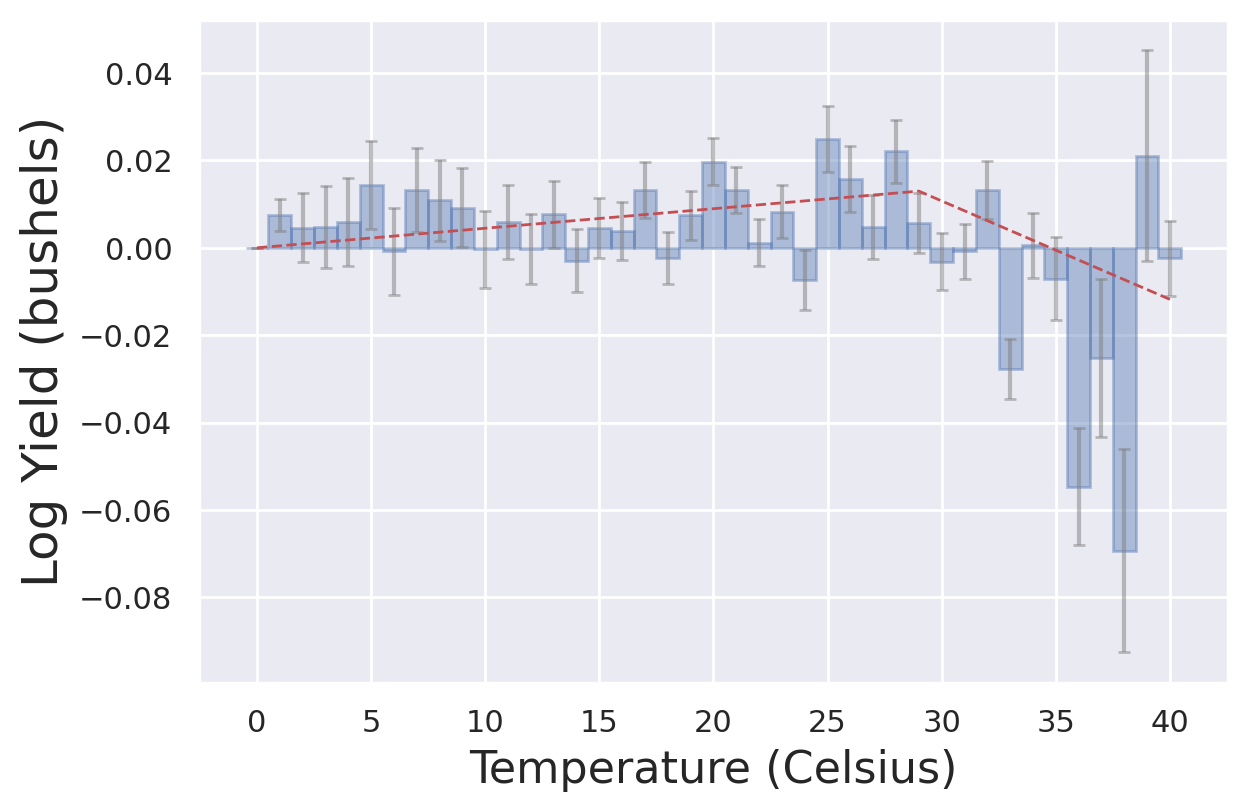}
         \caption{Soy, Short-Run Variation }
     \end{subfigure}
     % \hfill 
        \caption[OLS Estimates for Yearly Flexible, starting in 1970]{Comparison of relationship between temperature and yields with short-run and long-run variation, from 1970-1999. 
}
% \label{fig:OLS Estimates for Yearly Flexible}
\end{figure}

\begin{figure}[ht]
     \centering
     % \hfill 
      \begin{subfigure}[b]{0.4\textwidth}
         \centering
         \includegraphics[width=\textwidth]{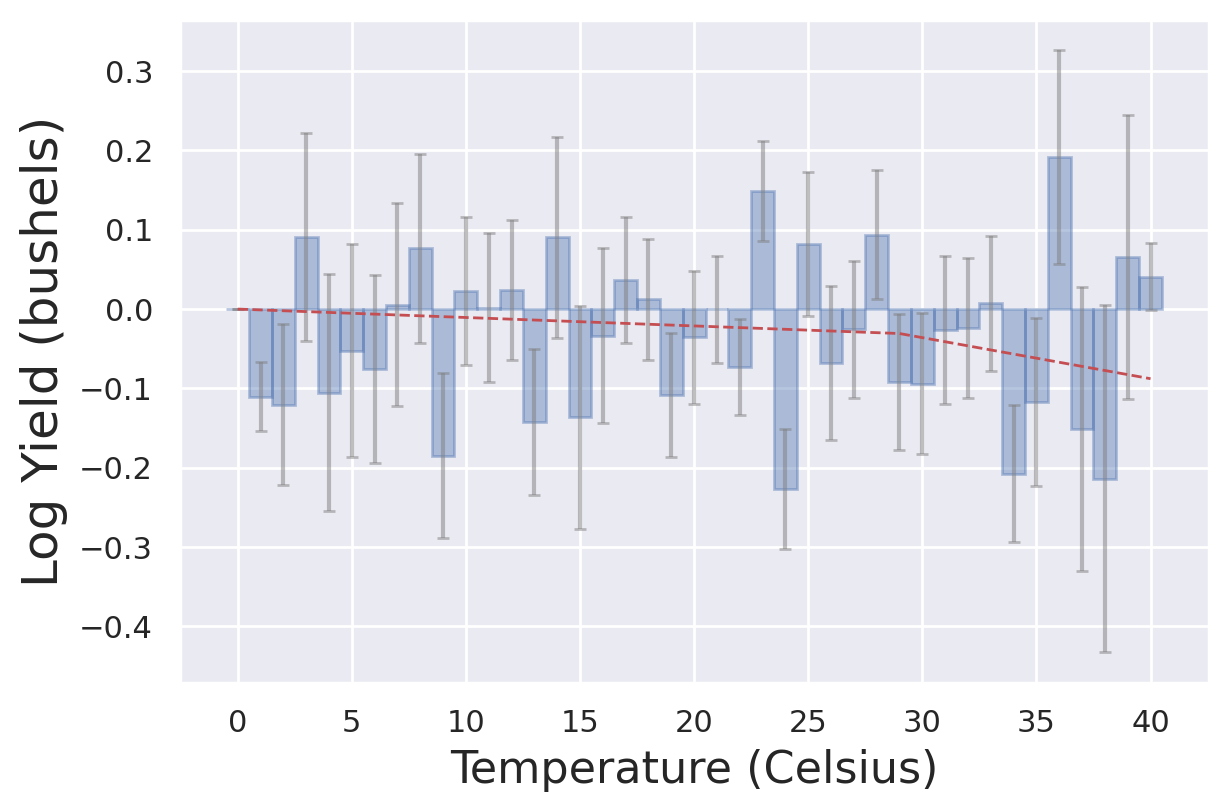}
         \caption{Corn, Long-Run Variation}
     \end{subfigure}
     \begin{subfigure}[b]{0.4\textwidth}
         \centering
         \includegraphics[width=\textwidth]{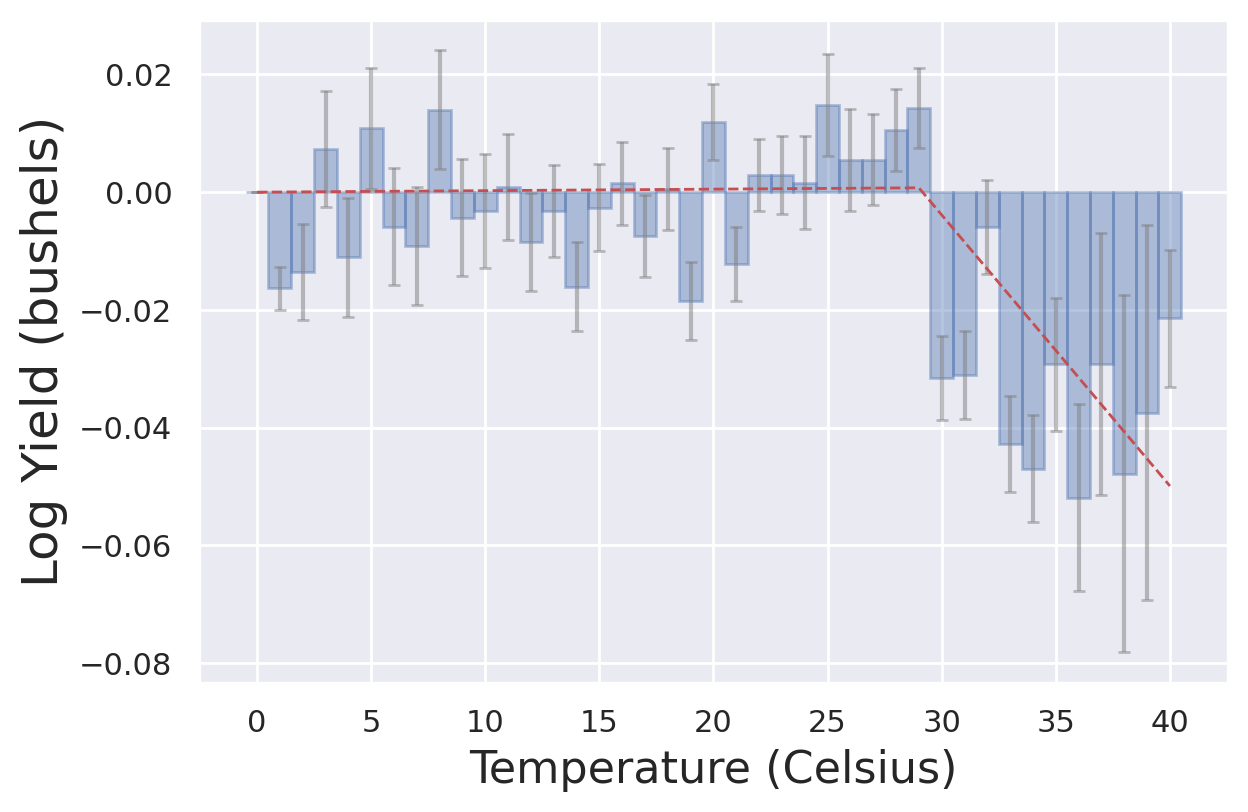}
         \caption{Corn, Short-Run Variation }
     \end{subfigure}
      \begin{subfigure}[b]{0.4\textwidth}
         \centering
         \includegraphics[width=\textwidth]{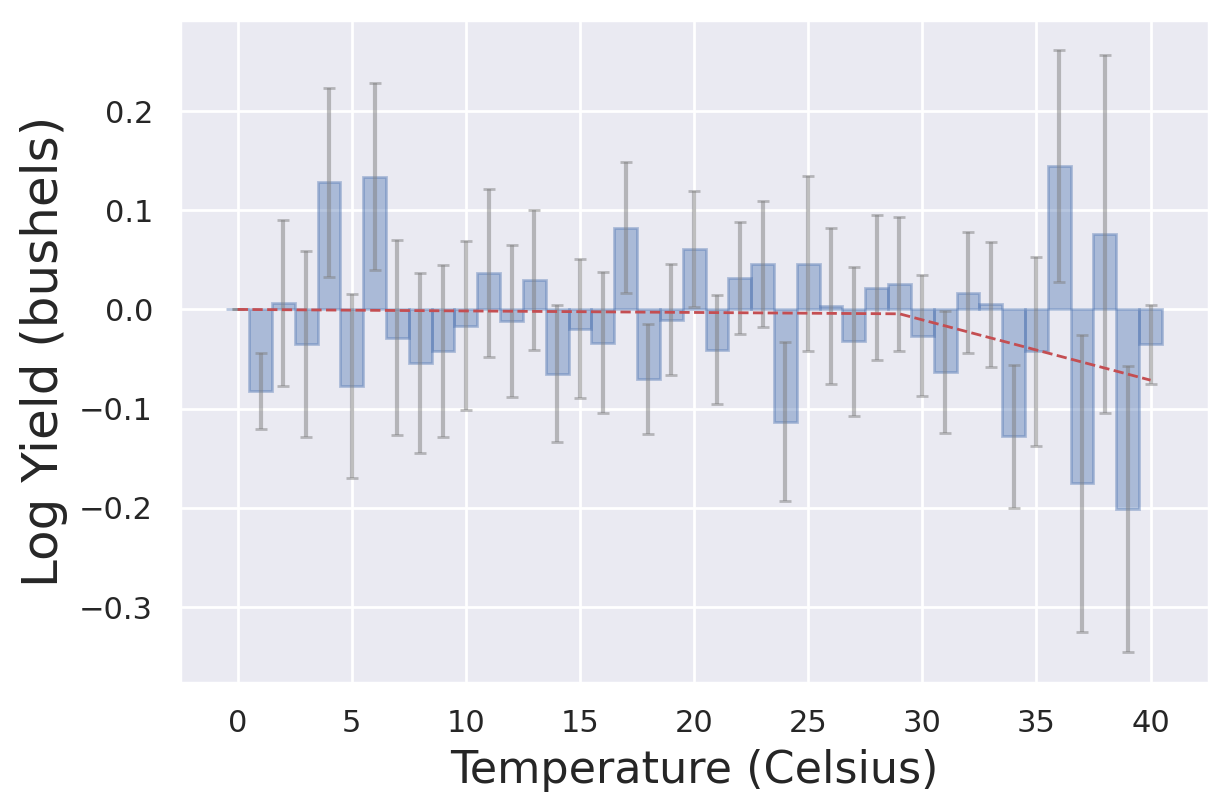}
         \caption{Soy, Long-Run Variation}
     \end{subfigure}
     \begin{subfigure}[b]{0.4\textwidth}
         \centering
         \includegraphics[width=\textwidth]{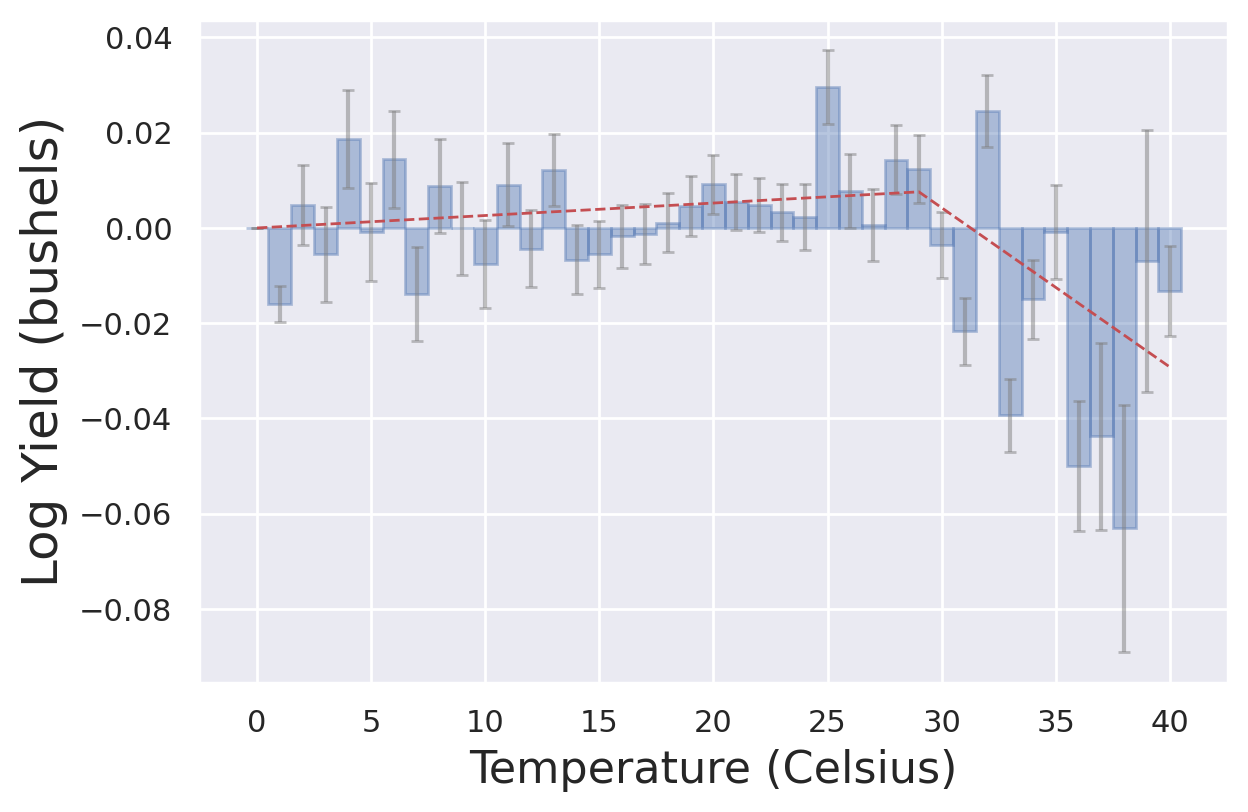}
         \caption{Soy, Short-Run Variation }
     \end{subfigure}
     % \hfill 
        \caption[OLS Estimates for Yearly Flexible, starting in 1980]{Comparison of relationship between temperature and yields with short-run and long-run variation, from 1980-2009. 
}
% \label{fig:OLS Estimates for Yearly Flexible}
\end{figure}

\begin{figure}[ht]
     \centering
     % \hfill 
      \begin{subfigure}[b]{0.4\textwidth}
         \centering
         \includegraphics[width=\textwidth]{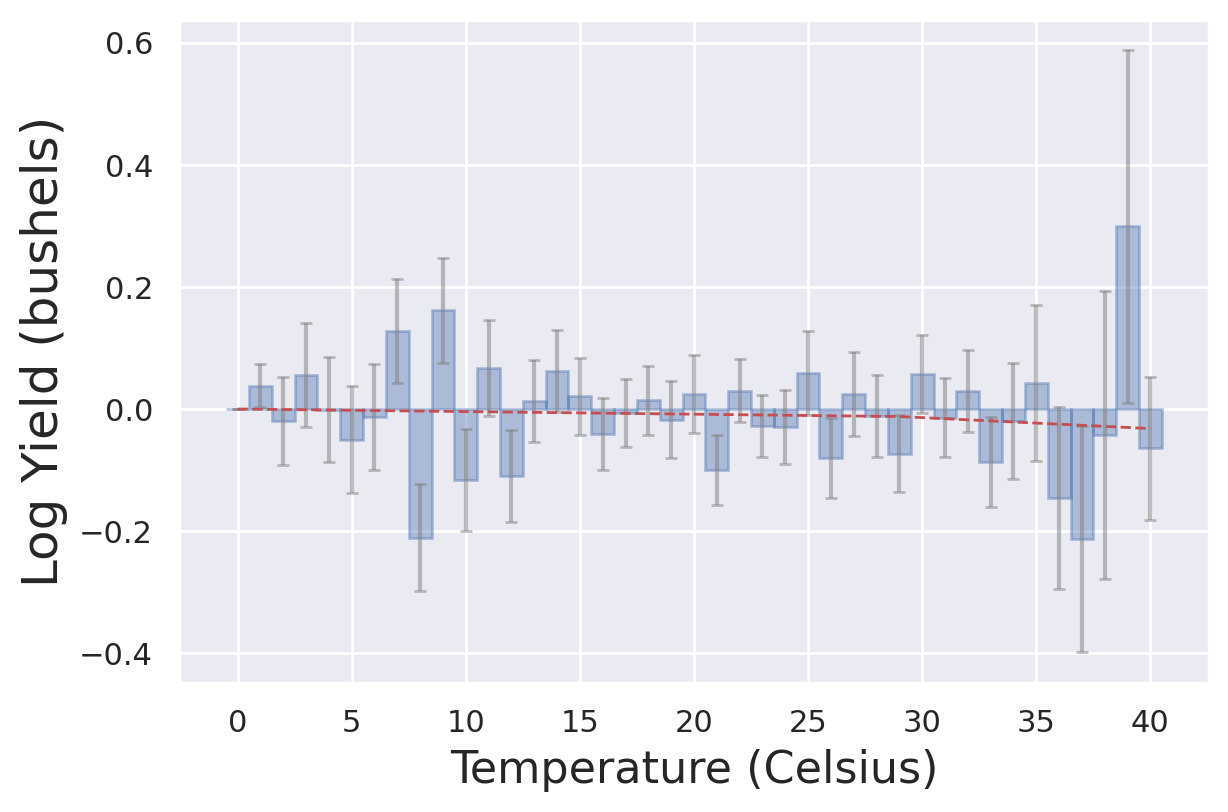}
         \caption{Corn, Long-Run Variation}
     \end{subfigure}
     \begin{subfigure}[b]{0.4\textwidth}
         \centering
         \includegraphics[width=\textwidth]{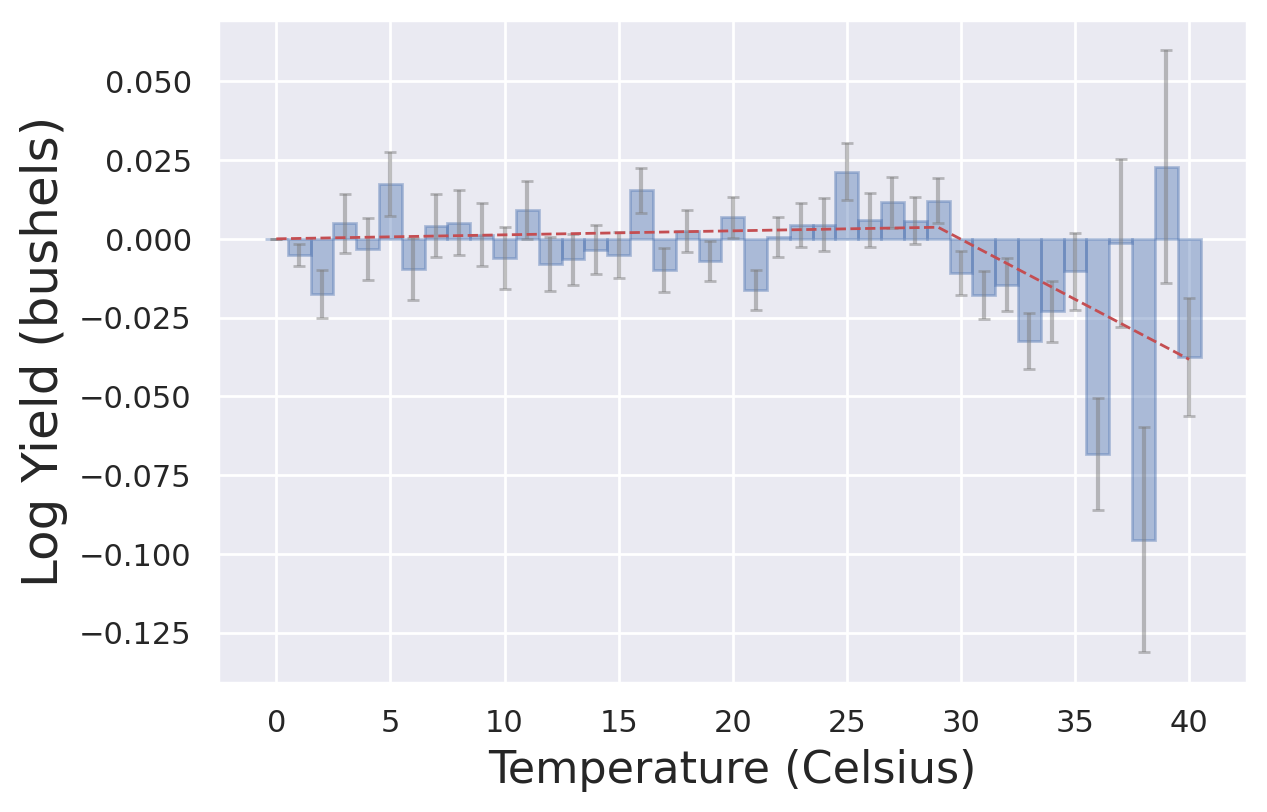}
         \caption{Corn, Short-Run Variation }
     \end{subfigure}
      \begin{subfigure}[b]{0.4\textwidth}
         \centering
         \includegraphics[width=\textwidth]{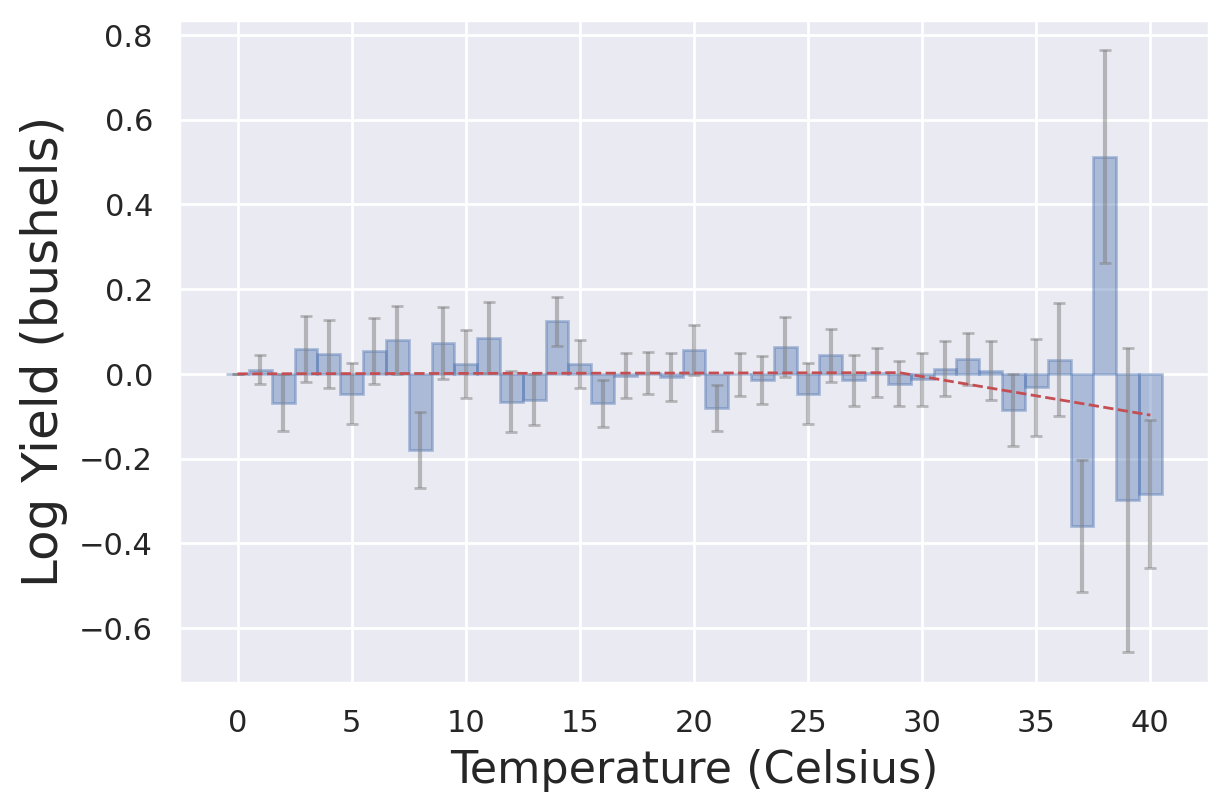}
         \caption{Soy, Long-Run Variation}
     \end{subfigure}
     \begin{subfigure}[b]{0.4\textwidth}
         \centering
         \includegraphics[width=\textwidth]{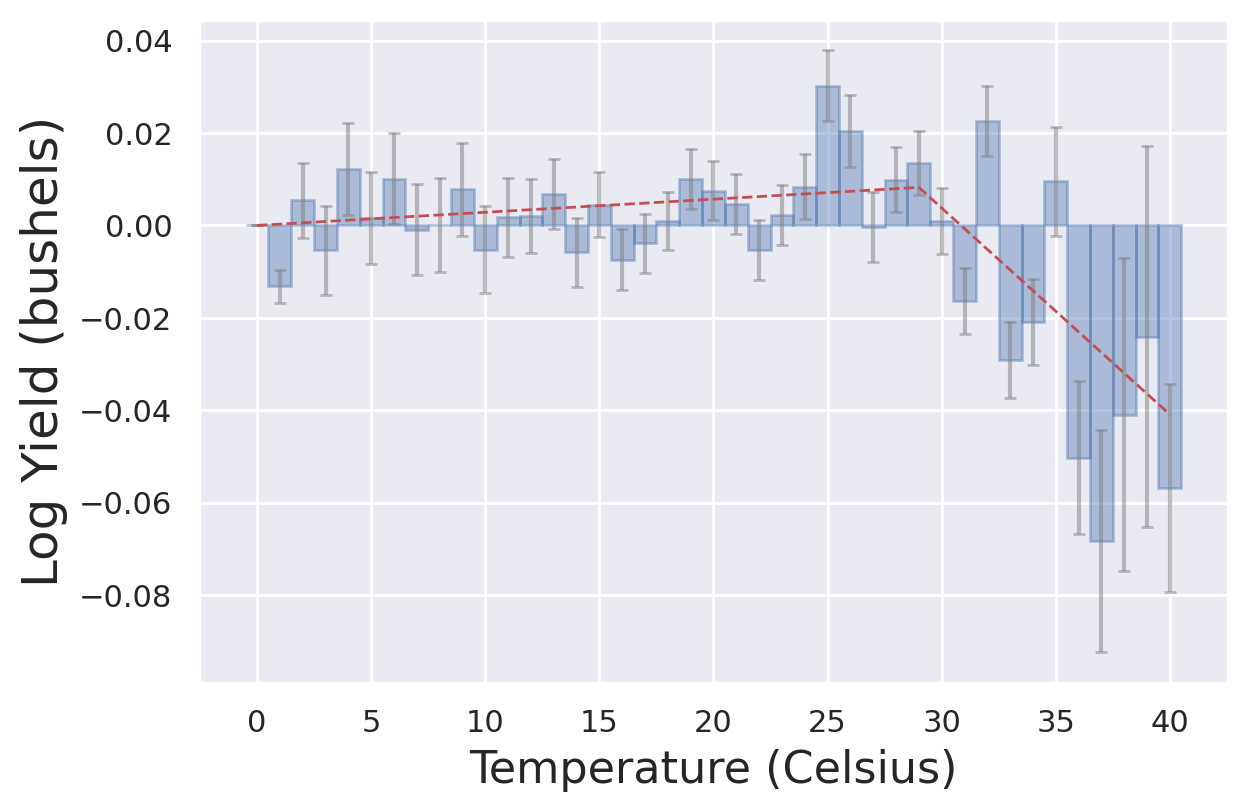}
         \caption{Soy, Short-Run Variation }
     \end{subfigure}
     % \hfill 
        \caption[OLS Estimates for Yearly Flexible, starting in 1990]{Comparison of relationship between temperature and yields with short-run and long-run variation, from 1990-2019. 
}
% \label{fig:OLS Estimates for Yearly Flexible}
\end{figure}

\end{document}